\documentclass[twocolumn]{aastex631}
\usepackage{gensymb}

\usepackage{amsmath}    
\usepackage{amssymb}    
\shorttitle{AASTeX v6.3.1 Sample article}
\shortauthors{Singh et al.}
\graphicspath{{./}{figures/}}

\begin{document}

\title{Accretion Funnel Reconfiguration during an Outburst in a Young Stellar Object: EX Lupi}

\author[0000-0002-7434-9681]{Koshvendra Singh}
\affiliation{Department of Astronomy and Astrophysics, Tata Institute of Fundamental Research, Mumbai 400005, India}
\author[0000-0001-8720-5612]{Joe P. Ninan}
\affiliation{Department of Astronomy and Astrophysics, Tata Institute of Fundamental Research, Mumbai 400005, India}
\author{Marina M. Romanova}
\affiliation{Department of Astronomy, Cornell University, Ithaca, NY 14853, USA}
\author[0000-0002-7004-9956]{David A. H. Buckley}
\affiliation{South African Astronomical Observatory, PO Box 9, Observatory 7935, Cape Town, South Africa}
\affiliation{Department of Astronomy, University of Cape Town, Private Bag X3, Rondebosch 7701, South Africa}
\affiliation{Department of Physics, University of the Free State, PO Box 339, Bloemfontein 9300, South Africa}
\author[0000-0001-9312-3816]{Devendra K. Ojha}
\affiliation{Department of Astronomy and Astrophysics, Tata Institute of Fundamental Research, Mumbai 400005, India}
\author[0000-0001-7650-1870]{Arpan Ghosh}
\affiliation{Aryabhatta Research Institute of Observational Sciences (ARIES), Manora Peak, Nainital 263001, India}
\author[0000-0002-0048-2586]{Andrew Monson}
\affiliation{Steward Observatory and Department of Astronomy, University of Arizona, 933 N. Cherry Avenue, Tucson, AZ 85721, USA}
\author[0000-0001-7825-0075]{Malte Schramm}
\affiliation{Universit\"at Potsdam, Karl-Liebknecht-Str. 24/25, D-14476 Potsdam, Germany}
\author[0000-0001-5731-3057]{Saurabh Sharma}
\affiliation{Aryabhatta Research Institute of Observational Sciences (ARIES), Manora Peak, Nainital 263001, India}
\author{Daniel E. Reichart}
\affiliation{Department of Physics and Astronomy, University of North Carolina at Chapel Hill, Campus Box 3255, Chapel Hill, NC 27599-3255, USA}
\author[0000-0003-3457-0020]{Joanna Mikolajewska}
\affiliation{Nicolaus Copernicus Astronomical Centre, Polish Academy of Sciences, Bartycka 18, PL-00716 Warsaw, Poland}
\author[0000-0002-8409-6534]{Juan Carlos Beamin}
\affiliation{Fundaci\'on Chilena de Astronom\'ia, El Vergel 2252, Santiago, Chile}
\author[0000-0002-5936-7718]{J. Borissova}
\affiliation{Instituto de F\'isica y Astronom\'ia, Facultad de Ciencias, Universidad de Valpara\'iso, ave. Gran Breta\~na, 1111, Casilla 5030, Valpara\'iso, Chile}
\affiliation{Millennium Institute of Astrophysics (MAS), Nuncio Monse\~nor Sotero Sanz 100, Of. 104, Providencia, Santiago, Chile}
\author{Valentin D. Ivanov}
\affiliation{European Southern Observatory, Karl-Schwarzschild-Strasse 2, D-85748 Garching bei München, Germany}
\author{Vladimir V. Kouprianov}
\affiliation{Department of Physics and Astronomy, University of North Carolina at Chapel Hill, Campus Box 3255, Chapel Hill, NC 27599-3255, USA}
\author[0000-0003-0125-8700]{Franz-Josef Hambsch}
\affiliation{Vereniging Voor Sterrenkunde, Oostmeers 122 C, 8000 Brugge, Belgium}
\affiliation{American Association of Variable Star Observers, 185 Alewife Brook Parkway, Cambridge, MA 02138, MA, USA}
\affiliation{Bundesdeutsche Arbeitsgemeinschaft f\"ur Ver\"anderliche Sterne e. V., Munsterdamm 90, D-12169 Berlin, Germany}
\affiliation{Groupe Europ\'een d'Observations Stellaires (GEOS), 23 Parc de Levesville, 28300 Bailleau l\'Ev\^eque, France}
\author{Andrew Pearce}
\affiliation{American Association of Variable Star Observers, 185 Alewife Brook Parkway, Cambridge, MA 02138, MA, USA}








\begin{abstract}

EX Lupi, a low-mass young stellar object, went into an accretion-driven outburst in March of 2022. The outburst caused a sudden phase change of $\sim$ 112$\degree$ $\pm$ 5$\degree$ in periodically oscillating multiband lightcurves. Our high resolution spectra obtained with HRS on SALT also revealed a consistent phase change in the periodically varying radial velocities, along with an increase in the radial velocity amplitude of various emission lines. The phase change and increase of radial velocity amplitude morphologically translates to a change in the azimuthal and latitudinal location of the accretion hotspot over the stellar surface, which indicates a reconfiguration of the accretion funnel geometry. Our 3D MHD simulations reproduce the phase change for EX Lupi. To explain the observations we explored the possibility of forward shifting of the dipolar accretion funnel as well as the possibility of an emergence of a new accretion funnel. During the outburst, we also found evidence of the hotspot's morphology extending azimuthally, asymmetrically with a leading hot edge and cold tail along the stellar rotation. Our high cadence photometry showed that the accretion flow has clumps. We also detected possible clumpy accretion events in the HRS spectra, that showed episodically highly blue-shifted wings in the Ca II IRT and Balmer H lines.

\end{abstract}




\section{Introduction} \label{sec:intro}

In the last three decades or so, magnetospheric accretion has been well-established in the low-mass young stellar objects \citep[hereafter, YSOs; see reviews in][]{2007prpl.conf..479B,2016ARA&A..54..135H}. The phenomenon of magnetospheric accretion links the accreting matter from the circumstellar disk to the stellar surface. Originally developed for the highly magnetized accreting stars like white dwarfs, neutron stars, etc., the theory of magnetospheric accretion was later applied to the YSOs \citep{1978ApJ...223L..83G,1991ApJ...370L..39K}. The observation of Zeeman splitting in the photospheric lines of the YSOs indicated a surface magnetic field strength of the order of 1-3 kG in the YSOs \citep{1999ApJ...510L..41J,2001ASPC..248..527J}. The magnetic field truncates the circumstellar disk at inner disk radius equal to a few stellar radii. The material from the inner disk follows the stellar magnetic field lines and gets accreted on the stellar surface near the magnetic poles. The pathway of magnetic field lines channeling the accretion matter is called the magnetic accretion funnel (or accretion funnel). This freely falling material impacts the star near the photosphere at supersonic speed, creating a shocked and hot region ($\sim$ 10$^{4}$ K) called a hotspot \citep{1998ARep...42..322L,2015SSRv..191..339R,2016ARA&A..54..135H}. The success of the magnetospheric accretion mechanism lies in its ability to explain the periodically varying lightcurves by the presence of the hotspot on the rotating star and the explanation of high-velocity
absorption in the spectral lines, like H$\alpha$, in terms of absorption by cool material falling through the accretion channel \citep{1994AJ....108.1906H}. The interaction of the stellar magnetosphere with the disk is so complex that it gives rise to all sorts of accretion phenomena with varying thermal and morphological structures of the accretion funnel and the hotspot. 

The three-dimensional (3D) and 2D magnetohydrodynamic (MHD) simulations have revealed different properties of the magnetospheric accretion. Simulations have shown that the disk is stopped by the magnetosphere at the magnetospheric radius $r_m$ where the magnetic pressure of the magnetosphere matches the matter pressure in the disk \citep{2002ApJ...578..420R,2009A&A...508.1117Z,2015SSRv..191..339R,2022ApJ...941...73T}.   3D simulations of accretion onto stars with the tilted dipole magnetosphere have shown that in many instances, matter accretes onto a star in two funnel streams (as predicted by theoretical work) and forms the banana-shaped spots which have the inhomogeneous density and flux distribution  \citep{2003ApJ...595.1009R,2004ApJ...610..920R,2021Natur.597...41E}, which can be azimuthally asymmetric \citep{2021Natur.597...41E}. The spots may shift along the surface of the star if the inner disk rotates more rapidly than the closed magnetosphere  \citep{2010MNRAS.403.1193B,10.1093/mnras/stt945}.  Simulations also have shown a possibility of unstable accretion where matter penetrates through the magnetosphere in the form of temporary equatorial filaments, or ``tongues"   
which create hotspots at lower latitudes \citep{2008MNRAS.386..673K,2009MNRAS.398..701K,2013MNRAS.431.2673K,2016MNRAS.459.2354B,2022A&A...657A.112B}. This regime is favorable if the magnetosphere rotates slower than the star and conditions for the magnetic Rayleigh-Taylor instability are satisfied \citep[e.g.,][]{1995MNRAS.275.1223S}.

The magnetic field of stars may be more complex than the tilted dipole \citep[e.g.,][]{1999MNRAS.302..437D}. 3D MHD models of accretion onto stars with complex magnetic fields have shown that 
the dipole component typically dominates in the disk-magnetosphere interaction, while the higher multipoles determine the shape of the spots on stellar surface \citep[e.g.,][]{2011MNRAS.413.1061L,2012NewA...17..232L}.

More recently, 3D simulations were performed in a general case where both the magnetic and rotational axes of the star are tilted about the rotational axis of the disk. Simulations have shown that the inner disk tends to be shifted towards the equatorial (rotational) plane of the star and other features associated with the tilt of the rotational axis. \citep{2021MNRAS.506..372R}.

For a single star, only the
mass accretion rate or magnetic field strength and topology can vary \citep{2022AAS...24040613J,2023MNRAS.526.4627F}, which would then change the accretion funnel geometry. Any change in the accretion funnel geometry would alter the thermal and morphological structure of the hotspot. \citet{2012NewA...17..232L,10.1093/mnras/stt945} predicted that there could be changes in size, shape, azimuthal and polar positions and thermal structure of the hotspot, during an increase in accretion rate. The magnitude of these changes depends upon the magnetic field structure, rotation period, magnetic obliquity, etc. \citep{10.1093/mnras/stt945}. Thus, the stars with varying accretion rates become an ideal laboratory for comparisons of simulations with observations.

A class of YSOs, FUors and EXors, undergo episodic accretion-driven outbursts where the accretion rate increases by a factor from a few to 100. FU Orionis is the prototype of the outbursting YSOs, FUors. FUors undergo outbursts on the timescale of years to decades followed by a quiescence for decades to century \citep{1977ApJ...217..693H}. The EXors undergo outbursts on the timescale of months followed by a quiescence for years to decades. EXors are more frequent repeaters of the outburst, with a lesser brightness increase than FUors \citep{2014prpl.conf..387A}. EX Lupi, the prototype of EXors, is a touchstone star as it has undergone multiple outbursts.
It underwent three major outbursts ($\Delta V$ $\sim$ 5 mag) along with several small outbursts since its first reported outburst observation in the 1940s \citep{1946AJ.....52..109M, 1977ApJ...217..693H, 1991PVSS...16...49B,2001PASP..113.1547H,2007AJ....133.2679H,Aspin_2010,2022ATel15271....1Z,2022RNAAS...6...52K,2023A&A...678A..88C,2023ApJ...957..113W}. Recently, EX Lupi went into an accretion-driven outburst in 2022 March \citep{2022ATel15271....1Z,2022RNAAS...6...52K,2023A&A...678A..88C,2023ApJ...957..113W}. The state of high accretion rate remained for about 100 days before returning to its pre-outburst photometric state. Photometrically, the star brightened by $\Delta$$g$ $\sim$ 2 mag. The wealth of high-cadence photometry and high-resolution spectroscopy during the outburst made EX Lupi an ideal laboratory to enhance our understanding of the magnetospheric accretion phenomenon.

EX Lupi is an M0-type star with M$_{*}$ = 0.6$M_{\odot}$ and  R$_{*}$ = 1.6$R_{\odot}$ and it is located in the Lupus 3 star-forming cloud \citep{2005A&A...443..541G,2008hsf2.book..295C,2009A&A...507..881S}. The distance to EX Lupi is 154.57$^{+0.33}_{-0.40}$ pc \citep{2021AJ....161..147B}. \citet{2020ApJ...904...37W} used a 3 kG surface magnetic field strength of EX Lupi citing the preliminary analysis of spectropolarimetric data by \citet{2020IAUGA..30..125K}. The magnetic obliquity of EX Lupi is 13$\degree$, as calculated from the radial velocity (RV) amplitude of He I 5875 \r{A} \citep{2015A&A...580A..82S,2020MNRAS.497.2142M}. The magnetized EX Lupi has a stable rotation period of 7.417$\pm$0.001 days \citep[see Appendix \ref{App:EXLupiPeriod} and][]{2014A&A...561A..61K}. This sets the corotation radius\footnote{The corotation radius $r_c$ is the radius where the angular velocity of the star matches the angular velocity in the disk} at 8.44R$_{*}$ (0.0628 au). 

Generally, the star's disk is inclined with respect to the line of sight by the inclination angle (\textit{i}). \citet{2009A&A...507..881S} performed a spectral energy distribution (SED) fit onto the quiescent state fluxes of EX Lupi. The authors found that the inclination angle between $i$=20$^\circ$ to $i$=40$^\circ$ produces similar results. \citet{2011ApJ...728....5G} modelled CO fundamental bands at 4.6-5.0 $\mu$m and found that $i$=40$\degree$-50$\degree$ produces good fits to the observation. Further \citet{2014A&A...561A..61K} and \citet{2015A&A...580A..82S} found $v$sin($i$) = 4.4 km/s which corresponds to an inclination angle of 25$\degree$.5. \citet{2018ApJ...859..111H} determined an inclination angle of 32$\degree$.4 $\pm$ 0$\degree$.9 by fitting a 2D elliptical Gaussian over the 1.29 mm continuum image of EX Lupi star-disk system taken with the Atacama Large Millimeter/submillimeter Array (ALMA). \citet{2018ApJ...859..111H} also calculated the disk inclination angle of 38$\degree$ $\pm$ 4$\degree$ from the modelling of CO isotopologues. This CO isotopologues modelling assumed a smaller value of stellar mass of 0.5$M_{\odot}$ for EX Lupi. Recent work by \citet{2023MNRAS.526.4885S} used an intermediate inclination angle of 20$\degree$-40$\degree$ to explain color modulation in lightcurves with a single hotspot model. The authors further modelled the width of Ca II 8662 \r{A} lines and found that an inclination angle of around 45$\degree$ best matches the observation. Though dust continuum traces the outer disk, we adopted an inclination angle of 32$\degree$.4 $\pm$ 0$\degree$.9 as it complies with \citet{2009A&A...507..881S,2015A&A...580A..82S}. \citet{2023ApJ...957..113W} also used a value of $i$=30$\degree$ to illustrate the two-component model of the hotspot. The inclination angle and the magnetic obliquity are such that (90$\degree$ - \textit{i} $>$ $\Theta$ = 13$\degree$) only one hotspot is visible. 

Since EX Lupi is located inside a star-forming region, it could be reddened by the dust extinction. \citet{2017A&A...600A..20A} found A$_V$ = 1.1 mag for EX Lupi by doing a chi-square minimization of the X-shooter spectra with a sum of the stellar photosphere and slab model for the hotspot. Other studies have suggested A$_V$ = 0 mag \citep[e.g.][]{2009A&A...507..881S}. Recently, \citet{2023ApJ...957..113W} estimated an extinction of A$_V$=0.1 mag towards EX Lupi. We used the conservative value of A$_V$ = 1.1 mag in our paper.

We elaborate on the observation and reduction of the spectroscopic and photometric data in Section \ref{sec:observation}. Analysis of the observed data is done in Section \ref{sec:AnalysisAndResults} and we conclude this section by listing the key observational results. In Section \ref{sec:DiscussionAndInterpretation}, we discuss the observational key results in detail and make further inferences about EX Lupi. We conclude our work in Section \ref{sec:Conclusion} by enumerating the key understandings about EX Lupi and the accretion process in general. Appendices describe the periodicity of EX Lupi, spectral line fittings, an alternate hypothesis of phase change in lightcurves and radial velocity and estimation of clumps' masses.

\section{Observations and Data Reduction}\label{sec:observation}
The spectroscopic and photometric observations we obtained from various telescopes are elaborated below.

\subsection{HRS-SALT}\label{subsec:HRS-SALT}
Beginning almost at the peak of the outburst ($\sim$ 2022 March 15), we observed EX Lupi spectroscopically on 16 epochs, over the next two and half months, with the High Resolution Spectrograph (HRS) in Medium Resolution (MR) mode (R = 37000; \cite{2010SPIE.7735E..4FB,2014SPIE.9147E..6TC}) on the 11m Southern African Large Telescope \citep[SALT;][]{2006SPIE.6267E..0ZB}. The spectra were collected under `The SALT Transient Programme' (2021-2-LSP-001) \citep{buckley_2017}. HRS covers a wavelength range of 3740 \r{A} - 8780 \r{A} using blue and red cameras. The wavelength coverage of the blue camera is 3740 \r{A} - 5560 \r{A} while the red camera covers 5450 \r{A} - 8780 \r{A}. The 2D to 1D extraction and wavelength calibration were done by the MIDAS pipeline \citep{2016MNRAS.459.3068K,2016arXiv161200292K}. Barycentric correction was applied to the spectra using a Python-based module: barycorrpy \citep{2018RNAAS...2....4K}. The instrumental response was estimated from the observation of the stars HD074000, HD38949 and HD205905 on 2022 March 02, 2022 April 12 and 2022 May 15 respectively. The HRS-SALT spectra of these standard stars were divided by the flux-calibrated spectra of the respective stars taken from the CALSPEC database \citep{2014PASP..126..711B}. The average of the three observations after respective median normalization was taken as the instrument response curve. EX Lupi spectra were divided by this instrument response curve to get the response corrected spectra. The spectra were flux-calibrated against the corresponding epoch's photometric flux. Since ASAS-SN \textit{g}-band observations were more densely sampled, we used it for interpolating the \textit{g}-band flux to the epochs of SALT HRS observations. Interpolation was done using a Gaussian Process with a combination of the Matern-5/2 and exponential-sine-squared kernel in the Python module, George \citep{2015ITPAM..38..252A}. \textit{B, V} and \textit{I}-band magnitudes from our TMMT multiband observations (see Section \ref{subsec:TMMT}) were found to correlate with ASAS-SN \textit{g} band observations. Hence, we used this correlation to convert the interpolated \textit{g}-band magnitudes to interpolated, \textit{B, V} and \textit{I} magnitudes. We found that flux calibration of spectra using the TMMT $I$-band flux produces spectral flux consistent with TMMT $B$-band in the blue part of the HRS spectra. Thus, the TMMT $I$-band interpolated fluxes were then used to scale the HRS spectra of EX Lupi. The spectra were further corrected for extinction \citep[A$_V$=1.1 mag, ][]{2017A&A...600A..20A}. The extinction curve was generated from \citet{1989ApJ...345..245C} via the Python module named \textit{dust-extinction} for $R_V$ = 3.1. The error for each pixel value in spectra was taken to be the square root of the effective number of photoelectrons detected. A typical signal-to-noise ratio (SNR) of 75 was attained at around 5000 \r{A}.

\subsection{TMMT}\label{subsec:TMMT}
We followed up EX Lupi during the 2022 March outburst and the post-outburst period using the Three-hundred MilliMeter (300 mm) Telescope \citep[TMMT,][]{2017AJ....153...96M} at Las Campanas Observatory in Chile. TMMT uses an Apogee Alta U42-D09 CCD Camera and we observed EX Lupi at a very high cadence for 52 nights with TMMT, collecting 1812, 1817 and 1817 frames in $B$, $V$ and $I$ bands, respectively. During the outburst, we observed for 43 nights starting from 2022 March 13 (JD = 2459651.7106) to 2022 June 20 (JD = 2459750.7426) and 1776, 1781 and 1782 frames were taken in $B$, $V$ and $I$ bands, respectively. The rest of the data were observed in 2023 January, covering 9 nights. The observation was conducted nearly at a one-day cadence, and inside each night, we conducted three-minute cadence exposures for about 4-5 hours. The photometric observation logs are tabulated in Table \ref{table:Photlogs}. Only a portion of the logs are provided here, while a complete log is available online in machine-readable format.

\subsection{LCRO}\label{subsec:LCRO}
Along with TMMT, we also followed up EX Lupi during the 2022 March outburst and the post-outburst period using the 305 mm Las Campanas Remote Observatory (LCRO) telescope at the Las Campanas Observatory in Chile. LCRO uses an FLI Proline 16803 CCD Camera with $g^{\prime}$, $r^{\prime}$, and $i^{\prime}$ photometric bands. We observed EX Lupi over 85 nights with a total of 933, 928 and 922 frames in $g^{\prime}$, $r^{\prime}$, and $i^{\prime}$ bands respectively. EX Lupi was observed on 11 nights during the outburst (2022 March 13 (JD = 2459651.7093) to 2022 March 25 (JD = 2459663.9104)) and for 74 nights in the post-outburst state starting from 2023 January 24 (JD = 2459972.8347) to 2023 May 14 (JD = 2460078.9175).

\subsection{CTIO}\label{subsec:CTIO}

We conducted multiband photometric observations of EX Lupi with an FLI camera on the Prompt6 telescope at Cerro Tololo Inter-American Observatory (CTIO), Chile. The telescope operates under the Skynet Robotic Telescope Network \citep{2019ApJS..240...12M}. Prompt6 is a 0.4m aperture telescope with a field of view of 15.1$^{\prime}$ $\times$ 15.1$^{\prime}$. It has \textit{U, B, V, R}, H$\alpha$, OIII and SII filters. We observed EX Lupi in \textit{B} and \textit{V} bands from almost the peak of the outburst, 2022 March 12 (JD=2459650.768) to the subsequent 60 days, 2022 May 11 (JD=2459710.5854), collecting 51 epochs of nearly one-day cadence photometric data. The images were corrected for dark currents and were also flat-corrected.

\subsection{ASAS-SN}\label{ASAS-SN}
High-cadence $g$ and $V$-band lightcurves are used from All-Sky Automated Survey for Supernovae \citep[ASAS-SN,][]{2014ApJ...788...48S}. The lightcurve we used in our study covers from 2017 May 27 (HJD = 2457900.7206\footnote{The HJD (Heliocentric Julian Date), BJD (Barycentric Julian Date), and JD (Julian Date) are not corrected to the same scale. The maximum difference of 8 min among them corresponds to an insignificant $\sim$ 0$\degree$.27 of EX Lupi's rotation. Our phase calculation on lightcurves has a gross error of $\sim$ 3-10$\degree$.}) to 2023 September 6 (HJD = 2460194.3491). For reference purposes, the time = 0 of all the calculations related to lightcurves is taken at HJD = 2457900.7206 (see Section \ref{sec:AnalysisAndResults}).

\subsection{AAVSO}\label{subsec:AAVSO}
We have used the high cadence multiband $U, B, V, R$ and $I$ lightcurves from the American Association of Variable Star Observers (AAVSO) data archive. The data used for our study covers the whole outburst starting from 2022 March 13 (JD = 2459651.9002) to the subsequent quiescence up to 2022 October 17 (JD = 2459870.4878). A significant amount of AAVSO data was contributed by the Remote Observatory Atacama Desert telescope \citep[ROAD;][]{2012JAVSO..40.1003H} and Perth Observatory's R-COP telescope. The ROAD telescope is a fully automated telescope located in Chile that obtains nightly photometry in Astrodon $B, V, I$ and Clear bands for a wide range of astronomical projects. It consists of a 40-cm f/6.8 Optimized Dall Kirkham telescope and uses an FLI ML16803 CCD camera with a 4k$\times$4k array with pixels of 9 $\mu$m in size. Data was reduced using a custom pipeline and then published in the AAVSO database. The R-COP is a telescope partnership that stands for
Remote Telescope Partnership: Clarion University - Science in Motion, Oil Region Astronomical Society, and Perth Observatory. The R-COP telescope is a part of the Skynet Robotic Telescope Network.

\subsection{TESS}\label{subsec:TESS}
Transiting Exoplanet Sky-Survey (TESS) observed EX Lupi in sector-12 from 2019 May 21 (JD = 2458624.9745) to 2019 June 18 (JD = 2458652.8701) and in sector-39 from 2021 May 27 (JD = 2459361.778) to 2021 June 24 (JD = 2459389.7169). We used sector-12 lightcurve from A PSF-based Approach to TESS High-Quality Data of Stellar Clusters \citep[PATHOS,][]{2019MNRAS.490.3806N} pipeline and sector-39 lightcurve from TESS Science Processing Operations Center pipeline \citep[TESS-SPOC,][]{2020RNAAS...4..201C}.

\begin{deluxetable*}{ccccccccc}
\tablecaption{The photometric and spectroscopic observation logs from TMMT, LCRO, CTIO and SALT-HRS observations. This table contains a portion of the whole data. The JDs of TMMT, LCRO, and CTIO data are rounded off to two decimal points to show all the multiband data points simultaneously for close epoched observations. The online table contains a complete data set.}\label{table:Photlogs}
\tablewidth{1pt}
\tablehead{
\colhead{JD} & \colhead{Telescope/Instrument} &
\colhead{$B$} & \colhead{$V$} & \colhead{$I$} & \colhead{$g^{\prime}$} & \colhead{$r^{\prime}$} & \colhead{$i^{\prime}$} & \colhead{Spectroscopy/} \\
\colhead{} & \colhead{} & \colhead{(mag)} & \colhead{(mag)} & \colhead{(mag)} & \colhead{(mag)} & \colhead{(mag)} & \colhead{(mag)} & \colhead{Photometry}
}
\startdata
2459654.489873 & SALT-HRS &  & & & & & & spec \\
2459659.475822 & SALT-HRS &  & & & & & & spec \\
2459682.412697 & SALT-HRS &  & & & & & & spec \\
.... & SALT-HRS & .... & .... & .... & .... & .... & ..... & spec \\
2459651.71 & TMMT & 12.81$\pm$0.02 & 12.15$\pm$0.01 & 10.71$\pm$0.02 &  &  &  & phot \\
2459651.71 & TMMT & 12.77$\pm$0.03 & 11.15$\pm$0.01 & 10.71$\pm$0.02 &  &  &  & phot\\
2459651.72 & TMMT & 12.77$\pm$0.02 & 11.14$\pm$0.01 & 10.70$\pm$0.02 &  &  &  & phot\\
2459651.72 & TMMT & 12.77$\pm$0.02 & 11.14$\pm$0.01 & 10.70$\pm$0.02 &  &  &  & phot\\
.... & TMMT & .... & .... & .... & .... & .... & ..... & phot\\
2459651.71 & LCRO &  &  &  & 12.14$\pm$0.01 & 11.68$\pm$0.01 & 11.89$\pm$0.01 & phot\\
2459651.71 & LCRO &  &  &  & 12.14$\pm$0.01 & 11.68$\pm$0.01 & 11.86$\pm$0.01 & phot\\
2459651.72 & LCRO &  &  &  & 12.09$\pm$0.01 & 11.67$\pm$0.01 & 11.85$\pm$0.01 & phot\\
2459651.72 & LCRO &  &  &  & 12.07$\pm$0.01 & 11.66$\pm$0.01 & 11.86$\pm$0.01 & phot\\
.... & LCRO & .... & .... & .... & .... & .... & ..... & phot \\
2459650.77 & CTIO & 12.31$\pm$0.003 & 11.56$\pm$0.002 &  &  &  &  & phot\\
2459651.69 & CTIO & 12.83$\pm$0.004 & 11.98$\pm$0.002 &  &  &  &  & phot\\
2459652.69 & CTIO & 12.54$\pm$0.004 & 11.85$\pm$0.002 &  &  &  &  & phot\\ 
2459653.82 & CTIO & 12.34$\pm$0.003 & 11.66$\pm$0.002 &  &  &  &  & phot\\
.... & CTIO & .... & .... & .... & .... & .... & ..... & phot \\
\enddata
\end{deluxetable*}

\subsection{Photometry and Radial Velocities from previous literatures}\label{subsec:ArchivalRVs}
We have also used the RV estimates for EX Lupi from the literature. \citet{2014A&A...561A..61K,2015A&A...580A..82S,2021MNRAS.507.3331C} presented the RVs of EX Lupi by cross-correlating the spectra with template spectrum and by fitting the emission spectral lines. These RVs cover an interval from the year 2007 to 2019.
The time evolution of the RV phase is taken from \citet{2021MNRAS.507.3331C}. We have used the RV amplitudes from \citet{2015A&A...580A..82S,2021MNRAS.507.3331C}. We have also used multiband high cadence photometry at the onset of the outburst from \citet{2022RNAAS...6...52K}.

\section{Analysis and Results}\label{sec:AnalysisAndResults}

The central part of this work is to understand the phase evolution of the various waveband lightcurves and the periodically modulated RVs. The techniques used to measure the phases of the lightcurves and the RV are discussed below in Sections \ref{subsec:PhaseOfLC} and \ref{subsec:PhaseOfRV}.

\subsection{Phases of the lightcurves}\label{subsec:PhaseOfLC}
A start time (t = 0) is needed to define a reference point for the calculation of the phase of the periodically varying lightcurves. We took t = 0 as the first epoch of our ASAS-SN lightcurve. It is JD = 2457900.7206 (hereafter startJD). All the phase estimates are done on the light curves after converting from magnitude to flux units. Lightcurve cutouts are made at an interval of 40 days. A straight line is fitted to each individual lightcurve cutout to remove any linear trend. Then the cutout is cross-correlated with a sine wave of the form $\sin(\frac{2\pi (t-t_{0})}{P} + \phi)$ where $P$ is the rotation period of EX Lupi and the $t$ is the epoch of observation in JD, and $t_{0}$ is startJD. The phase $\phi$ is varied over [0, 2$\pi$) at an interval of 1$\degree$, and the corresponding Pearson correlation coefficient is calculated. The value of $\phi$ which corresponds to the maximum estimate of the Pearson correlation coefficient is the phase of the lightcurve cutout. This phase $\phi$ is associated with the first epoch of the cutout. 

We calculated the phase along with the phase error estimates for every lightcurve cutout by performing the Monte Carlo method described below. Random light curves are generated by sampling from a Gaussian noise distribution centered at the original light curve values of each epoch. The standard deviation of this Gaussian distribution is taken as a quadrature sum of the lightcurve flux error and stellar activity error. We assigned the stellar activity error to the robust standard deviation of the residual of a sine curve fit onto the original lightcurve cutout. The standard deviation of the Gaussian distribution is dominated by the stellar activity error in almost all the cases. Then, phase is calculated on this simulated lightcurve cutout as described in the previous paragraph. The simulation is run 1000 times. The mean and standard deviations of the phase estimation from 1000 runs are taken as phase and phase error, respectively.

To confirm our phase estimates are not driven by any flares, for our phase analysis of the ASAS-SN lightcurve, we also manually selected clean durations of the lightcurve devoid of any flares and other variability. Three such epochs were identified in the quiescent state. Three more epochs of phases are estimated on manually selected regions during the state of outburst.

Similarly, the phase estimation is done for other lightcurves of EX Lupi. The AAVSO $U, B, V, R$, and $I$-band lightcurves are first converted into flux units (erg/s/cm$^{2}$/\r{A}) following zero-point flux in \citet{1998A&A...333..231B} and then the phase calculation is done. The phase calculations on the CTIO $B$ and $V$-band lightcurves are also done after converting them into the flux unit. The CTIO lightcurves cover 60 days, and so a single phase is calculated onto the whole lightcurve. TESS lightcurves in sector-12 and sector-39 have also been used to estimate the phases. We calculated the phases of the $g^{\prime}$, $r^{\prime}$, $i^{\prime}$ and $z^{\prime}$ band lightcurves, taken from \citet{2022RNAAS...6...52K}, at the onset of the 2022 March outburst. The phases are calculated one year after the outburst using LCRO $g^{\prime}$, $r^{\prime}$ and $i^{\prime}$ band lightcurves. The Sloan, $g^{\prime}$, $r^{\prime}$, $i^{\prime}$ and $z^{\prime}$ band, magnitudes are converted into flux unit by converting them to AB magnitudes and then using AB zeropoint flux density of 3631 Jy.

\begin{figure*}
\gridline{\fig{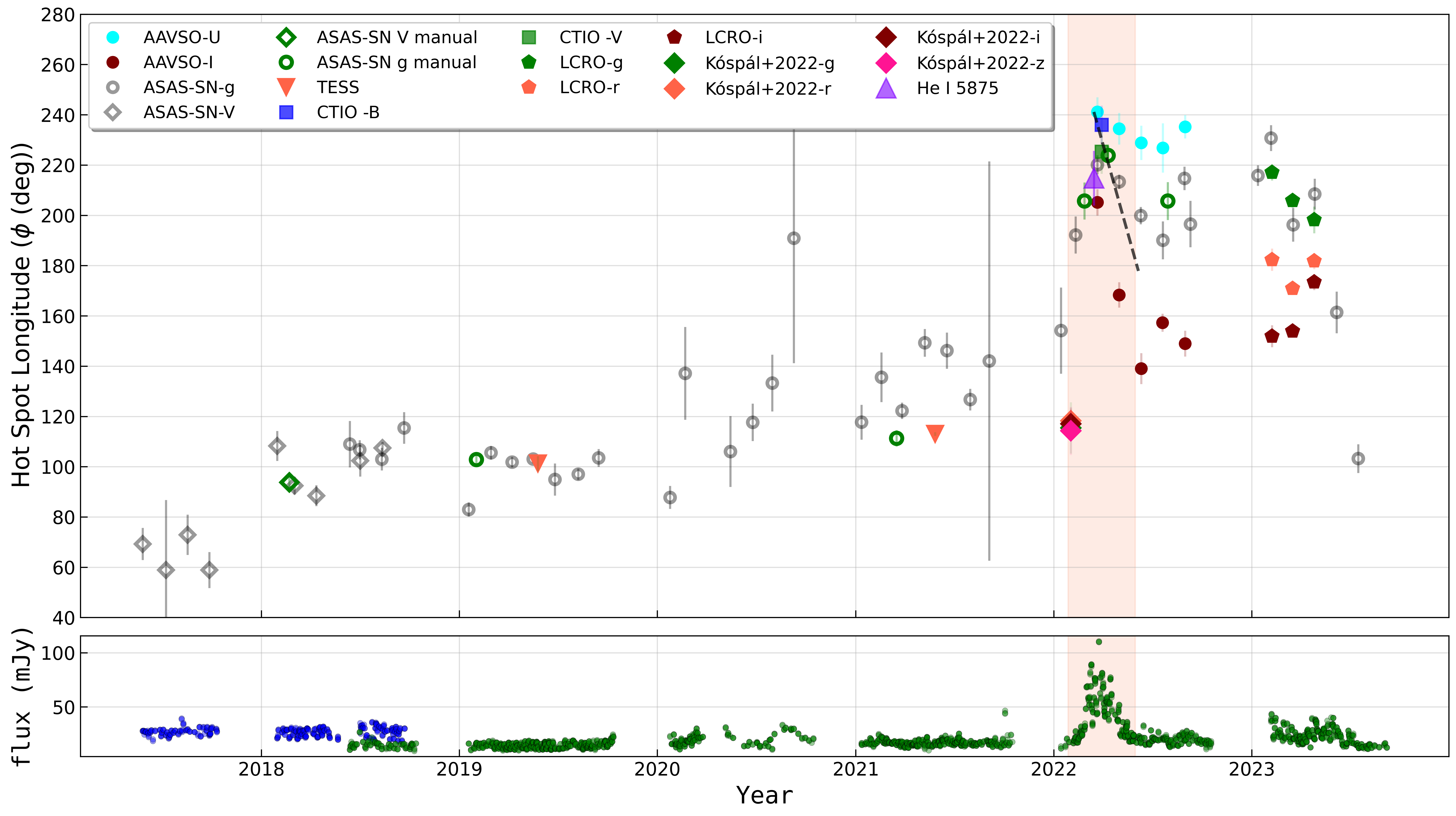}{1\textwidth}{}}
\caption{\textit{Top panel}: The Longitude of the Hotspot estimated with respect to the \textit{fixed frame} is plotted. Various colors and markers represent estimated hotspot longitudes from the multiband lightcurves and the RVs. Longitudes inferred from AAVSO lightcurves are shown for $U$ and $I$-bands only. Longitude inferred from RV is plotted only for He I 5875 to make the figure less crowded; however, all the RVs are plotted in Figure \ref{fig:RV_phasechange}. A straight black dashed line is the model-predicted decay rate of the phase, starting with the AAVSO-U band phase at the outburst. \textit{Bottom panel}: ASAS-SN $V$ and $g$-band lightcurves are plotted by blue and green dots respectively. The time period of 2022 March outburst is shaded by light red color. A more detailed post-outburst phases of multiple wavelengths are shown in Figure \ref{fig:AAVSOphaseEvolution}}
\label{fig:phaseVariation_AllLC}
\end{figure*}

The ASAS-SN $g$-band lightcurve phases during the quiescence, prior to the 2022 March outburst, are similar at around 100$\degree$-110$\degree$, as shown in the upper panel of Figure \ref{fig:phaseVariation_AllLC}. The ASAS-SN g-band phase (green circle) around 2019 is 102$\degree$.9 $\pm$ 2$\degree$.7 and the close-by sector-12 TESS lightcurve (inverted orange triangle) has a phase value of 103$\degree$.5 $\pm$ 0$\degree$.4 while the phases of ASAS-SN g-band lightcurve and TESS sector-39 lightcurves, around 2021, are 111$\degree$.3 $\pm$ 3$\degree$.2 and 113$\degree$.0 $\pm$ 0$\degree$.7, respectively. In Figure \ref{fig:phaseVariation_AllLC}, the grey circles are the phases from the ASAS-SN $g$-band lightcurve while the green circles are phases estimated on manually selected cleaner flare-free ASAS-SN $g$-band lightcurve cutouts. Thus, the phases from the TESS lightcurves are consistent with the nearby epoch ASAS-SN phases. The multiband phases from \citet{2022RNAAS...6...52K} lightcurves (diamonds) at the onset of the outburst have values consistent with that of the TESS sector-39 lightcurve. $g^{\prime}$-band phase from \citet{2022RNAAS...6...52K} is 115$\degree$.4 $\pm$ 4$\degree$.2 while maximum phase difference among the $g^{\prime}$, $r^{\prime}$, $i^{\prime}$ and $z^{\prime}$ bands is 10$\degree$.0. 

At the peak of the outburst, the phase of the ASAS-SN lightcurve increased to 223$\degree$.9 $\pm$ 3$\degree$.8. The $B$ (236$\degree$.0 $\pm$ 7$\degree$.6) and $V$-band (225$\degree$.3 $\pm$ 8$\degree$.8) phases derived from CTIO (squares) during the peak of the outburst are consistent with the ASAS-SN g-band phase, supporting the phase increase. AAVSO multiband phases, $U$ (241$\degree$.1 $\pm$ 5$\degree$.8), $B$ (231$\degree$.9 $\pm$ 3$\degree$.6), $V$ (221$\degree$.0 $\pm$ 3$\degree$.4), $R$ (214$\degree$.0 $\pm$ 3$\degree$.5) and $I$ (205$\degree$.2 $\pm$ 5$\degree$.2),  also agree well with the corresponding band phases from CTIO and ASAS-SN, during the peak of the outburst. We estimated the phase change during the outburst by comparing the ASAS-SN phase values (green circle, in Figure \ref{fig:phaseVariation_AllLC}) from close to the peak of the outburst (223$\degree$.9 $\pm$ 3$\degree$.8) and around 2021 (111$\degree$.3 $\pm$ 3$\degree$.2). The ASAS-SN lightcurve yields an increase in phase of 112$\degree$.6 $\pm$ 5$\degree$.0. 

During the decay of the outburst (see Section \ref{subsec:ChromaticPhaseChange}), the phases decayed differentially in multiband AAVSO (dots) and ASAS-SN lightcurves. Among AAVSO multibands, the $U$-band phases decreased smallest while $I$-band phases decreased by largest phase values. From the peak of the outburst to the subsequent post-outburst state, the $U$-band phases remained largest followed by $B$, $V$, $R$ and $I$ band phases, sequentially (see Figure \ref{fig:phaseVariation_AllLC}). The $U$ and $B$ band phases stayed close to their peak outburst phase values, after the outburst was over. Further, the lightcurve phases could not regain their pre-outburst value even after one year of the outburst. The first LCRO phases (pentagon) estimated on 2023 February 7 showed a larger than pre-outburst phase value in the $g^{\prime}$-band. Further, the LCRO $g^{\prime}$, $r^{\prime}$ and $i^{\prime}$ bands showed a phase difference among the bands: $g^{\prime}$ (217$\degree$.0 $\pm$ 3$\degree$.2), $r^{\prime}$ (182$\degree$.3 $\pm$ 4$\degree$.4) and $i^{\prime}$ (151$\degree$.9 $\pm$ 4$\degree$.4), The ASAS-SN $g$-band phase values are within $\sim$10$\degree$ to the close epoch LCRO $g^{\prime}$-band phases. However, the latest ASAS-SN phase value from 2023 July to 2023 September has returned to the pre-outburst quiescence phase value (see grey circles in Figure \ref{fig:phaseVariation_AllLC}).

\subsection{Phase of the radial velocity}\label{subsec:PhaseOfRV}

We took 16 epochs of high-cadence spectra with HRS on SALT, during the 2022 March outburst. Some of the high energetic emission spectral lines of He I, He II, Fe I, Fe II, Mg I, and Si II, which are visibly distinguishable, are modelled with multi-Gaussians + a constant continuum (see Table \ref{table:SineFit_on_RV} for energy levels of the lines). The modelled spectral lines have an SNR as low as $\sim$15 on a few epochs of quiescence. These high-energy lines are expected to originate in the hotspot and hence carry information about the hotspot dynamics. As inferred from the periodically varying lightcurve, the hotspot comes and goes out of our view as the star rotates
\footnote{For EX Lupi, magnetic obliquity and inclination angles are small, and the hotspot is visible at all rotational phases (see Section \ref{subsec:3DMHDsimulation} and Figure \ref{fig:3DMHDrotation}).}, and thus a periodicity in the RVs is expected for the spectral lines originating in the hotspot. From the fitted lines, the RV of the Gaussian representing the narrow component (NC) of the spectral lines are phase-folded and fitted with a sine wave of the form $A \sin(\frac{2\pi (t-t_{0})}{P} + \phi) + V_{0}$ where $A$, $P$, $\phi$ and $V_{0}$ are the amplitude of the RV, rotation period of the star (7.417 days), phase of the RV and an arbitrary constant respectively. Here, $t$ is the epoch of the spectra (in JD) and $t_{0}$ is startJD. The $\phi$ of this sine fit gives the RV phase at our startJD. To avoid outliers affecting the sine fit, we did a leave-one-out analysis. One RV point is removed at a time, and sine is fitted. This is repeated for each of the 16 RV points. For each spectral line, the median of each of the fit parameters ($A, \phi$ and $V_{0}$) and best-fit errors are taken as the value of the parameter and respective error. The phases cover a range of $\sim$ 110$\degree$ - 155$\degree$ except for the He II 4686 \r{A}, which has a phase of 176$\degree$. Except He II 4686 \r{A}, all the spectral line RVs are clumped around in a small phase band as shown in Figure \ref{fig:RV_phasechange}. We also found that the RV amplitudes, inferred from the aforementioned sine fit onto the RVs, have increased consistently for all the spectral lines with respect to the quiescent state RV amplitudes (see Table \ref{table:SineFit_on_RV}). RV amplitude is also largest for He II 4686 \r{A}. We do not understand the cause behind the large phase and large RV amplitude of He II 4686 \r{A}.

Since we do not have immediate pre-outburst spectra from HRS-SALT, we compared our RV phases with RV estimates from the literature to calculate any pre- to post-outburst phase changes in periodically varying RVs, similar to a phase change observed in the photometry.

\subsection{RV phase change in between the 2008 outburst and the 2022 outburst}\label{subsec:PhaseofKospal_RV}

\begin{figure*}
\gridline{\fig{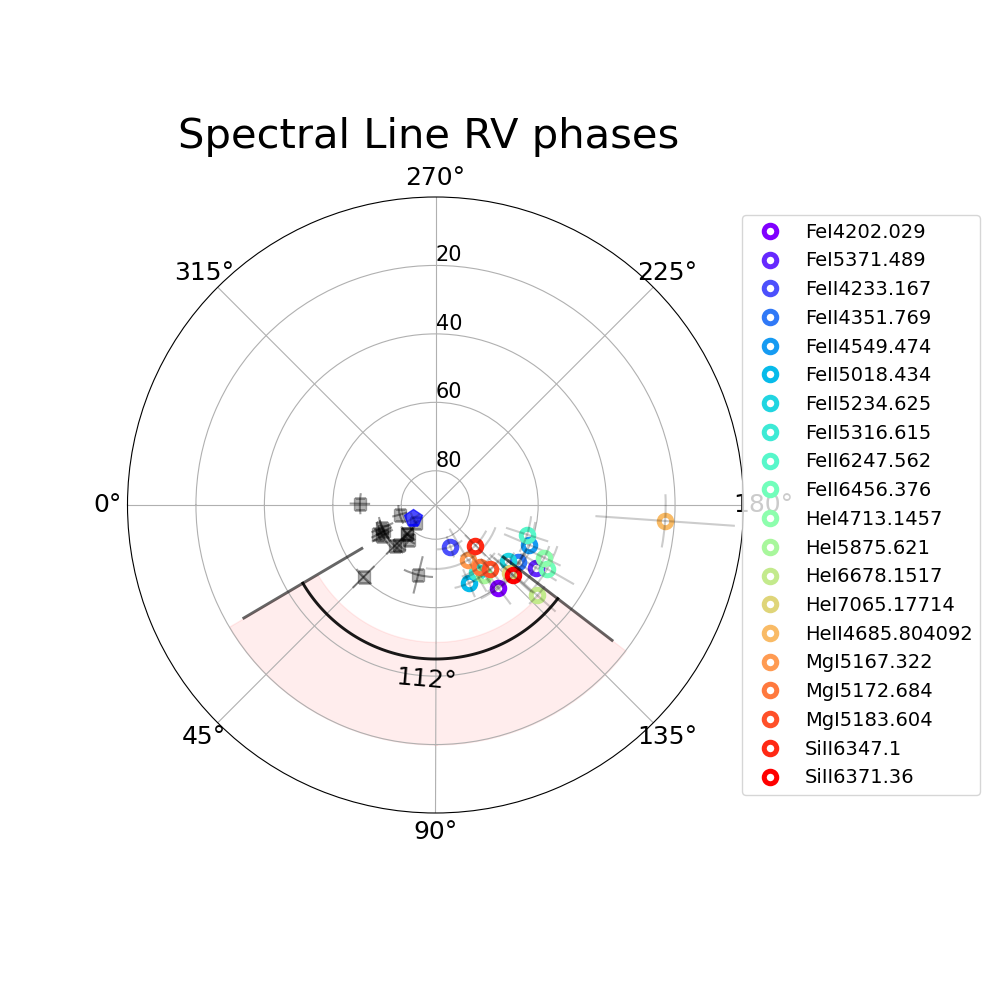}{0.7\textwidth}{}}
\caption{RV phases estimated for various spectral lines are plotted. The circles correspond to the RV phases from HRS-SALT spectra of EX Lupi during the 2022 Outburst. The grey squares are from \citet{2021MNRAS.507.3331C} for the time-interval JD = 2457548.9 - 2457563.9. Blue pentagon is from \citet{2014A&A...561A..61K}, for emission lines. The azimuthal position of the markers corresponds to the RV phase (at our startJD), and their radial position corresponds to the latitude inferred from the respective RV amplitude (see Section \ref{subsec:hotspotLatitudeChange}). The much-deviated orange-yellow color circle is He II 4686 \r{A}. A grey curve of 112$\degree$ is plotted anti-clockwise from the phase of the blue pentagon, corresponding to the phase change estimated from the lightcurves. The same region is shaded with a light red color.}
\label{fig:RV_phasechange}
\end{figure*}

\citet{2014A&A...561A..61K,2015A&A...580A..82S,2021MNRAS.507.3331C} have presented RVs of EX lupi for around a decade starting from the year 2007. \citet{2014A&A...561A..61K} presented the absorption and narrow emission line RVs, covering 5 years across July 2007 and July 2012. This time period covers a major outburst of the year 2008. The authors estimated emission line RVs by the multi-Gaussian fittings of 133 lines identified by \citet{2012A&A...544A..93S} in the 2008 outburst spectra. There is no apparent phase difference among the RVs from pre- to post-2008 outburst \citep[see Figure 2 in ][]{2014A&A...561A..61K}. We estimated the phases of the absorption and emission line RVs by fitting a sine curve to this dataset as done in Section \ref{subsec:PhaseOfRV} except that the zero point (t=0) is taken as JD = 2454309.6147 (in order to be consistent with \citet{2021MNRAS.507.3331C}, see paragraph below). For the emission line RVs, the phase is 327$\degree$.3 $\pm$ 13$\degree$.3 while for the absorption line, the RV phase is 114$\degree$.0 $\pm$ 3$\degree$.1. 
Considering EX Lupi's rotation period of 7.417 days, these RV phases are translated from JD = 2454309.6147 to our startJD. The respective RV phases for the emission and absorption lines are 30$\degree$.5 $\pm$ 13$\degree$.3 and 176$\degree$.0 $\pm$ 3$\degree$.1 respectively (at our startJD). This emission line phase is plotted in Figure \ref{fig:RV_phasechange} as a blue pentagon. For comparing the phase shift inferred from the emissions lines with that of the photometric lightcurves, we have overlaid a black curve representing the 112$\degree$ phase shift inferred from the photometric lightcurves on the same plot. \citet{2012A&A...544A..93S} estimated the RVs for metallic lines during the state of outburst in the year 2008. The outburst RVs were estimated on 5 epochs spread over 56 days in 2008 April - 2008 June. We could not fit a sine curve onto these RVs due to insufficient data points and thus we do not have any direct measure of phase during the major outburst of the year 2008.

Furthermore, \citet{2021MNRAS.507.3331C} did multi-Gaussian fittings to various narrow metallic and He I spectral lines from the spectra observed for EX Lupi during its quiescence. The authors also computed the phase of the RVs by fitting a sine function (the t=0 is taken as JD = 2454309.6147) to the RVs of various time intervals to understand the time-evolution of the RV phases and they found that the phases had some scatter during the time interval of the recorded data (JD = 2454874.8 - 2457563.9) but remained confined to a 90$\degree$ quadrant \citep[see Figure 8 in ][]{2021MNRAS.507.3331C}. To compare the quiescent state RV phase with that of the 2022 March outburst state RV phase, we take the latest RV phases (JD = 2457548.9 - 2457563.9) from \citet{2021MNRAS.507.3331C} and re-evaluate them at our startJD. A typical quiescence phase value for He I 5016 \r{A} is 42$\degree \pm$ 7$\degree$. RV phases for other lines are tabulated in Table \ref{table:SineFit_on_RV}. These values are consistent with the emission line RV phase of \citet{2014A&A...561A..61K}. For a few spectral lines, \citet{2021MNRAS.507.3331C} also gave RV phases for the time period JD = 2458634.9 - 2458644.0, which translates to FeII 4924 (152$\degree \pm$ 60$\degree$), Fe II 5018 (181$\degree \pm$ 28$\degree$) and HeII 4646 (-29$\degree \pm$ 110$\degree$), at our startJD. These phase values differ significantly among various spectral lines in contrast to the consistency among various spectral line phases for JD = 2457548.9 - 2457563.9 and hence we considered the RV phase values for time-interval JD = 2457548.9 - 2457563.9 as pre-outburst RV phases. All these quiescent state RV phases are plotted by grey squares in Figure \ref{fig:RV_phasechange}. This suggests that RV also showed a phase change during the 2022 March outburst, consistent with the photometric phase shift.

\subsection{Mass accretion rate}\label{subsec:MassAccretionRate}
\begin{figure*}
\gridline{\fig{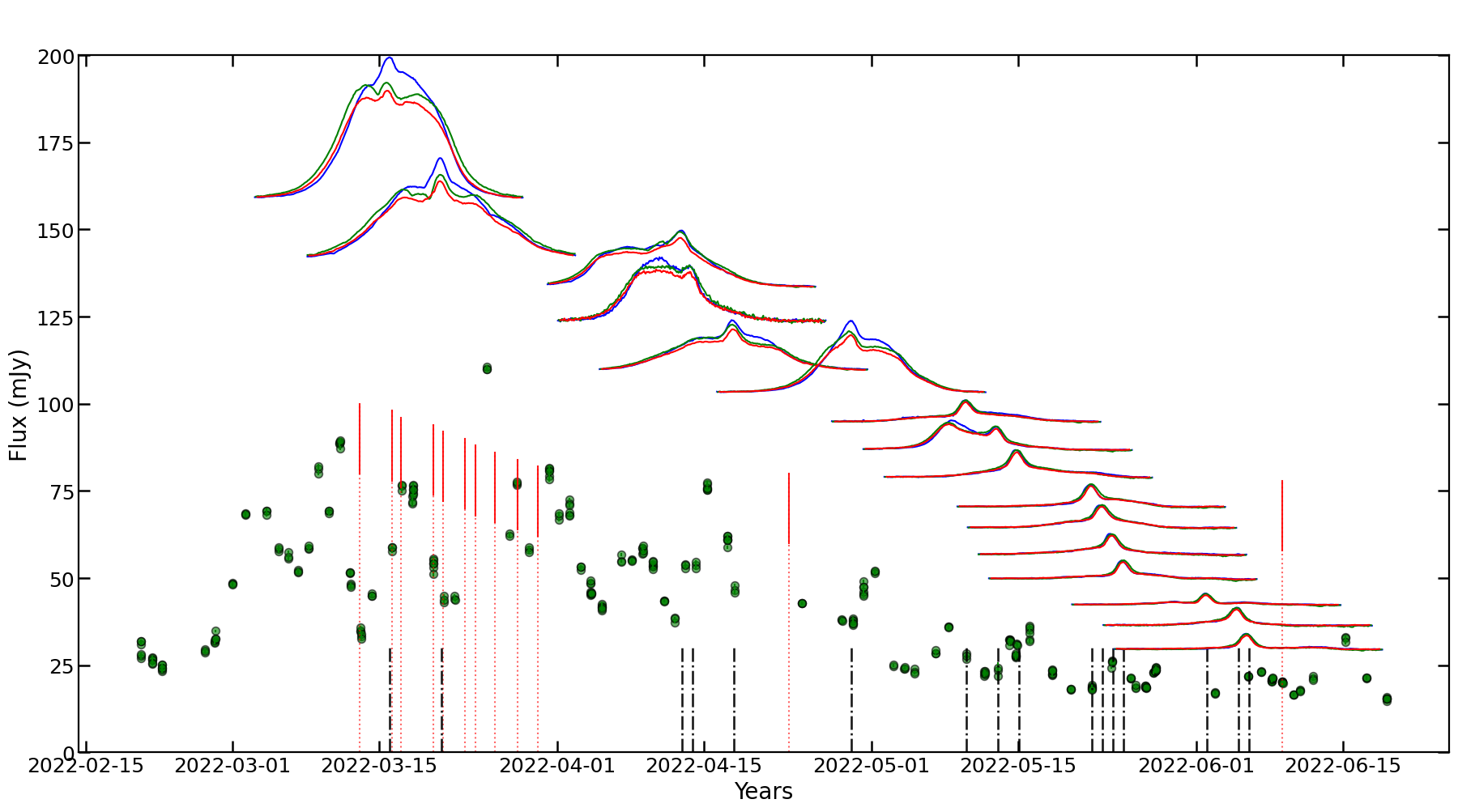}{1\textwidth}{(3)}}
\caption{The Ca II IRT spectra are plotted over ASAS-SN $g$-band lightcurve during the outburst. Each spectrum is plotted on the respective epoch of observation, shown by a vertical dashed grey line. Ca II lines at 8498.02 \r{A}, 8542.09 \r{A} and 8662.14 \r{A} are shown by red, green and blue colors, respectively. The broad component disappeared from the Ca II IRT lines with the decay of the excess continuum from the outburst, around after 8th epoch. The red dotted vertical lines are the epochs of observation of hour timescale photometric enhancement in the TMMT lightcurves (see Section \ref{subsec:TMMT_clumpy})}.
\label{fig:CaII_onLC}
\end{figure*}

Our multi-epoch high-resolution HRS-SALT spectra of EX Lupi are used to estimate the mass accretion rate evolution during- and post-outburst. The evolution of the spectral line over the outburst is shown by plotting the Ca II infrared triplet (hereafter, Ca II IRT) line over the ASAS-SN $g$-band lightcurve in figure \ref{fig:CaII_onLC}. The Ca II IRT lines are plotted on the epochs of observed spectra. The broad component (BC) of Ca II IRT lines faded in tandem with the decay of the outburst (see Figure \ref{fig:CaII_onLC}). The mass accretion rate is calculated by estimating the line luminosities of good accretion estimator lines \citep[as marked in][]{2017A&A...600A..20A}. They are Ca II (K) (3933.66 \r{A}), H8 (3889.049 \r{A}), H$_{\beta}$ (4861.325 \r{A}), H$_{\gamma}$ (4340.464 \r{A}), H$_{\delta}$ (4101.734 \r{A}), He I (4026.191 \r{A}), He I (4471.48 \r{A}), He I (5875.621 \r{A}), He I (6678.151 \r{A}) and He I (7065.19 \r{A}). A spectral cutout is prepared by manually selecting a wavelength window around every spectral line to avoid any nearby lines. The local continuum of the spectral line is estimated by fitting a straight line on the 2 \r{A} window at both edges of the spectral cutout. Then, the local continuum subtracted spectral line is integrated. The integrated flux is scaled by 4$\pi$d$_{*}^{2}$ to get the line luminosity at the stellar surface, where d$_{*}$ is the distance to EX Lupi. The line luminosity is converted to accretion luminosity following the relations developed in \citet{2017A&A...600A..20A}.
Assuming that the energy of the infalling matter is completely converted into radiation, the accretion luminosity is converted into a mass accretion rate following the Equation \ref{Equ:accretionRate2}.

\begin{equation}\label{Equ:accretionRate2}
    \dot{M} = \frac{L_{acc} R_{*}}{GM_{*} \left( 1 - \frac{R_{*}}{R_{in}}  \right) } ~,
\end{equation}
where $\dot{M}$ and $L_{acc}$ are mass accretion rate and accretion luminosity. $R_{*}$ and $M_{*}$ are the radius and mass of EX Lupi and $G$ is the gravitational constant. $R_{in}$ is inner disk radius which is taken to be 5$R_{*}$. The estimated mass accretion rates from various spectral lines are shown in Table \ref{table:MassAccretionRates}.

The mass accretion rates for all the aforementioned good accretion tracer lines are similar except for Ca II (K). So, for calculations concerning accretion rates, the average accretion rate for all lines except Ca II (K) will be used. The mass accretion rate estimated from various lines has a spread of about a factor of 2-3 (see Table \ref{table:MassAccretionRates}). The average accretion rates on JD = 2459654.4899 (around the peak of the outburst) and 2459736.539 (almost around quiescence) are 3.2$\pm$0.3 $\times$ 10$^{-8}M_{\odot}$/yr and 1.0$\pm$0.1 $\times$ 10$^{-8}M_{\odot}$/yr, respectively. An average from the last three epochs gives a quiescence mass accretion rate of 7.7$\pm$0.4 $\times$ 10$^{-9}M_{\odot}$/yr that indicates a factor of 4 increase in mass accretion rate during the outburst. \citet{2023ApJ...957..113W} estimated a mean mass accretion rate of 1.74 $\times$ 10$^{-8}M_{\odot}$/yr during the peak of the outburst and 3.33 $\times$ 10$^{-9}M_{\odot}$/yr after the completion of the outburst. The authors used $U$-band excess flux over the photospheric level to estimate the mass accretion rate. The authors used a different extinction of A$_V$=0.1 mag, and this caused a small difference in mass accretion rates estimates from that of ours. \citet{2023A&A...678A..88C} performed slab modelling and estimated a factor of seven change in the mass accretion rate from quiescence to the peak of the outburst which is consistent with that of our result and from that of \citet{2023ApJ...957..113W}.

From the detailed analysis of the long-term multiband photometry and multi-epoch high resolution spectra, the main observational conclusions are enumerated below.

\begin{itemize}
    \item[\textbf{O1}:] The RV phase change was not observed among the spectra from pre- and post-outburst of the year 2008.
    \item[\textbf{O2}:] Lightcurve and RV phases changed during the 2022 March outburst in multiband data. 
    \item[\textbf{O3}:] Phase increased by $\sim$112$\degree \pm$ 5$\degree$ in the forward direction of EX Lupi rotation.
    \item[\textbf{O4}:] While the multi-band lightcurves had similar phase values during pre-outburst, they all started showing different phase values among themselves during the outburst.
    \item[\textbf{O5}:] $U, B$, and $V$ band phases didn't revert after the completion of the outburst, while $R$ and $I$ did revert immediately.
    \item[\textbf{O6}:] Phase decreased differentially in optical and infrared lightcurves but not in $U$ and $B$ bands.
    \item[\textbf{O7}:] The ASAS-SN $g$-band phase decreased to pre-outburst value during the time interval of 2023 July 16 (JD = 2460140.7206) to 2023 September 06 (JD = 2460190.7206)
    \item[\textbf{O8}:] The RV amplitude of the emission lines increased during the outburst.
\end{itemize}

The following sections shall scrutinize these observational results.

\section{Discussions and Interpretation}\label{sec:DiscussionAndInterpretation}

To aid the discussion, we start by defining coordinate systems to
characterize the phenomenon of phase change.

\subsection{Coordinate System}\label{subsec:CoordinateSystem}

We define two coordinate systems, one fixed on the rotating star ((X,Y,Z) $\equiv$ (r,$\theta$,$\varphi$), called \textit{fixed frame}) and the other one fixed with respect to the hotspot ((X$^{'}$,Y$^{'}$,Z$^{'}$) $\equiv$ (r$^{'}$,$\theta^{'}$,$\varphi^{'}$), \textit{variable frame}). Z and Z$^{'}$ axes are aligned to the rotation axis of the star, and the X-Y and X$^{'}$-Y$^{'}$ planes coincide. The angle between X and X$^{'}$ axes is the phase of the lightcurve at any given epoch, which is $\sim$70$\degree$ at our startJD (in $V$-band, see the first diamond grey phase point in Figure \ref{fig:phaseVariation_AllLC}).

\begin{figure}[!ht]
    \centering
    \includegraphics[scale=0.35]{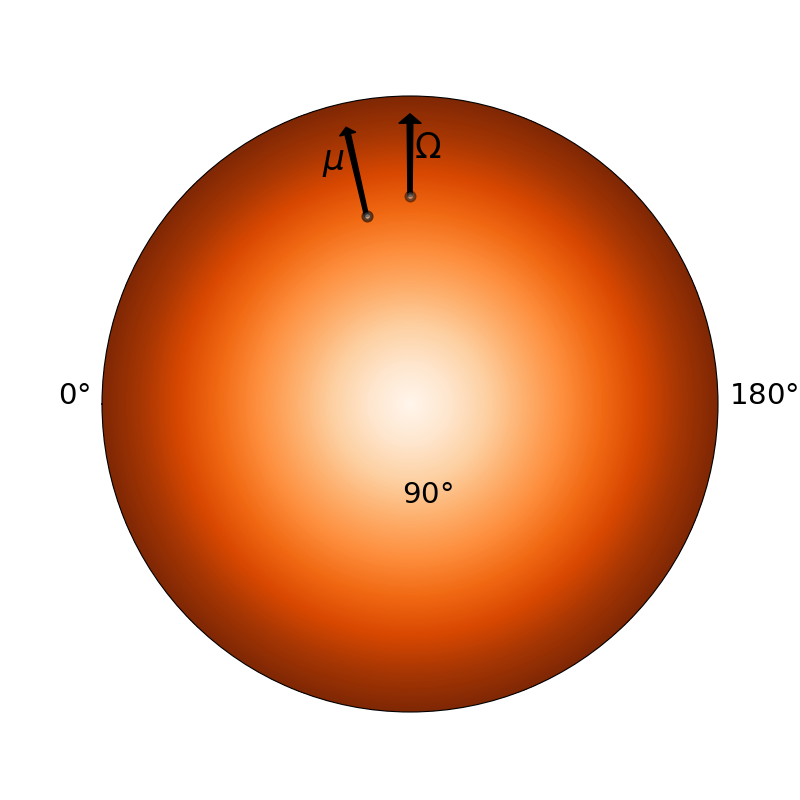}
    \caption{Azimuthal angle sign convention to be followed in this work. 0$\degree$ and 180$\degree$ are the tangential left and right points, respectively, while 90$\degree$ is facing us. The two arrows corresponding to $\Omega$ and $\mu$ are rotational and magnetic axes, respectively. The star is shown to be rotating anticlockwise when viewed from above the $\Omega$ axis.}
    \label{fig:phasedefinition}
\end{figure}

The physical interpretation of the RV phases and lightcurve phases are different. The same value of RV and lightcurve phases do not indicate the same location of the hotspot on the stellar surface. The photometric phase is maximum (90$\degree$ of the sine fit) when the hotspot is facing us (90$\degree$ in Figure \ref{fig:phasedefinition}) while the RV phase is maximum (90$\degree$ of the sine fit to RVs) when the hotspot is at the tangential right side point (180$\degree$ in Figure \ref{fig:phasedefinition}). Redshifted RV is positive in our convention. Thus, the hotspot location is given by the photometric lightcurve phase value, or equivalently, the RV phase + 90$\degree$.

\subsection{Detection of the azimuthal movement of hotspot via phase measurement}\label{subsec:hotspotAzimuthmovement}

The phase of a lightcurve depends upon the location of the hotspot with respect to the observer's line of sight. An increase in phase implies that the hotspot is observed at an azimuthal position larger than 90$\degree$, at a time when it is expected to be at 90$\degree$. Since the rotation period of EX Lupi is found to be consistent with 7.417 days (see Appendix \ref{App:EXLupiPeriod}), the location of the hotspot should move forward over the stellar surface to explain the phase change in the lightcurve. In terms of the coordinate system, (X$^{'}$,Y$^{'}$,Z$^{'}$) frame rotated anticlockwise with respect to (X,Y,Z) frame when viewed from the top (i.e., from above in Figure \ref{fig:phasedefinition}, this rotation direction is a convention used in this paper). 

The hotspot is an observable component of matter carried by the funnel streams. The freely falling accreting matter along the accretion funnel defines the location of the hotspot over the stellar surface. A change in the hotspot location over the stellar surface indicates a change in the accretion funnel geometry, which depends on the relative rotation of the magnetosphere and the inner disk and also on the stellar magnetic field configuration. For EX Lupi, \citet{2015A&A...580A..82S} favour a dipolar magnetic field configuration, since it is a low mass \citep{2014IAUS..302...44G}, fully convective star. The magnetic field configurations of typical YSOs are observationally found to be dipolar and also multipolar \citep[e.g.,][]{2014MNRAS.437.3202J}. Owing to the unavailability of the magnetic field configuration of EX Lupi, the discussions concerning the accretion funnel geometry shall assume a dipolar magnetic field configuration. 

In the subsection below, we propose a hypothesis to interpret the azimuthal movement of the hotspot over the stellar surface. 

\subsection{Hotspot's movement due to shifting of the accretion funnel}\label{subsec:Shifting Accretion Funnel}

The accretion onto EX Lupi, with dipolar magnetic field topology, is expected to be dominated by two accretion funnels which dump the infalling matter near magnetic pole, at high latitudes. The hotspot's azimuthal position on the stellar surface depends upon the inner disk radius (which is $r_{m}$) and $r_{c}$, i.e., effectively on the relative angular velocity between the inner disk and the star. It can be understood as follows: the Keplerian angular velocity falls as $\propto$ $r^{-3/2}$. The inner disk radius decreases when the accretion rate increases (see Equation \ref{Equ:MagnetosphericRadius}), and the infalling matter will have higher angular velocity than the prior accretion state. Since the star and its magnetosphere rotate slower, they will start lagging behind the faster rotating infalling matter. Thus, the matter falls at a larger longitude, along the disk rotation, when the accretion rate increases. This creates the hotspot at a larger longitude along the disk rotation. Observationally, it would appear as a phase change (increase) in the periodic lightcurve and RV.

As the hotspot moves to a larger longitude over the stellar surface, it would result in an azimuthal forward shifting of the dipolar accretion funnel. Hence, we name this phenomenon `shifting of the accretion funnel'.


\citet{10.1093/mnras/stt945} performed 3D MHD simulations of magnetospheric accretion and combined a range of parameters to derive relations governing the shape and size of the hotspot. We used Equation 6 from \citet{10.1093/mnras/stt945} to estimate the azimuthal location of the hotspot at various instances of the outburst in EX Lupi:

\begin{equation}\label{Equ:TheoreticalPhase}
    \phi = 300\degree - 57\degree \frac{r_{m}}{R_{*}}~,
\end{equation}
where $r_{m}$, the magnetospheric radius is defined by the balance of disk ram pressure and magnetic field pressure. Since the azimuthal position of the hotspot depends upon the $r_{m}$ (Equation \ref{Equ:TheoreticalPhase}), we estimate $r_{m}$ for EX Lupi at the epochs we have the accretion rates from our HRS-SALT by equating the disk ram pressure to the magnetic field pressure as follows

\begin{equation}\label{Equ:MagnetosphericRadius}
    \frac{r_{m}}{R_{*}} =    k7.1B^{\frac{4}{7}}_{3}\dot{M}^{\frac{-2}{7}}_{-8}M^{\frac{-1}{7}}_{0.5}R^{\frac{5}{7}}_{2}~,
\end{equation}
where $r_{m}$ and $R_{*}$ are pre-defined parameters. $B_{3}$ is 
the magnetic field strength in units of kG. $\dot{M}_{-8}$ is mass accretion rate in units of 10$^{-8}$ $M_{\odot}$/yr. $M_{0.5}$ and $R_{2}$ are stellar mass and radius in units of 0.5$M_{\odot}$ and 2$R_{\odot}$, respectively \citep{2007prpl.conf..479B}. $k$ is a factor for correcting the spherical to disk-mediated accretion, where $k \approx$ 0.5 (e.g., \citealt{2005ApJ...634.1214L})\footnote{ Note that \citet{2005ApJ...634.1214L} and \citet{10.1093/mnras/stt945} found that the magnetosphere is typically slightly compressed and is not perfectly dipole, and dependencies in Equation \ref{Equ:MagnetosphericRadius} are somewhat different. \citet{10.1093/mnras/stt945} have shown that for small magnetospheres Equation \ref{Equ:MagnetosphericRadius} is still valid, if $k\approx 0.5-0.7$.}. 

The values of the stellar parameters are described in Section \ref{sec:intro}. The mass accretion rates used in Equation \ref{Equ:MagnetosphericRadius} are estimated from the spectral lines, for 16 epochs (see Section \ref{subsec:MassAccretionRate} and Table \ref{table:MassAccretionRates}). Substituting the value of $r_{m}/R_{*}$ in Equation \ref{Equ:TheoreticalPhase}, we calculated the model predicted azimuthal positions of the hotspot on 16 epochs.

\subsubsection{Hotspot location during the rise of the outburst}\label{subsub:TheoreticalHotspot_riseOfOutburst}

We estimated the predicted relative change in the hotspot's azimuthal position by subtracting the model predicted quiescent state azimuthal position from that of the model predicted outburst state position. Using the accretion rate from our first epoch spectra (JD = 2459654.4899), we estimated the hotspot's azimuthal position predicted by the model in the outburst state as 74$\degree$.3 $\pm$ 6$\degree$.3, as this epoch is close to the outburst peak (see Figure \ref{fig:CaII_onLC}). To estimate the quiescent state azimuthal position of the hotspot, we averaged the hotspot's azimuthal position inferred from the last three epoch spectra (-43$\degree$.2 $\pm$ 5$\degree$.5) as they correspond close to the quiescence (JD = 2459732.5452, 2459735.5317, 2459736.539). Thus, the calculated change in the hotspot's azimuthal position from quiescence to the peak of the outburst is 117$\degree$.5 $\pm$ 8$\degree$.3. This is consistent with observations of phase increase by 112$\degree$ (see Section \ref{subsec:PhaseOfLC} and Figures \ref{fig:phaseVariation_AllLC} and \ref{fig:RV_phasechange}).

We would like to note that, due to the periodic variability in accretion rate towards the tail end of the outburst decay, the choice of how we calculate the accretion rate of the quiescent state slightly alters this calculation. For instance, if we choose the azimuthal position estimate from our last spectroscopic observation as the quiescence hotspot location, then the phase change is 88$\degree$.9 $\pm$ 10$\degree$.6.  Due to this uncertainty of the true quiescence accretion rate, the above-estimated phase change (using equation \ref{Equ:TheoreticalPhase}) would have a difference of 15$\degree$-30$\degree$, depending upon the choice of the quiescence accretion rate.

\subsubsection{Hotspot location during the decay of the outburst}\label{subsub:TheoreticalHotspot_decayOfOutburst}

In order to understand the evolution of the hotspot location during the outburst decay, a straight line is fitted onto the model predicted azimuthal locations ($\phi$) estimated from Equation \ref{Equ:TheoreticalPhase}. This model predicted phase decay curve is plotted in Figure \ref{fig:phaseVariation_AllLC} as a black dashed line. For visual clarity, the first epoch of the predicted phase decay line is shifted to match the closest AAVSO-$U$ band phase. One can see in Figure \ref{fig:phaseVariation_AllLC} that the AAVSO-$U$ band phases did not revert as per the dashed line while $R$ and $I$ bands followed the model predicted decay curve. 

\smallskip 
Our hypothesis of `shifting of the accretion funnel' predicts a phase variation in tandem with the accretion rate variation following Equation \ref{Equ:TheoreticalPhase}. Hence, this hypothesis could explain \textbf{O1}: The increased phase could have reverted back after the year 2008 outburst was over. The first
quiescent state spectrum was taken nearly five months after the completion of the 2008 outburst \citep{2014A&A...561A..61K}, and thus, the RVs estimated among the pre-and post-outburst states would have missed any phase change. This hypothesis also explains the \textbf{O2} and \textbf{O3}: hotspot's longitude position changed during the outburst in the forward direction of disk rotation. However, this hypothesis fails to explain \textbf{O5} and \textbf{O6} completely: the $U, B$, and $V$ band phases did not revert back after the completion of the outburst. 

\smallskip 
To explain \textbf{O5}, we hypothesize a \textit{thicker and warmer} disk post-outburst. Outbursts are expected to heat up inner parts of the disk. The detections of 2.29 micron CO band heads in emission in some YSOs are believed to be an indication of this phenomenon \citep{2009ApJ...693.1056L}. There has also been evidence of outbursts heating up outer disks and annealing silicates (in EX Lupi) \citep{2009Natur.459..224A}, as well as moving the water snowline outward (in V883 Ori, albeit this is much more powerful FUor) \citep{2016Natur.535..258C}. The heated inner disk will be puffed up, and the disk will also start receding from the star with a decrease in the accretion rate. However, in the geometrically thicker disk, the funnel matter flows from higher altitudes of the disk, where the lifting force is smaller. This may help to support a high azimuthal shift of the hotspot for a longer time, compared with the case of a thinner disk. The funnel would carry most of the accreted mass and hence the corresponding hotspot will emit mostly in shorter wavelengths. This could explain a delayed phase decay in the AAVSO-$U, B$, and $V$ bands. In this scenario, the phases will decay on the cooling timescale of the heated inner disk.

The observed phenomenon of persistent phase shift after the outburst (O5) is one of the most difficult observations to explain. The above scenario of the disk heating and thickening is one of the possibilities. In Appendix \ref{app:otherPhasechangeHypothesis} we discuss a few other hypotheses that may explain the persistent phase shift in $U, B$ and $V$ bands after the outburst. \\

\subsection{3D MHD simulation of the accretion onto EX Lupi}\label{subsec:3DMHDsimulation}

We simulated the accretion onto a star with parameters of  EX Lupi to understand the phase shifts of the hot spots. We used our earlier developed 3D MHD Godunov-type code with a cubed sphere grid, which has been developed earlier and used to study the magnetospheric accretion \citep[e.g., ][]{10.1093/mnras/stt945,2016MNRAS.459.2354B}. Here, we used this code to model funnel streams and spots in EX Lupi. We developed a model of a star with parameters of EX Lupi as discussed in Section \ref{sec:intro}: mass and radius of the star, 0.6$M_{\odot}$, 1.6$R_{\odot}$, its period: 7.417 days and corresponding corotation radius: 8.44R$_{*}$. The tilt of the magnetospheric axis $\Theta=13^\circ$  and the stellar magnetic field $B$ = 3 kG.

\begin{figure*}
\gridline{\fig{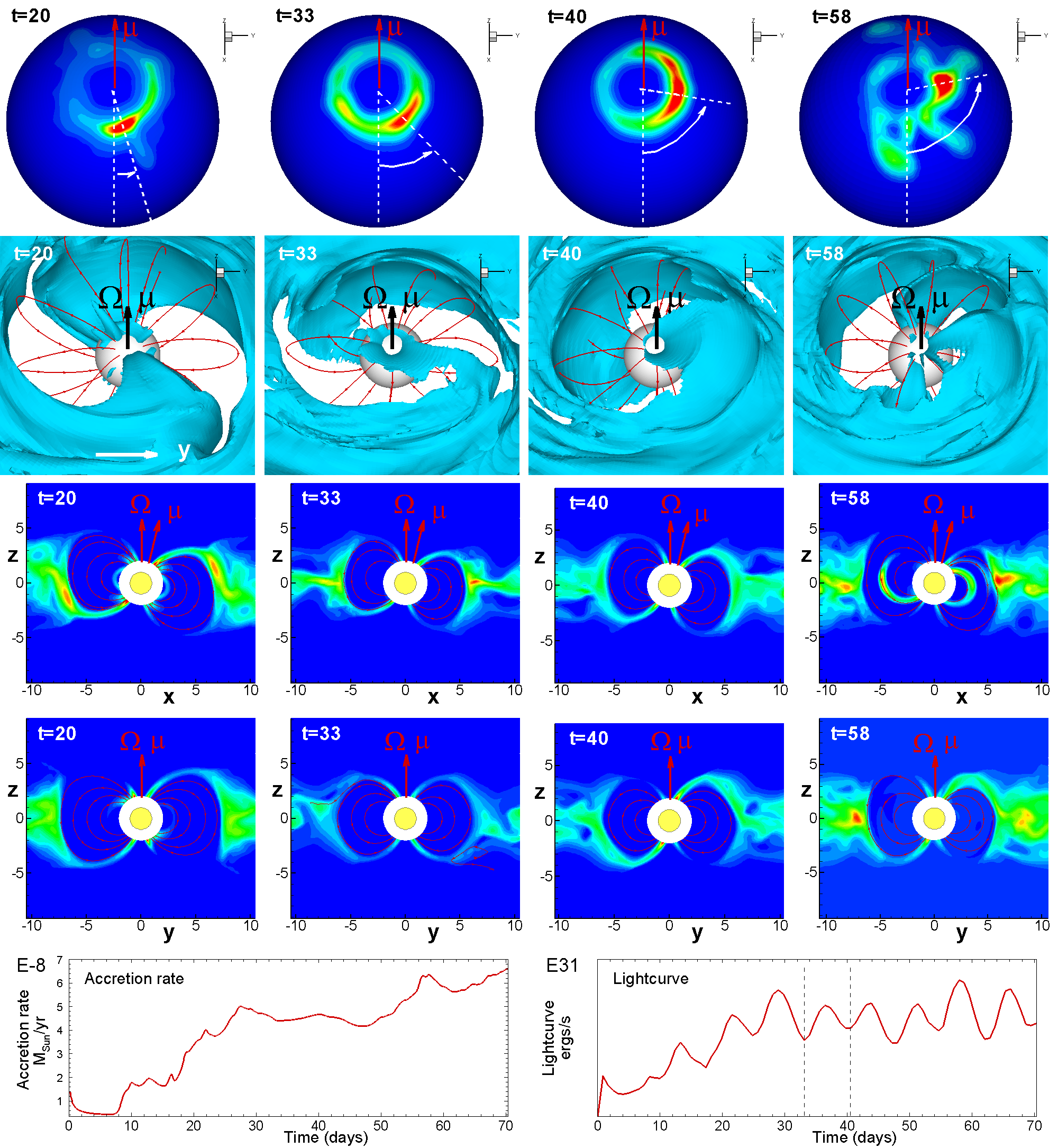}{0.95\textwidth}{}}
\caption{Results of 3D MHD simulations of the model with parameters of EX Lupi. \textit{1st row from the top:} The distribution of the kinetic energy of the infalling matter at the surface of the star 
at different moments of time $t$ counted from the beginning of simulations. Time is measured in days. Red and dark blue colors show areas of the highest and lowest kinetic energy, respectively. Vectors $\mu$ and $\Omega$ show the directions of the magnetic and rotational axes of the star. The white arrow shows the phase shift. Spots are shown at the inclination angle of the observer $i=32^\circ$ and at the zero phase of stellar rotation. \textit{2nd row:}  3D views of matter flow at different moments of evolution. One of the density levels is shown for clarity. Lines are selected magnetic field lines.
 \textit{3rd row:} Density distribution in the $\mu-\Omega$ plane.   \textit{4th row:} Density distribution in the plane perpendicular to $\mu-\Omega$ plane. Red and dark blue colors show the highest and lowest densities, respectively.
 \textit{Bottom left panel:} the accretion rate onto the surface of the star. 
  \textit{Bottom right panel:} The lightcurve calculated for inclination angle $i=32^\circ$. }
\label{fig:3DMHDsimulation}
\end{figure*}

The outburst in accretion rate is caused by some processes that are not well understood and which we cannot presently model numerically due to the unknown physics of the process and its possible complexity. Instead, we modelled parts of the process corresponding to different moments of time in the outburst.  The result of simulations (the location and the shape of spots) depends on several dimensionless parameters: the truncation (magnetospheric) radius $r_m/R_*$, the corotation radius $r_{\rm c}/R_*$ (or, $\omega_s$), and the tilt of the magnetosphere $\Theta$. Earlier, \citet{10.1093/mnras/stt945} investigated shapes and phases of spots in models at a wide range of these parameters and derived different dependencies and correlations. Typically, they combined results for a wide range of magnetospheric tilts, taking, e.g.,  values in the range from  $\Theta=2^\circ-5^\circ$ up to   $\Theta=20^\circ-30^\circ$.  Here, we recalculated the model using a tilt of the magnetosphere  $\Theta=13^\circ$ corresponding to EX Lupi and at several values of  $r_m/R_*$ which may correspond to different stages of EX Lupi outburst.  We also used twice as fine a numerical grid compared with \citet{10.1093/mnras/stt945} to ensure even more accurate simulations. To speed up the simulation, the star is placed inside the inner boundary (see Figure \ref{fig:3DMHDsimulation}) such that the inner boundary of the simulation region is 2$R_{*}$.

Initially, we placed the inner disk away from the star at the corotation radius. This provided more precise initial equilibrium \citep[e.g.,][]{2003ApJ...595.1009R}. Subsequently, the disk matter moved towards the star and started accreting. This led to the gradual rise of the accretion rate shown in the bottom left panel of Fig. \ref{fig:3DMHDsimulation}.

Simulations show that when the disk matter starts approaching the star, it is stopped by the magnetosphere at the distance of $r_m\approx 6-7 R_*$ and formed two funnel streams  (see 3D views of matter flow in Figure \ref{fig:3DMHDsimulation}; second row from the top, leftmost panel at $t=20$. At this time, the magnetosphere is relatively large, and this moment may correspond to the state which is close to the pre-outburst state. The top left panel shows that the hotspot shifted only slightly from the $\mu-\Omega$ plane (which is the plane of the tilt of the magnetic axis). Subsequently, at $t=33$, the accretion rate increased, the inner disk moved closer to the star, and the spot was shifted by a larger angle, $\sim 35^\circ$ (see 3D plot and spot at $t=33$ ). Later, at $t=40$, the accretion rate increased, and the inner disk moved the base of the funnel stream to a larger angle such that the spot was shifted by $\sim 85^\circ-100^\circ$. 

At moments of enhanced accretion rate at $t>50$), we observed the transition from stable to the mildly unstable regime where both funnels and tongues were observed (see the rightmost panels in the same figure at $t=58$.). Instability started to appear when 
the magnetospheric radius decreased to values of $r_m\approx$ 5.5-5.8 and the fastness parameter $\omega_s$ = ($r_{m}$/r$_{c}$)$^{3/2}$ $\approx$ 0.53-0.57 which corresponds to mildly unstable regime \citep{2016MNRAS.459.2354B}\footnote{In \citet{2016MNRAS.459.2354B} the boundary between stable and unstable regimes corresponds to $\omega_s$=0.6 and $\omega_s$=0.54 in cases of the dipole obliquities of theta=5$^\circ$ and 20$^\circ$, respectively.}. In this regime, matter flows through different channels however most energetic part of the spot with the highest kinetic energy of the flow was determined by the modified funnel stream and was shifted by $\sim 100^\circ$. The 3rd and 4th rows from the top show the slices of density distribution in the $\mu-\Omega$ plane and in the perpendicular plane (where the flow is expected in case of strong phase shift). One can see that at $t=20$ (small shift), the funnel streams are stronger in the $\mu-\Omega$ plane, while at $t=58$ (strong shift), streams are stronger in the perpendicular plane.

\begin{figure*}
    \centering
    \includegraphics[scale=0.22]{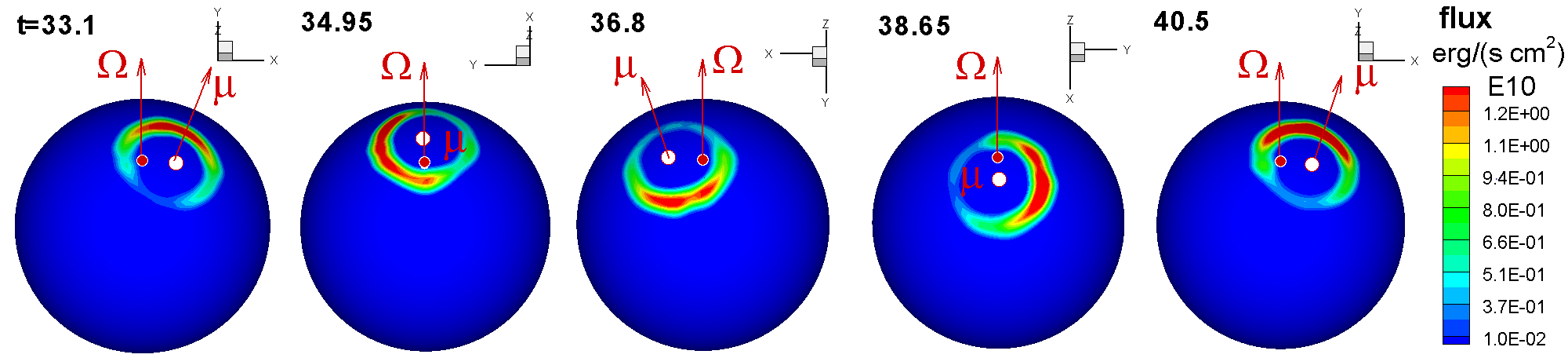}
    \caption{Hot spots at different phases of stellar rotation during one rotational period of the star. The color background shows the flux of the kinetic energy density on the surface of the star. $\mu$ and $\Omega$ are vectors of the magnetic and rotational axes of the star.}
\label{fig:3DMHDrotation}
\end{figure*}

We calculated the lightcurve for observer located at the inclination angle $32^\circ$ (see bottom right planel of Figure \ref{fig:3DMHDsimulation}). Here, we suggested that all kinetic energy of matter falling to the surface of the star is re-radiated isotropically.  The rotation of the spot leads to variation of the amount of radiation towards the observer, and the lightcurve shows maxima and minima.  We chose one period of stellar rotation (shown between dashed vertical lines in Figure \ref{fig:3DMHDsimulation} for lightcurve) and show the location and shape of spots at different moments in time and different phases of stellar rotation (see Figure \ref{fig:3DMHDrotation}). One can see that at the inclination angle of $32^{\circ}$ the observer will always see the spot. The maximum in the lightcurve corresponds to the time $t=36.8$ when the spot is located closer to the center of the star, facing us, and the two minima correspond to the times $t=33.1$ and $t=40.5$ when the spot is closer to the edge of the star.

\subsection{Chromatic phase shift during the outburst period}\label{subsec:ChromaticPhaseChange}

The analysis of Section \ref{subsec:PhaseOfLC} showed that the TESS and nearby epoch ASAS-SN phases were consistent during the quiescence of pre 2022 March outburst. Further at the onset of the outburst, $g^{\prime}$, $r^{\prime}$, $i^{\prime}$ and $z^{\prime}$ band phases are all consistent within 10$\degree$ (see Section \ref{subsec:PhaseOfLC} and Figures \ref{fig:phaseVariation_AllLC} and \ref{fig:AAVSOphaseEvolution}). This implies that the azimuthal location is the same for the regions emitting across the wavebands. Therefore, the hotspot was azimuthally confined and nearly single-temperature on the EX Lupi during the pre-outburst quiescence. However, the AAVSO $U, B, V, R$ and $I$-band lightcurves observed during the peak of the outburst (see Section \ref{subsec:PhaseOfLC}) showed finite differences among the hotspot longitude position inferred for $U, B, V, R$ and $I$-bands (see \textbf{O4} and Figure \ref{fig:AAVSOphaseEvolution}). The inferred hotspot longitude positions for wavebands among AAVSO $U, B, V, R$ and $I$ have a decreasing trend. $U$-band lightcurve peaks first followed by $B, V, R$ and $I$-band lightcurves, sequentially. The maximum phase difference between $U$ to $I$ band is 36$\degree$. The longitude inferred from the CTIO $B$-band lightcurve is also larger than that of the CTIO $V$-band lightcurve by 10$\degree$.7. Similarly, \citet{2021Natur.597...41E} also found phase differences among UV and optical lightcurves of the star GM Aurigae where the peak of the UV lightcurve precedes that of the optical lightcurve. The authors suggested it to be a consequence of azimuthal asymmetry in the hotspot. For EX Lupi, this indicates a temperature gradient in the hotspot where the front end, along the rotation, is hottest, subsequently followed by cooler and cooler regions. Spread in multiband phases during the outburst could also refer to a chromatic azimuthal enlargement of the hotspot. This could further be interpreted as a broadening of the accretion funnel to channel a large amount of mass during the outburst.

This chromatic phase difference could also be a consequence of the appearance of many azimuthally nearby funnels during the outburst. \citet{2016MNRAS.459.2354B} showed that the accretion becomes unstable when $\omega_{s} \leq$ 0.6 for magnetic obliquity of 5$\degree$ (close to that of EX Lupi). Under this condition, many equatorial matter-carrying tongues penetrate the magnetosphere and form chaotic spots near the magnetic equator. But under this condition, the lightcurves are expected to be irregular. The lightcurves become periodic with smaller periods at other extremes when $w_{s} \leq$ 0.45. We did a periodogram with the declining lightcurves after the outburst, and we didn't find any statistically significant period value smaller than the 7.417 days. We also calculated the $w_{s}$ at the epochs of our spectra and found that the accretion lies in the mildly unstable regime for the initial epochs and then turns to the stable regime of accretion in the later epochs. Upon varying the value of `k' in between 0.5 and 0.7 in Equation \ref{Equ:MagnetosphericRadius}, EX Lupi stayed around mild unstable (appearance of a few equatorial tongues with profound dipolar funnel) to the stable regime (accretion is channelled by dipolar funnel).

Not only did the inferred longitude from multiband lightcurves become different during the outburst, but in the subsequent post-outburst state, the inferred longitude of the wavebands with longer effective wavelengths decreased much faster than that of the wavebands with shorter effective wavelengths.

\subsection{Evolution of chromatic phase shift post-outburst}\label{subsec:AAVSOPhaseEvolution}

\begin{figure}[!ht]
    \centering
    \includegraphics[scale=0.33]{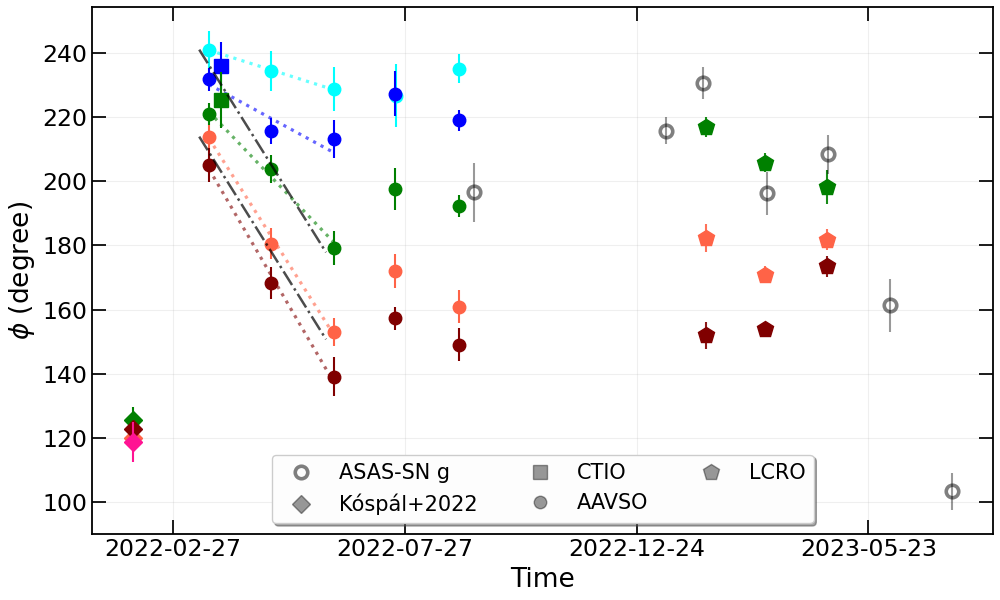}
    \caption{The evolution of multiband phases from the onset of the outburst to the subsequent post-outburst are shown. A straight line fit (dotted line) is plotted on the first three phase points of AAVSO multiband phases (during outburst decay) as they follow a linear trend. The phase values of the CTIO lightcurves are shown near the peak of the outburst. Other phases are plotted from \citet{2022RNAAS...6...52K}, ASAS-SN, and LCRO lightcurves. The grey dash-dot lines are the model predicted phase decay curve estimated from the accretion rate during the outburst, as discussed in Section \ref{subsec:Shifting Accretion Funnel}. The model-predicted phase decay curve is shifted to match the AAVSO phases at the epoch of the first spectrum observation, shown for $U$ and $R$ bands. The color-code for the various bands are: cyan ($U$), blue ($B$), green ($V, g^{\prime}$), orange ($R, r^{\prime}$), brown ($I, i^{\prime}$) and pink ($z^{\prime}$).}  
    \label{fig:AAVSOphaseEvolution}
\end{figure}

During the decay of the outburst, the phases decreased differentially among $U, B, V, R$, and $I$-bands (see Figure \ref{fig:AAVSOphaseEvolution}). The slopes of the multiband phase decays are- $U:$-0.153, $B:$-0.263, $V:$-0.504, $R:$-0.758, $I:$-0.824 (degree/day). As we discussed in Section \ref{subsub:TheoreticalHotspot_decayOfOutburst}, $R$ and $I$ band lightcurves followed the model predicted phase decay but the warmer and thicker inner disk could have delayed/slowed the phase decay among the shorter wavelengths. Interestingly, EX Lupi maintained chromatic phase differences even after the outburst was over. We found chromatic phase differences in our LCRO lightcurves taken from 2023 February to 2023 May. We suspect that the occasional small bursts and variability seen in the ASAS-SN lightcurve during the post-2022 March outburst kept the inner disk hot and thick. A hot disk could maintain the larger amount of matter flow along the `shifted funnel' and support the hotspot at a larger azimuth that corresponds to emission at a shorter wavelength. However, a smaller amount of matter might have been flown by the azimuthally lagging funnels showing the phase decay at the longer wavelengths. These chromatic phase differences showed long-term sustainability of a temperature gradient in the hotspot as it was observed among LCRO $g^{\prime}$, $r^{\prime}$ and $i^{\prime}$ band lightcurves, one year after the outburst. It explains the persistent chromatic phase differences in the post-outburst state. If this hypothesis is true, it suggests that the chromatic phases should return to quiescent values in the relaxation timescale of the puffed-up hot inner disk. To verify this, we recommend a further high-cadence multiband observation of EX Lupi during its quiescence, akin to that of 2019.

\begin{figure*}
\gridline{\fig{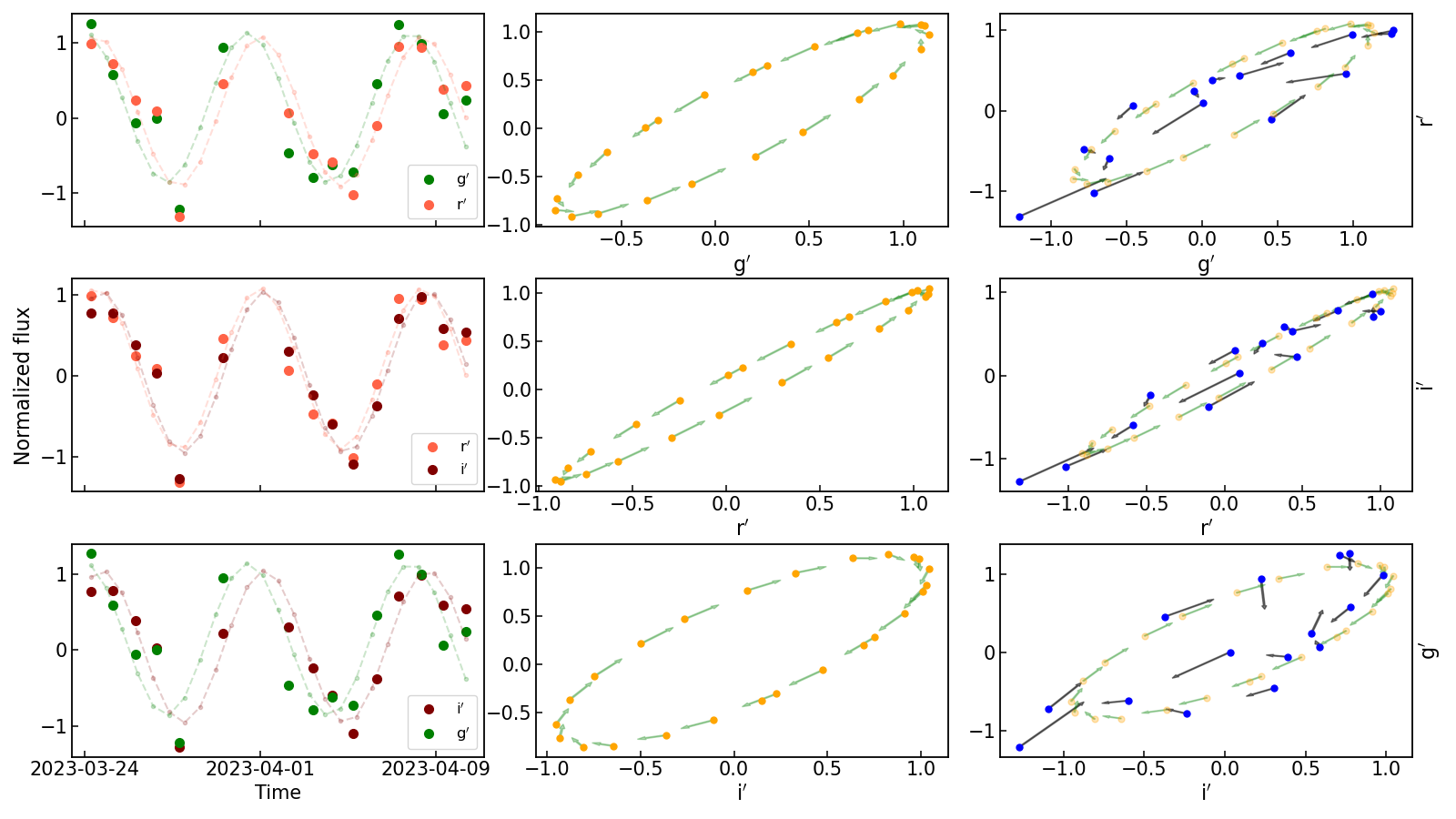}{1\textwidth}{}}
\caption{Lissajous figures for the LCRO $g^{\prime}$, $r^{\prime}$ and $i^{\prime}$ band lightcurves during JD = 2460127 - 2460145. \textit{Left column}: Normalized lightcurve cutouts along with their best sine fit are plotted for $g^{\prime}$, $r^{\prime}$ and $i^{\prime}$ band by green, orange and brown colors, respectively. The photometry is 1-day averaged. \textit{Middle column}: The yellow dots are the Lissajous figure for the sine fit onto the respective lightcurves, with green arrows pointing to the nearest next epoch. \textit{Right column}: The normalized flux-flux plots (Lissajous figures) for combinations among $g^{\prime}$, $r^{\prime}$ and $i^{\prime}$ bands (blue dots). Black arrows connect to the nearest next epoch observation data. The Lissajous figures from the middle panel are shown in this panel for comparison.}
\label{fig:LCOBOT_lissagous}
\end{figure*}

Further, to visualize the phase difference among $g^{\prime}$, $r^{\prime}$ and $i^{\prime}$ bands, we introduce Lissajous figures for $g^{\prime}$-$r^{\prime}$, $r^{\prime}$-$i^{\prime}$ and $g^{\prime}$-$i^{\prime}$ band lightcurves. The lightcurves are normalized before plotting the respective Lissajous figures. The data were averaged for one-day observations before fitting the sine curve for the plots in Figure \ref{fig:LCOBOT_lissagous}. Two sine curves with a $\delta$ phase difference, when plotted onto the X- and Y-axes, produce an ellipse. The ratio of the semi-minor to semi-major axis varies between 0 ($\delta$ = 0$\degree$, the ellipse becomes a straight line) and 1 ($\delta$ = 90$\degree$, the ellipse becomes a circle). The ratio of the semi-minor to semi-major axis is maximum for g$^{\prime}$-i$^{\prime}$ than the other two combinations, indicating a larger phase difference between $g^{\prime}$ and $i^{\prime}$ bands (see Figure \ref{fig:phaseVariation_AllLC}) than the phase differences between $g^{\prime}$-$r^{\prime}$ and $r^{\prime}$-$i^{\prime}$.

During the final stages of this manuscript preparation, an independent study by \citet{2023ApJ...957..113W} also reported the chromatic phase shift in EX Lupi post-outburst.

\subsection{Larger phases after one year of the outburst: LCRO and ASAS-SN lightcurves}\label{subsec:LCOBOTphase}

The phase values from the multiband LCRO and ASAS-SN lightcurves after many months of the outburst are essential to understanding the long-timescale evolution of the system. Our LCRO observations in $g^{\prime}$, $r^{\prime}$ and $i^{\prime}$ taken one year after the 2022 March outburst show the light curve phases were still around 200$\degree$, 180$\degree$ and 150$\degree$ in $g^{\prime}$, $r^{\prime}$ and $i^{\prime}$ band lightcurves respectively (see Figures \ref{fig:phaseVariation_AllLC} and \ref{fig:AAVSOphaseEvolution}). These LCRO multiband phases are comparable to the respective band phase values of the AAVSO observations in September 2022 ($g^{\prime} \sim V$, $r^{\prime} \sim R$ and $i^{\prime} \sim I$; see Figure \ref{fig:phaseVariation_AllLC}).

These elevated phases for more than a year after the March 2022 outburst are in sharp contrast to the major outburst in 2008. After the 2008 outburst, there was no indication of an elevated phase after about 5 months of the outburst \citep{2014A&A...561A..61K}. If a similar phase change had happened in the 2008 outburst, it relaxed to the quiescent phase much faster. ASAS-SN $g$-band lightcurve shows the flux from the EX Lupi did not fully return to pre-outburst level after the March 2022 outburst (See lower panel of Figure 9). Between March 2022 and August 2023, there have been numerous low amplitude brightening. Sometimes brightening as high as $\sim$1.2 mag in $g$-band during February and March 2023. If our hypothesis is true, these higher levels of accretion activity might have kept the inner disk warm and thicker to support the elevated azimuth of the accretion funnel.

\subsection{Phase decay after one and half year of the outburst: ASAS-SN view}\label{phaseDecay}

\begin{figure*}
\gridline{\fig{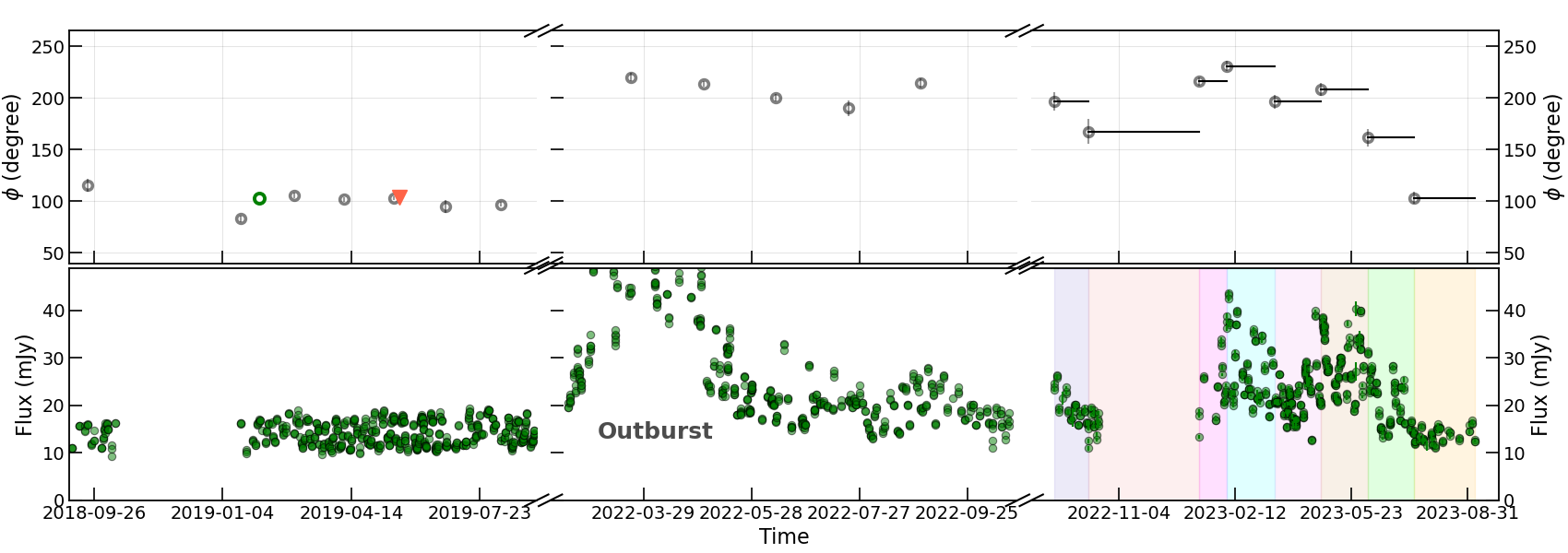}{1\textwidth}{}}
\caption{\textit{Top panel}: ASAS-SN g-band phase evolution is shown during the pre- and post-2022 March outburst. The Y-axis is the phase in degrees, and the X-axis is time. The grey circles are phases from the ASAS-SN $g$-band lightcurve while the green circle is a manually estimated phase from ASAS-SN $g$-band lightcurve cutout (see Section \ref{subsec:PhaseOfLC} and Figure \ref{fig:phaseVariation_AllLC}). The orange-colored inverted triangle is the phase from the TESS sector-12 lightcurve. The horizontal line attached to the grey circles shows the range of epochs used for phase estimation. \textit{Bottom panel}: ASAS-SN $g$-band lightcurve is plotted with sections of colors showing the data used for phase estimation for the respective phase values in the top panel. The Y-axis is flux in mJy, while the X-axis is time in years. Only a portion of the outburst is shown. The outburst flux increased to $\sim$ 90 mJy.}
\label{fig:ASAS_SNPhaseDecay}
\end{figure*}

The ASAS-SN $g$-band lightcurve of EX Lupi covering from 2023 January 12 (JD = 2459956.8644) to 2023 September 06 (JD = 2460194.3491) showed the larger phases. The phase of the lightcurve remained similar to the value calculated just after the outburst (see Figures \ref{fig:phaseVariation_AllLC}, \ref{fig:AAVSOphaseEvolution} and \ref{fig:ASAS_SNPhaseDecay}). But, the latest 50 days of the ASAS-SN g-band lightcurve from 2023 July 16 (JD = 2460140.7206) to 2023 September 06 (JD = 2460194.3491) shows the phase to have decreased to the pre-outburst value (see Figure \ref{fig:ASAS_SNPhaseDecay}). This latest phase value is 103$\degree \pm$ 6$\degree$ while the pre-outburst phase values estimated for the manually selected clean lightcurve cutouts (see Section \ref{subsec:PhaseOfLC}), epochs close to the TESS sector 12 and sector 39 observations, are 102$\degree \pm$ 3$\degree$ and 111$\degree \pm$ 3$\degree$, respectively. The bottom panel of the Figure \ref{fig:ASAS_SNPhaseDecay} shows the brightness level variation of the EX Lupi in the pre outburst, during the outburst and post-outburst state. EX Lupi did not completely return to the pre-outburst brightness level for more than a year of the post-outburst state. The ASAS-SN $g$-band lightcurve shows the phase of EX Lupi's lightcurve returned to the pre-outburst level in tandem with the flux (see Figure \ref{fig:ASAS_SNPhaseDecay}) during September 2023.

\subsection{Change in the radial velocity amplitudes: 
 Possible latitudinal movement of the hotspot}\label{subsec:hotspotLatitudeChange}

A sine curve model fit the phase folded RVs estimates the phase and amplitude of the RV for the specific spectral line (see Section \ref{subsec:PhaseOfRV}). The fitted parameters are tabulated in Table \ref{table:SineFit_on_RV}. The RV amplitudes from previous existing literature are also tabulated in the same table. In Table \ref{table:SineFit_on_RV}, columns 8 and 10 contain the ratio of the RV amplitudes from our estimation to that of  \citet{2015A&A...580A..82S} and \citet{2021MNRAS.507.3331C} respectively. The ratios are larger than 2 in column 8 and mostly larger than 1.5 in column 10. Thus, there is a significant RV amplitude enhancement during the outburst. For a hotspot on a rotating sphere with a period $P$ and radius $R$, the RV amplitude can be modelled as:

\begin{equation}\label{Equ:RVamplitude}
    A_{RV}  = \frac{2\pi R \sin(i)\cos(\Lambda)}{P} ~,
\end{equation}
where $i$ and $\Lambda$ are the inclination angle of the rotation axis with respect to the line of sight and latitude of the hotspot, respectively.

Considering Equation \ref{Equ:RVamplitude}, an increase in the RV amplitude can be a consequence of four possible causes: 1) the hot region where lines form is at a higher altitude from the stellar surface, along the accretion funnel. 2) The accreted matter is channeled through funnels to the magnetic pole, along with the equatorial tongues dumping matter near the equator. 3) The hotspot is more longitudinally confined during the outburst than during the quiescence. 4) The hotspot's latitude decreased. We scrutinize each of the possibilities below. \\

\textit{Can `hot region' at a higher altitude explain the increased RV amplitude?}\\

Taking He I 5875 \r{A} as a reference, the RV amplitude increased by a factor of 2.38 (see column 8 in Table \ref{table:SineFit_on_RV}). Considering other parameters are the same in Equation \ref{Equ:RVamplitude}, the hot region where spectral lines form would have to be at a radial distance of 2.38$R_{*}$ along the accretion funnel (if the quiescent state spectral lines formed at the stellar surface ($R_{*}$)). In contradiction to this hypothesis, the increased accretion rate would decrease the size of the pre-and post-shock region by the increase of infalling matter's ram pressure during the outburst state. We calculated the size of the pre-shock of the accretion-shocked region to be 0.0025$R_{*} << R_{*}$ \citep[Equation 10 in][]{2016ARA&A..54..135H}; where we used mass accretion rate of 2 $\times$ 10$^{-8}$ $M_{\odot}$/yr and the hotspot's fractional area of 5 percent of the total stellar surface. Thus, the explanation of the RV amplitude increase by the spectral line formation at higher altitudes along the funnel is highly unlikely. \\

\textit{Can an unstable accretion regime explain the increased RV amplitude?} \\

The Rayleigh-Taylor instability could develop at the magnetosphere-inner disk boundary during the states of high accretion rate, which produces equatorial matter-carrying tongues and dumps matter near the stellar equator \citep{2016MNRAS.459.2354B}. These equatorial tongues can co-exist with the dipolar accretion funnel. If equatorial tongues carry a substantial amount of accreted mass, it could result in hotspots near the equator. Thus, the effective RV amplitude could be an amalgamation of the lines forming at the near-polar and near-equator hotspots. It will effectively increase the RV amplitude. Our analysis in Section \ref{subsec:3DMHDsimulation} found that the equatorial tongues could have developed on the initial few epochs of spectra observation around the outburst peaks. However, our RV amplitude is an aggregate estimate of 16 epochs over 80 days, and those one or two epochs would not alter the estimates much. Also, the most energetic flow of matter is determined by dipolar funnel on these epochs (see top right panel in Figure \ref{fig:3DMHDsimulation}) but we found consistent RV amplitude increase in all the high excitation energy lines, like He I, as well as in low excitation energy lines, like Fe I (see Table \ref{table:SineFit_on_RV}). Thus, an unstable accretion regime could not explain the increased RV amplitude during the small outburst of 2022 March. However, this phenomenon could produce apparent RV amplitude variations in stars under a highly unstable accretion regime. \\

\textit{Can a varying size of the hotspot explain the increased RV amplitude?}\\

\begin{figure*}
\gridline{\fig{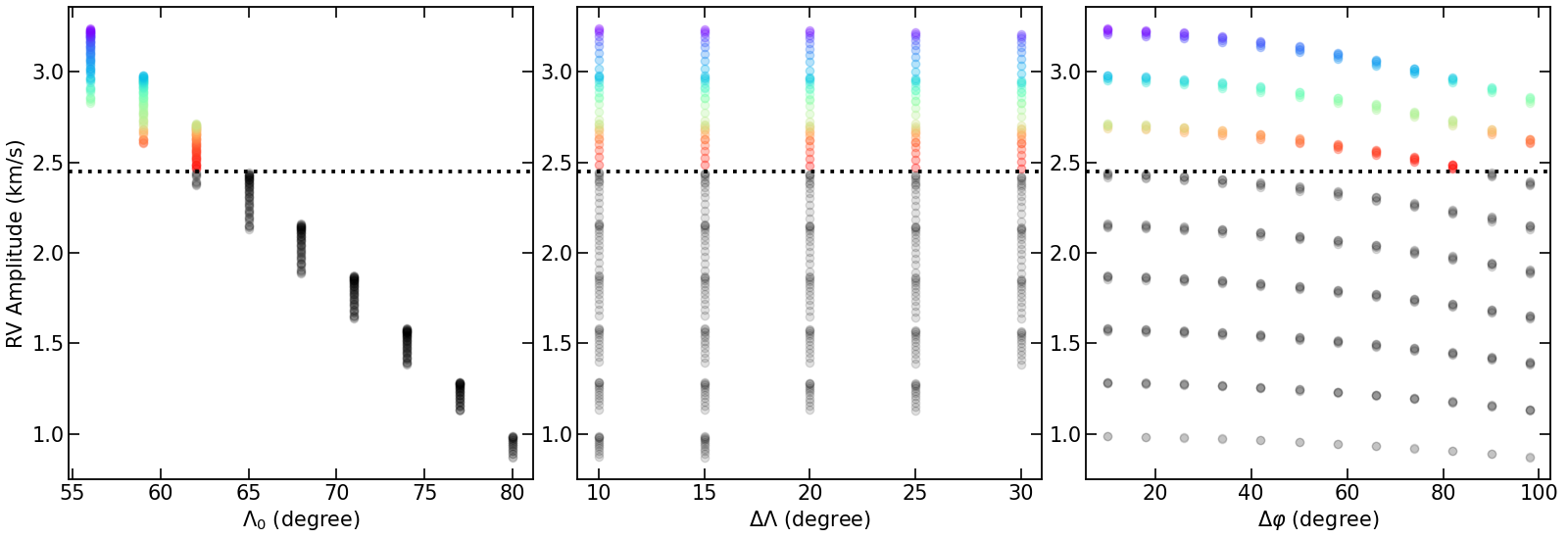}{1\textwidth}{}}
\caption{RV amplitudes are plotted against the hotspot's latitude location (\textit{left panel}), latitude size (\textit{middle panel}), and longitude size (\textit{right panel}). Each plot of RV amplitudes vs a parameter has other two parameters as variables. The scatter in RV amplitude, say around 80$\degree$ is because of various latitudinal (10$\degree$ - 30$\degree$) and azimuthal (10$\degree$ - 100$\degree$) sizes acquired by the hotspot. A horizontal dotted line represents RV amplitude of 2.45 km/s, for He I 5875 \r{A} (see Table \ref{table:SineFit_on_RV}). The colored and grey circles represent RV amplitude larger and smaller than 2.45 km/s, respectively. The X-axis is in degree, and Y-axis is in km/s. }
\label{fig:RV_SpotSizeSimulate}
\end{figure*}

A change in hotspot size will change the average RV amplitude. For instance, if the hotspot is spread uniformly across all azimuth, due to symmetry, any redshift from one side of the star will be cancelled by the blueshift from the other side of the star, effectively showing no RV changes. But, if the hotspot is confined within a small azimuthal region, the spectral lines produced in it will show redshift and blueshift as it rotates with the star. To understand the relation between the size of the hotspot and the RV amplitude, we simulated hotspots of various sizes on the stellar surface with three parameters: azimuthal size ($\Delta \varphi$), latitudinal size ($\Delta \Lambda$) and latitude location ($\Lambda _{0}$). The hotspot is sampled onto a grid (latitude and azimuth angle) with a grid size of 0.5$\degree$. Each grid point of the hotspot is assigned an RV ($A_{RV}\sin(\varphi)$), where $\varphi$ is the longitude of the grid. We approximated the RV of any spectral line forming in this hotspot as a uniformly weighted mean of RVs calculated over the hotspot grid. To simulate the peak RV amplitude of the RV time series, we placed the hotspot at the tangent of the stellar surface to our line of sight (at 180$\degree$ in Figure \ref{fig:phasedefinition}). By placing the hotspot at the tangential point, only half of the hotspot's azimuthal size contributed to the RV. The azimuthal and latitude sizes of the hotspot were varied between 10$\degree$-100$\degree$ and 10$\degree$-30$\degree$ respectively. And we further varied the hotspot's latitude location from 55$\degree$ to 80$\degree$ (see Figure \ref{fig:RV_SpotSizeSimulate}). Hotspot's latitudinal size is limited by the latitude location. Each panel in Figure \ref{fig:RV_SpotSizeSimulate} shows RV amplitude variation as a function of one parameter (X-axis label) while keeping the other two parameters free. 

We found that the increase in the azimuthal size decreases the RV amplitude (right panel in Figure \ref{fig:RV_SpotSizeSimulate}). The RV amplitude of around 1.0 km/s at $\Delta \varphi$=10$\degree$ (see right panel of Figure \ref{fig:RV_SpotSizeSimulate}) decreases as azimuthal size is increased to 100$\degree$. Similarly, an increase in latitudinal size decreases the RV amplitude but is not apparent in the middle panel of Figure \ref{fig:RV_SpotSizeSimulate} because of a smaller range of latitude sizes. But, one can see in the left panel that decreasing the latitude position of the hotspot, increases the RV amplitude. Taking He I 5875 \r{A} as a reference (RV amplitude = 2.45 km/s from our HRS spectra is plotted as a dotted horizontal line in Figure \ref{fig:RV_SpotSizeSimulate}), we found that the hotspot latitude location below $\sim$65$\degree$ is able to reproduce the observed RV amplitudes in our analysis. The RV amplitude of $\sim$1 km/s, as shown in \citet{2015A&A...580A..82S}, can be reproduced with the hotspot's latitude location of $\sim$80$\degree$ along with the latitude and longitude sizes of 10$\degree$-15$\degree$ and $\sim$20$\degree$, respectively (see Figure \ref{fig:RV_SpotSizeSimulate}). A larger inclination angle of 45$\degree$ \citep{2011ApJ...728....5G} does not change our qualitative understanding of hotspot's latitude change by 10$\degree$-15$\degree$.

During an outburst, an accretion funnel is expected to become broader and enlarged to transport more mass. Hence, the outburst should have increased the hotspot size, effectively decreasing the RV amplitude. Thus, the scenario of increasing RV amplitude due to tighter confinement of the accretion funnel is also highly unlikely. \\

\textit{Can a hotspot at a lower latitude explain the increased RV amplitude?}\\

The simulation in the previous section showed that the decrease in the hotspot's latitude could explain the increment of the RV amplitude. Considering a decrease in the hotspot's latitude while keeping the other two parameters invariant during the outburst, we calculated the hotspot's latitude using equation \ref{Equ:RVamplitude}. The HRS-SALT spectra covered $\sim$ 80 days, and hence our calculated hotspot latitude in Table \ref{table:SineFit_on_RV} is an aggregate of all epochs. The calculated hotspot latitude from HRS-SALT spectra and the prior available RV amplitudes are tabulated in Table \ref{table:SineFit_on_RV}. The average latitude differences from that of \citet{2015A&A...580A..82S} and \citet{2021MNRAS.507.3331C} to our HRS-SALT data are 16$\degree$ and 6$\degree$, respectively. Latitude estimates from our Mg I 5167 \r{A} and Si II 6347 \r{A} lines showed a larger values than that in \cite{2021MNRAS.507.3331C}, removing these two points gives an average latitude decrease of 9$\degree$ during the 2022 March outburst. 

The dipolar magnetic field configuration also predicts a decrease in the hotspot's latitude during the state of increased accretion rate. The dipolar field lines follow a parametric equation \( r/\sin^{2}(\theta) = const\), where $\theta$ is measured from the magnetic axis and $r$ is radial distance \citep{10.1093/mnras/stt945}.
The dipolar magnetic field structure is such that the inner closed field lines converge on the stellar surface at lower latitudes. The increased accretion would cause the inner disk to come closer to the stellar surface (see Equation \ref{Equ:MagnetosphericRadius}). The decreased inner disk radius would let the infalling disk matter to follow the inner magnetic field lines, eventually dumping matter at the lower latitudes. Thus, a decrease in latitude during an accretion outburst is consistent with the dipolar accretion funnel. Assuming a dipolar magnetic field topology, EX Lupi's hotspot latitude is calculated using \citep[Equation 3 in][]{10.1093/mnras/stt945}

\begin{equation}\label{Equ:SpotLatitude}
    Latitude = 90\degree - \Theta - \sin^{-1}(\cos(\Theta)\sqrt{\frac{R_{*}}{r_{m}}}) ~,
\end{equation}
where $\Theta =$ 13$\degree$ for EX Lupi. $R_{*}$ and $r_{m}$ are already defined. Since the disk would also compress the magnetic field lines, forcing them to converge at some higher latitude, the spot latitude inferred from Equation \ref{Equ:SpotLatitude} is a lower limit. Using Equation \ref{Equ:SpotLatitude}, we calculated the lower limit of the spot latitude among the spectral lines listed in Table \ref{table:MassAccretionRates} to be 49$\degree$.2 $\pm$ 2$\degree$.0. We also calculated spot latitudes from prior estimates \citep[e.g.,][]{2012A&A...544A..93S} of the accretion rate during the post-outburst quiescence of the year 2008. \citet{2023A&A...678A..88C,2023ApJ...957..113W} showed that the prior mass accretion rate estimates using 10$\%$ of H$\alpha$ line-width are underestimated. We used a new mass accretion rate estimate of 1.34 $\times$ 10$^{-9}$ $M_{\odot}$/yr from \citet{2023ApJ...957..113W} to calculate the hotspot latitude on 2010 May 4. This accretion rate is calculated from the line luminosity of Br $\gamma$ in X-Shooter spectra, using relations from \citet{2017A&A...600A..20A}, and hence it is best suited for comparison with our mass accretion estimates. The hotspot latitude during the quiescence on 2010 May 4 is 59$\degree$. According to this theoretical dipolar magnetic field model, the expected latitude change between the quiescence of the  post-2008 outburst and during the 2022 March outburst is $\sim$ 10$\degree$. This is consistent with the observed difference in the hotspot latitude between the 2022 March outburst and that of the quiescence from \citet{2015A&A...580A..82S} and \citet{2021MNRAS.507.3331C}. Thus, a decrease in the hotspot's latitude explains the increase in the RV amplitude. However, a detailed simulation study
is required to understand the variation of hotspot latitude during the state of variable accretion rate. \\

Thus, we conclude that the RV amplitude increment is a consequence of an effective decrease in the stellar hotspot's latitude.

\subsection{A possible azimuthal asymmetry in the accretion funnel: Blue shifted wing in the Ca II IRT lines}\label{subsec:CaII_blueWings}

\begin{figure}[!ht]
    \centering
    \includegraphics[scale=0.3]{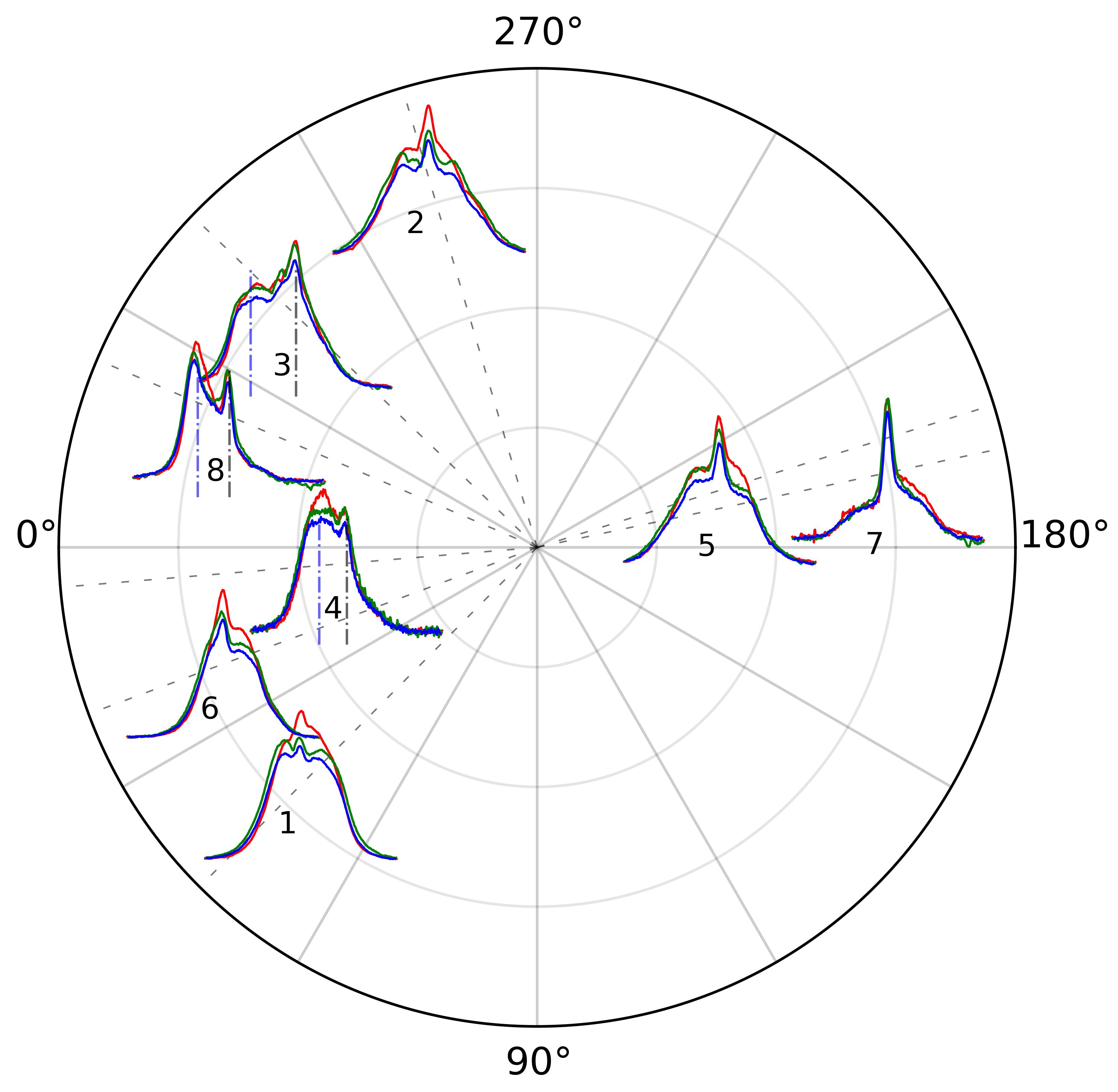}
    \caption{Polar plot showing the Ca II IRT spectra at the azimuthal location of the hotspot on EX Lupi's surface. Grey sparsely dashed radial spikes show the hotspot location at the epoch of spectra. Black-colored numerals show the order of the spectra observation. A blue dashed vertical line shows the mean position of the blue wing, as estimated from the 3-component Gaussian fit to the spectral lines. The velocities of the blue shifted wing on spectra marked 3, 4 and 8 are -116.6 km/s, -71.4 km/s and -81.4 km/s, respectively. Ca II IRT lines at 8498.02 $\AA$, 8542.09 $\AA$ and 8662.14 $\AA$ are shown by blue, green, and red colors respectively. Grey dash-dot lines on spectra 3, 4, and 8 mark the zero velocity position. The continuous radial spikes are at an interval of 30$\degree$. The Ca II IRT lines are plotted at different radial positions for visual clarity.}
    \label{fig:CaII_PolarPlot}
\end{figure}

During the outburst, the Ca II IRT lines are seemingly composed of three Gaussian components: a Narrow Component (NC), and two Broad Component (BC) Gaussians; one on the blue side and the other on the red side of the NC (see Figure \ref{fig:CaII_onLC}). The chromospheric origin of the NC is suggested by \citet{1992ApJS...82..247H,1996ApJS..103..211B}. The broad component is believed to be formed in a highly turbulent, physically extended region outside the stellar surface \citep{1992ApJS...82..247H,1996ApJS..103..211B,2015A&A...580A..82S}. There is a high correlation between the existence of broad components in Ca II IRT and the high accretion rate/optical continuum, as illustrated in Figure \ref{fig:CaII_onLC}. We modelled the Ca II IRT lines with a sum of three Gaussians: two broad Gaussians each leftward and rightward of a central narrow and small Gaussian. On three epochs of JD = 2459682.4127, 2459683.4179 and 2459712.604 (30.2 and 29.2 days apart, $\approx$ an integer multiple of EX Lupi's rotation period, 7.417 days), the Ca II IRT lines showed the presence of a prominent blue shifted wing (see 3rd, 4th and 8th epoch spectra in Figures \ref{fig:CaII_onLC} and \ref{fig:CaII_PolarPlot}). We see similar structures in H-Balmer lines also during these epochs. The velocity of the blue-shifted wings are -116.6 km/s, -71.4 km/s and -81.4 km/s on JD = 2459682.4127, 2459683.4179 and 2459712.604, respectively. To visualize rotational phase-based dynamics, the Ca II IRT line profiles are plotted in a polar diagram (see Figure \ref{fig:CaII_PolarPlot}). The polar position of the Ca II IRT in the diagram is the hotspot's azimuthal location at the respective epochs of observation. Since the photometric phase from the AAVSO $U$-band lightcurve is nearly stable even after 5 months of the outburst, EX Lupi's surface is mapped onto the phase values using the AAVSO $U$-band phase during the outburst. The 3rd, 4th, and 8th epoch spectra (epochs, where the intense blue shifted wings are seen) are observed when the hotspot is approaching us and is almost tangential to our line of sight (0$\degree$ in the Figures \ref{fig:phasedefinition} and \ref{fig:CaII_PolarPlot}, pictorially, the star is shown to be rotating anti-clockwise,). As we discussed in Section \ref{subsec:3DMHDsimulation}, the accretion is dominated by a dipolar funnel, and it approaches us with maximum velocity when the hotspot approaches us. Thus, the blue shifted emission components in spectral lines could imply a hot blob/material coming towards us with maximum projected RV at specific phases (around 0$\degree$ in Figure \ref{fig:CaII_PolarPlot}, when the funnel approaches us with maximum speed). The freefall timescale of material falling onto the stellar surface from the inner disk is only a few hours. However, the signature of a hot blob appeared even after 4 rotation cycles of the star and hence the blob cannot be expected to be fixed in the accretion funnel. One possibility could be that the mass accretion itself is clumpy. And, we see the extreme blue wings in Ca II IRT lines when the clump is falling through the accretion funnel and the funnel is approaching us at the highest projected velocity. The clumpiness being a random event could explain why not every Ca II IRT line has the same level of extreme blue-shifted wings.

Moreover, the blob has to be hot enough to produce an emission comparable to the rest of the Ca II IRT line emission regions. A detailed analysis is needed to understand the thermo-physical dynamics of the funnel, which is beyond the scope of this paper. We plan to understand the dynamics by doing spectral line analysis in our future work. However, we can estimate the distance of the hot blob from the stellar surface, assuming the measured velocity of the blob to be a sum of radial infall velocity and Keplerian rotation velocity.

\begin{equation}\label{Equ:HotBlobRadius}
\begin{split}
    V_{blob}^{2} = 2GM_{*}[\frac{1}{r} - \frac{1}{r_{m}}]\sin^{2}(i)\sin^{2}({\varphi}) + \\ \frac{GM_{*}}{r}\sin^{2}(i)\cos^{2}({\varphi}) ~,
\end{split}
\end{equation}
Since the infalling matter doesn't exactly follow the Keplerian velocity but slows down, we calculated the lower limit of the radial distance by assuming the blob to be co-rotating with the star:
\begin{equation}\label{Equ:HotBlobRadius_lowLimit}
\begin{split}
    V_{blob}^{2} = 2GM_{*}[\frac{1}{r} - \frac{1}{r_{m}}]\sin^{2}(i)\sin^{2}({\varphi}) + \\  \frac{GM_{*}r^{2}}{R_{*}^{3}}\sin^{2}(i)\cos^{2}({\varphi}) ~,
\end{split}
\end{equation}
where $r$ is the radial distance of the hot blob. $r_{m}$ and $i$ are pre-defined parameters. $\varphi$ is the phase of the hotspot. Any velocity component perpendicular to the disk is neglected. $r_{m}$ and $\varphi$ are estimated for the given epoch. Solving the equations \ref{Equ:HotBlobRadius} and  \ref{Equ:HotBlobRadius_lowLimit}, we get a bound on the radial distance of the hot blob. The radial distances are 0.80-1.52$R_{*}$, 0.50-4.03$R_{*}$ and 0.54-3.08$R_{*}$ for the epochs JD = 2459682.4127, 2459683.4179 and 2459712.604, respectively. A larger inclination angle would reduce these radial distances of the hot blobs. The estimated radial distances for the hot blobs are smaller than the inner disk radius and thus it further supports the idea that the blue-shifted wings in Ca II IRT lines could be representing the blobs of matter falling through the accretion funnel. 

\subsection{Accretion is not streamlined but clumpy: Photometric perspective}\label{subsec:TMMT_clumpy}

\begin{figure*}
\gridline{\fig{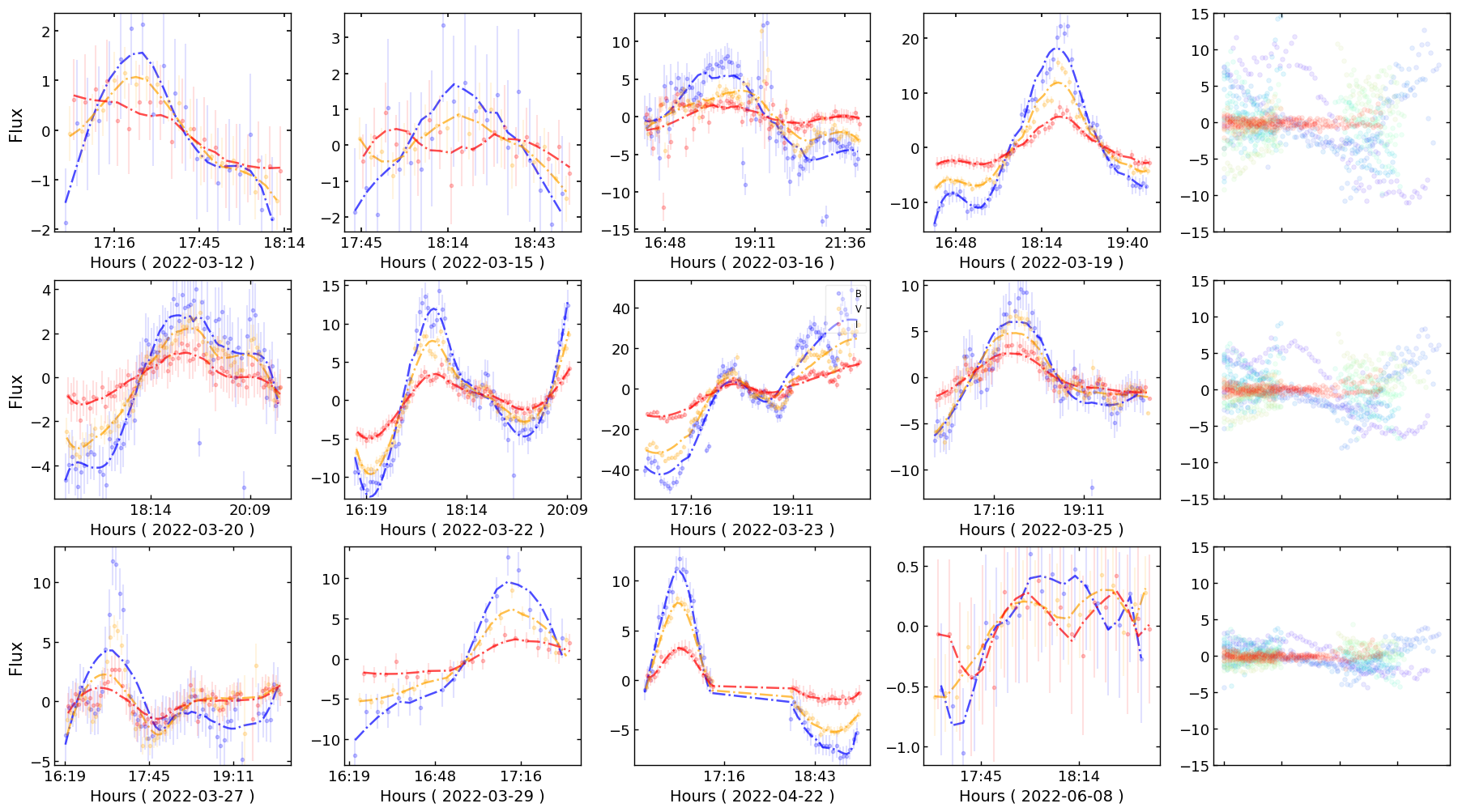}{1\textwidth}{}}
\gridline{\fig{lcobot_Clump_ApJreview.png}{0.2\textwidth}{}
          \fig{lcobot_Clump_2023_ApJreview.png}{0.2\textwidth}{}
          }
\caption{\textit{Top panel:} TMMT lightcurve cutouts showing hourly photometric flux enhancement. The X-axis is in Hours for a particular date and the Y-axis is scaled and shifted by mean of respective band fluxes. $B, V,$ and $I$ band lightcurve cutouts are plotted by blue, orange and red dots respectively. The Savitzky-Golay filter is applied over $B, V$, and $I$ cutouts and is plotted with respective color lines. The last column shows the scaled and shifted fluxes from the rest of the observations where no clumps were detected. The colors of the data points represent the epoch of observation, with violet being the first and red being the last epoch of TMMT observation. The top panel of the last column shows $B$-band fluxes while the middle and bottom panels show $V$ and $I$ band observations respectively.
\textit{Bottom panels:} LCRO $g^{\prime}$, $r^{\prime}$ and $i^{\prime}$ band lightcurves showed similar hour timescale photometric enhancement on 2022 March (left, during outburst) and 2023 March (right, quiescence of post-outburst).} 
\label{fig:TMMT_clump}
\end{figure*}

\begin{figure}[!ht]
    \centering
    \includegraphics[scale=0.55]{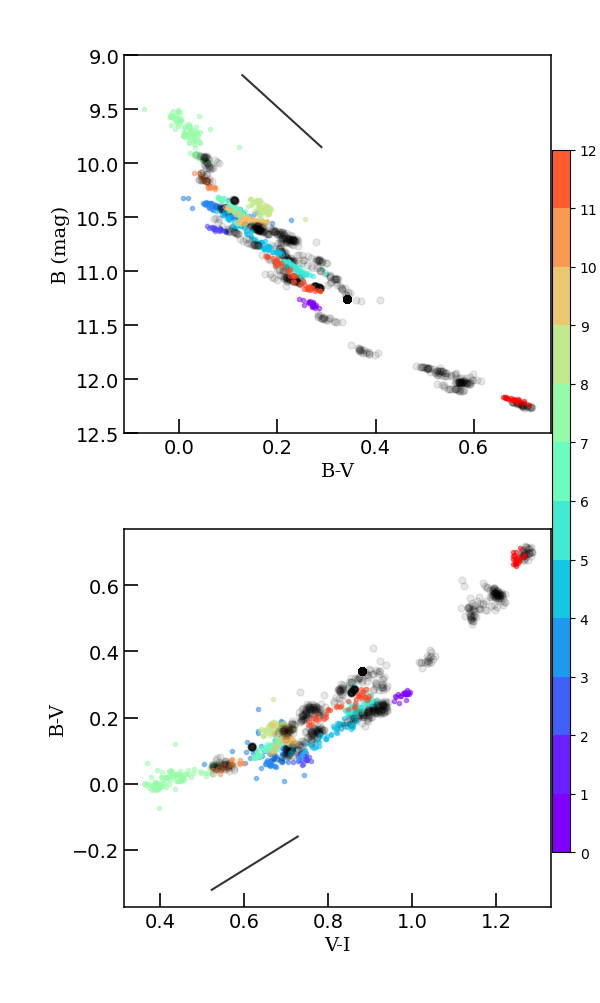}
    \caption{Color-color (top) and color-magnitude (bottom) plot during the outburst are represented by grey dots. Colored dots represent the hour timescale events. All 12 visually identified events are shown by color spectrum. The color bar shows the order number of the events. The solid black line is reddening vectors plotted at an interval of $A_{V}$ = 0.5 mag \citep{1989ApJ...345..245C}}
    \label{fig:Clump_CMD}
\end{figure}

\begin{figure}[!ht]
    \centering
    \includegraphics[scale=0.42]{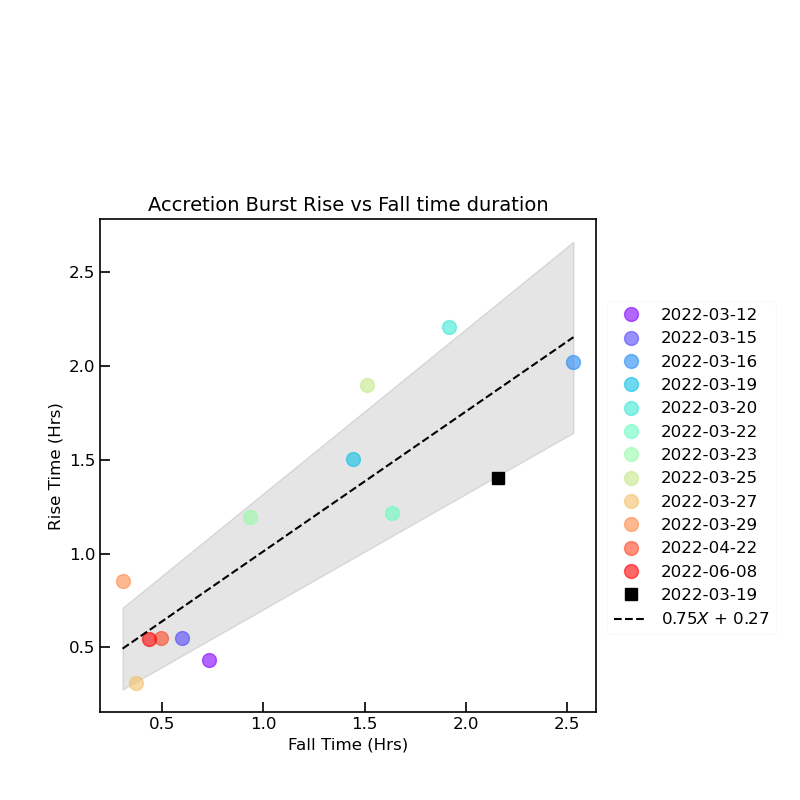}
    \caption{The rise timescale (hours) is plotted against the fall timescale (hours) and a linear fit is shown by the dashed line. The various colors mark the various epochs of the event. The circular and square markers denote the events as observed by TMMT and LCRO telescopes, respectively.}
    \label{fig:ClumpRiseFallTime}
\end{figure}

During the outburst, on multiple nights, we conducted high-cadence multiband photometric observations spanning a few hours. These observations revealed the presence of sudden enhancement in brightness structures on the time scales of a few hours (see Figure \ref{fig:TMMT_clump} for a few examples). These few hour-timescale brightness enhancements follow the same path in color-color and color-magnitude diagrams as followed by the EX Lupi in its quiescent state to outburst state transition (see Figure \ref{fig:Clump_CMD}). It suggests that these hour timescale events are related to the accretion process. \citet{1996A&A...307..791G} also reported similar hourly time scale events for T Tauri star BP Tauri. It was understood to be a consequence of accretion of the mass clumps \citep{1996A&A...307..791G}. Figure \ref{fig:TMMT_clump} shows that the time scales are not the same for each event, which may suggest the difference in the clumps' masses. We estimated the duration of the rise and fall of these events by visual inspection of the $B$-band lightcurve. In some of the cases, the estimated duration is a lower limit, as the whole event couldn't be observed. To estimate the asymmetry of the profiles, we measured the rise and fall times for each individual event. The rise and fall timescales show a linear relation with the slope and intercept of 0.75$\pm$0.13 and 0.27$\pm$0.18, respectively (see Figure \ref{fig:ClumpRiseFallTime}). The relation shows that the fall timescale of the events is longer than that of the rise timescale by a factor of $\sim$1.3. 


We further estimated the masses of these clumps by converting the brightness increase to accretion luminosity and then to the total mass (see Appendix \ref{App:PhotoMassAccRate}). The estimated masses of the clumps are given in Table \ref{tab:massAccClumps}. The clumps' masses range from 0.1 to 19.2 $\times$ 10$^{-5}$ lunar mass (an equivalent of $\sim$ 300 to 57000 Halley's comet mass)

Assuming a random distribution, 12 such events observed out of 42 high cadence photometric observations using TMMT implies that the clumps were $\sim$ 29$\%$ of the time. In contrast, out of 74 epochs of observations with LCRO one year after the outburst from January 2023 to May 2023, only a very few observations hint at such events (one such event is shown in the bottom right plot in Figure \ref{fig:TMMT_clump}). The LCRO observations in 2023 are relatively sparse, which makes it hard to find and characterize the clumpy events. The other possibility could be that the clumpiness of the accretion is dependent upon the mass accretion rate. During the quiescent state mass accretion rate, the accretion could be less clumpy. The clumps could start appearing as random events as the accretion rate increases. In extremely large accretion rates, the falling clumps could become too frequent to be termed continuous mass-infall. We calculated the rotational phase of EX Lupi during the epochs of these clumpy events to search for any rotational-phase correlation. The clumpy accretion is observed at all rotational phases of EX Lupi.

\section{Summary and Conclusions}\label{sec:Conclusion}
The 2022 March outburst of EX Lupi was observed at an unprecedented high-cadence multiband photometry as well as high-resolution multi-epoch spectroscopy. We searched for a change in the hotspot's azimuthal and latitudinal position during the 2022 March outburst and compared it with the hotspot's position during the outburst of the year 2008. The conclusions of our work are enumerated below:

\begin{enumerate}
    \item The hotspot's azimuthal position has been stable for about one and a half decades, starting from the epochs preceding the year 2008 outburst. The hotspot's stability for such a long time has been established with the study of archival RVs preceding to the 2022 March outburst. The long baseline high-cadence ASAS-SN $g$-band photometry also consistently established the stability of the hotspot's position prior to the 2022 March outburst.
    \item The accretion-driven 2022 March outburst in the EX Lupi caused a forward movement of the hotspot by 112$\degree \pm$ 5$\degree$. This forward movement is understood as a consequence of the high accretion rate. The increased accretion rate brings the disk closer to the stellar surface, thus increasing the Keplerian angular velocity of the inner disk material to be accreted onto the star. The slowly rotating star and its magnetosphere start to lag behind the faster-rotating infalling matter. The matter falls on the stellar surface at a larger longitude. Thus, the hotspot moves forward during the outburst.
    The forward movement of the hotspot is consistently observed in our high cadence, high resolution spectroscopic, and multiband photometric monitoring with HRS-SALT, LCRO, TMMT, and CTIO, along with the available archival high cadence multiband photometry from AAVSO and ASAS-SN. 
    \item The predicted accretion rate-dependent hotspot's azimuthal position is not observed in the outburst decay state. The red lightcurves ($R$ and $I$) show the phase decaying to the pre-outburst state value, as per the prediction. But, the shorter wavelength waveband's ($U, B$, and $V$) phases stay at higher values. One particular promising hypothesis is that the disk remained heated and thick for a longer period after the outburst, resulting in sustaining the higher longitude of the accretion hotspot.
    
    \item Our analysis shows a predominantly confined single hotspot in the pre-outburst state gets spread azimuthally and covers a large azimuthal area during the outburst. The hotspot in the post-outburst state has an azimuthal temperature gradient. The front end of the hotspot is hotter, while the lagging end has a cooler temperature. 
    \item Our SALT-HRS spectra show an increase in the RV amplitudes during the outburst. We discuss in detail multiple hypotheses to explain this observation in Section \ref{subsec:hotspotLatitudeChange}. The decrease in the hotspot's latitude from $\sim$ 75$\degree$ to $\sim$ 65$\degree$ explains the RV amplitude increase. Outburst in EX Lupi caused the hotspot to move azimuthally forward and latitudinally down.
    
    \item  Our high cadence photometric observations using TMMT and LCRO show short timescale variabilities, indicating clumpiness in the accretion. The clumpiness is observed $\sim$ 29$\%$ of the time during the outburst state. We suspect the highly blue-shifted emission wing in the Ca II IRT and H$\alpha$ lines are also related to these clumpy accretion events. The heavily blue-shifted components are observed only during the rotational phases corresponding to the funnel moving towards us with maximum projected RV.
   
    \item The phase of the EX Lupi's lightcurves ($g^{\prime}$,$r^{\prime}$ and $i^{\prime}$ bands from LCRO and $g$-band from ASAS-SN) remained almost constant to the outburst state phase value up to almost one and half years after the outburst. But, the ASAS-SN g-band phase value for the lightcurve from 2023 July 16 (JD = 2460140.7206) to 2023 September 06 (JD = 2460194.7206) decreased to the pre-outburst phase value. 
    
\end{enumerate}

We recommend further follow-up of this touchstone star EX Lupi to understand the long-term evolution of EX Lupi's magnetosphere. Further simulations of the inner accretion disk and the magnetosphere need to be developed to understand the complex behaviours seen in these systems.

\begin{longrotatetable}
\begin{deluxetable*}{lllllllllll}
\tablecaption{Sine fit parameters to radial velocities derived from the HRS-SALT spectra of EX Lupi during its 2022 March Outburst. $^{a}$ indicates that the RV phases are re-estimated at our startJD. $^{b}$ and $^{c}$ indicate the RV amplitude ratio from \textit{Sicilia-Aguilar et al. (2015)} and \textit{Campbell-White et al. (2021)} data with respect to our HRS-SALT data, respectively. The detected spectral lines are cross-checked against the lines found during the earlier outburst by \textit{Sicilia-Aguilar et al. (2012)} and the respective energies are taken from the NIST Atomic Spectra Database \citep{NIST_ASD}.
   \label{table:SineFit_on_RV}}
\tablewidth{700pt}
\tabletypesize{\scriptsize}
\tablehead{
\colhead{Line} & \colhead{E$_{k}$ - E$_{i}$} & \colhead{RV amplitude} & 
\colhead{RV phase} & \colhead{RV offset} & 
\colhead{Spot Latitude } & \colhead{RV phase$^{a}$} & \colhead{RV ratio$^{b}$} & \colhead{Spot Latitude } & \colhead{RV ratio$^{c}$} &
\colhead{Spot latitude} \\ 
\colhead{} & \colhead{} & \colhead{} & 
\colhead{} & \colhead{} & 
\colhead{} & \colhead{(C-W)} & \colhead{} & \colhead{(S-A)} & 
\colhead{} & \colhead{(C-W)} \\ 
\colhead{} & \colhead{(eV)} & \colhead{(kms$^{-1}$)} & \colhead{(degree)} & \colhead{(kms$^{-1}$)} & 
\colhead{(degree)} & \colhead{(degree)} & \colhead{} & \colhead{(degree)} & \colhead{} & \colhead{(degree)}
} 
\startdata
{\bf Fe I 4202.029} & 4.435-1.485 & 2.95$\pm$0.46 & 127.2$\pm$11.4 & -5.01$\pm$0.37 & 59.6$\pm$6.1 & \nodata & \nodata & \nodata & \nodata & \nodata \\
{\bf Fe I 5269.537} & \nodata & \nodata & \nodata & \nodata & \nodata & 30$\pm$14 & \nodata & \nodata & \nodata & 72.1$\pm$3.1 \\
{\bf Fe I 5371.489} & 3.266-0.958 & 3.32$\pm$0.52 & 148.0$\pm$9.5 & -5.03$\pm$0.44 & 55.5$\pm$7.2 & \nodata & 2.21 & 78.7$\pm$2.1 & \nodata & \nodata \\
{\bf Fe II 4233.167} & 5.511-2.583 & 1.31$\pm$0.30 & 109.6$\pm$17.5 & -5.71$\pm$0.24 & 77.0$\pm$3.3 & 15$\pm$15 & \nodata & \nodata & 1.19  & 79.1$\pm$2.2 \\
{\bf Fe II 4351.763} & 5.553-2.704 & 2.87$\pm$0.30 & 145.1$\pm$6.7 & -3.69$\pm$0.24 & 60.6$\pm$4.4 & \nodata & 2.71 &79.6$\pm$1.9 & \nodata & \nodata \\
{\bf Fe II 4549.474} & 5.553-2.828 & 2.89$\pm$0.64 & 157.0$\pm$13.4 & -1.32$\pm$0.50 & 60.4$\pm$7.8 & 27$\pm$12 & 2.63 &79.2$\pm$2.0 & 1.70 & 73.1$\pm$2.6  \\
{\bf Fe II 4923.9216} & \nodata & \nodata & \nodata & \nodata & \nodata & 46$\pm$11 & \nodata & \nodata & \nodata & 75.2$\pm$2.0 \\
{\bf Fe II 5018.434} & 5.361-2.891 & 2.46$\pm$0.34 & 113.2$\pm$10.2 & -3.58$\pm$0.30 & 65.1$\pm$4.4 & 46$\pm$14 & \nodata & \nodata & 2.05  & 78.1$\pm$3.2  \\
{\bf Fe II 5234.625} & 5.589-3.221 & 2.65$\pm$0.25 & 142.2$\pm$6.7 & -4.25$\pm$0.2 & 63.1$\pm$3.8 & 45$\pm$11 & \nodata & \nodata & 1.56   & 73.1$\pm$3.4 \\
{\bf Fe II 5316.615} & 5.484-3.153 & 2.28$\pm$0.34 & 122$\pm$11.3 & -2.77$\pm$0.30 & 67.0$\pm$4.2 & 27$\pm$6 & 2.30 & 80.2$\pm$1.8 & 1.2 & 71.0$\pm$2.1 \\
{\bf Fe II 6247.562} & 5.876-2.892 & 2.76$\pm$0.53 & 162.1$\pm$11.2 & -5.34$\pm$0.38 & 61.9$\pm$6.5 & \nodata & \nodata & \nodata & \nodata & \nodata\\
{\bf Fe II 6456.376} & 5.823-3.903 & 3.57$\pm$0.66 & 150.2$\pm$11.1 & 3.37$\pm$0.48 & 52.4$\pm$9.0 & \nodata & \nodata & \nodata & \nodata & \nodata \\
{\bf He I 4471.480} & \nodata & \nodata & \nodata & \nodata & \nodata & 53$\pm$9 & \nodata & \nodata & \nodata & 77.2$\pm$2.0 \\
{\bf He I 4713.1457} & 23.594-20.964 & 3.38$\pm$0.31 & 154.1$\pm$6.0 & -0.79$\pm$0.26 & 54.6$\pm$5.3 & \nodata & 1.94 & 72.7$\pm$3.1 & \nodata & \nodata \\
{\bf He I 5015.678} & \nodata & \nodata & \nodata & \nodata & \nodata & 42$\pm$7 & \nodata & \nodata & \nodata & 82.1$\pm$1.0 \\
{\bf He I 5875.621} & 23.074-20.964 & 2.45$\pm$0.36 & 124.8$\pm$10.9 & 0.90$\pm$0.30 & 65.2$\pm$4.6 & \nodata & 2.38 &  79.8$\pm$2.1 & \nodata & \nodata \\
{\bf He I 6678.1517} & 23.074-21.218 & 3.74$\pm$0.44 & 138.4$\pm$5.0 & 0.09$\pm$0.3 & 50.2$\pm$7.1 & \nodata & 4.61 & 82.0$\pm$1.4 & \nodata & \nodata\\
{\bf He I 7065.17714} & 22.718-20.964 & 2.88$\pm$0.67 & 137.6$\pm$16.7 & 0.88$\pm$0.57 & 60.4$\pm$8.1 & \nodata & \nodata & \nodata & \nodata & \nodata\\
{\bf He II 4685.804092} & 51.016-48.371 &  5.39$\pm$0.64 & 176.0$\pm$6.7 & -6.18$\pm$0.48 & 22.7$\pm$20.4 & 45$\pm$6 & 1.76 & 58.4$\pm$4.5 & 1.86 & 60.3$\pm$4.5 \\
{\bf Mg I 5167.322} & 5.108-2.709 & 1.87$\pm$1.02 & 120.5$\pm$39.1 & -2.16$\pm$ 0.82 & 71.3$\pm$10.7 & -1$\pm$8 & 2.25 & 81.8$\pm$1.8 & 0.85 & 67.9$\pm$3.0 \\
{\bf Mg I 5172.684} & 5.108-2.712 & 2.22$\pm$0.50 & 125.6$\pm$15.4 & -5.43$\pm$0.41 & 67.7$\pm$5.7 & 23$\pm$11 & 3.11 & 83.0$\pm$1.4 & 1.30 & 73.1$\pm$3.4 \\
{\bf Mg I 5183.604} & 5.108-2.717 & 2.43$\pm$0.30 & 130.2$\pm$9.5 & -5.27$\pm$0.26 & 65.4$\pm$4.0 & 48$\pm$9 & 2.29 & 79.6$\pm$1.7 & 1.52 & 74.1$\pm$2.5 \\
{\bf Si II 6347.1} & 10.073-8.12 & 1.67$\pm$0.40 & 134.5$\pm$18.7 & -5.25$\pm$0.33 & 73.4$\pm$4.3 & 76$\pm$12 & \nodata & \nodata & 0.80 & 69.0$\pm$5.6 \\
{\bf Si II 6371.36} & 10.066-8.12 & 2.98$\pm$0.62 & 138.0$\pm$15.0 & -1.77$\pm$0.51 & 59.3$\pm$7.8 & \nodata & \nodata & \nodata & \nodata & \nodata \\
\enddata
\end{deluxetable*}
\end{longrotatetable}

\begin{longrotatetable}
\begin{deluxetable*}{lllllllllll}
\tablecaption{Mass accretion rate estimated from the HRS-SALT spectra of EX Lupi for 16 epochs. The mass accretion rates are given in units of 10$^{-9}$ $M_{\odot}$/yr.
    \label{table:MassAccretionRates}}
\tablewidth{700pt}
\tabletypesize{\scriptsize}
\tablehead{
\colhead{Epoch} & \colhead{He I} & \colhead{He I} & 
\colhead{He I} & \colhead{He I} & 
\colhead{He I} & \colhead{H$\beta$} & \colhead{H$\gamma$} & \colhead{H$\delta$} &
\colhead{H8} & \colhead{Ca II (K)} \\ 
\colhead{} & \colhead{4026} & \colhead{4471} & 
\colhead{5875} & \colhead{6678} & 
\colhead{7065} & \colhead{4861} & \colhead{4340} & 
\colhead{4101} & \colhead{3889} & \colhead{3933} \\
} 
\startdata 
{\bf 2459654.489873} & 30.0$\pm$8.2 & 20.2$\pm$5.5 & 30.7$\pm$7.8 & 45.9$\pm$18.8 & 26.8$\pm$9.7 & 37.4$\pm$7.2 & 30.9$\pm$5.9 & 34.0$\pm$7.3 & 32.8$\pm$7.1 & 143.76$\pm$29.0 \\
{\bf 2459659.475822} & 28.0$\pm$7.6 & 9.0$\pm$2.5 & 19.8$\pm$5.1 & 22.8$\pm$9.6 & 14.3$\pm$5.3 & 32.0$\pm$6.2 & 22.8$\pm$4.4 & 23.3$\pm$5.1 & 19.5$\pm$4.3 & 153.5$\pm$31.0\\
{\bf 2459682.412697} & 15.9$\pm$4.4 & 22.6$\pm$6.1 & 24.0$\pm$6.2 & 27.3$\pm$11.4 & 18.1$\pm$6.6 & 27.3$\pm$5.3 & 23.9$\pm$4.6 & 28.8$\pm$6.2  & 17.3$\pm$3.9 & 77.8$\pm$16.1\\
{\bf 2459683.417894} & 7.2$\pm$2.0 & 25.2$\pm$6.8 & 30.4$\pm$7.8 & 23.5$\pm$9.8 & 24.3$\pm$8.8 & 28.5$\pm$5.6 & 22.9$\pm$4.4 & 34.6$\pm$7.4 & 6.0$\pm$1.4 & 51.6$\pm$10.9\\
{\bf 2459687.397465} & 13.5$\pm$3.8 & 16.9$\pm$4.6 & 20.0$\pm$5.2 & 19.4$\pm$8.2 & 17.6$\pm$6.4 & 23.1$\pm$4.5 & 19.6$\pm$3.8 & 25.2$\pm$5.5 & 21.0$\pm$4.6 & 51.0$\pm$10.8\\
{\bf 2459698.620602} & 19.0$\pm$5.3 & 26.5$\pm$7.1 & 26.5$\pm$6.8 & 34.8$\pm$14.4 & 22.8$\pm$8.3 & 26.6$\pm$5.2 & 29.1$\pm$5.6 & 36.2$\pm$7.8  & 24.8$\pm$5.4 & 70.0$\pm$14.6\\
{\bf 2459709.608854} & 7.2$\pm$2.1 &  9.0$\pm$2.5 & 10.0$\pm$2.7 & 6.2$\pm$2.7 & 9.2$\pm$3.4 & 10.1$\pm$2.1 & 10.2$\pm$2.0 & 13.9$\pm$3.1 & 10.2$\pm$2.3 & 14.5$\pm$3.2\\
{\bf 2459712.603958} & 14.5$\pm$4.1 & 14.4$\pm$4.0 & 12.3$\pm$3.2 & 11.8$\pm$5.1 & 12.9$\pm$4.8 & 14.0$\pm$2.8 & 13.8$\pm$2.7 & 18.5$\pm$4.1 & 10.2$\pm$2.3 & 29.0$\pm$6.2\\
{\bf 2459714.582326} & 10.2$\pm$2.9 & 14.0$\pm$3.8 & 17.0$\pm$4.4 & 17.9$\pm$7.6 & 14.7$\pm$5.4 & 15.1$\pm$3.0 & 17.9$\pm$3.5 & 23.1$\pm$5.0 & 17.0$\pm$3.8 &19.0$\pm$4.2\\
{\bf 2459721.560833} & 13.7$\pm$3.8 & 14.9$\pm$4.1 & 14.9$\pm$3.9 & 15.7$\pm$6.7 & 12.3$\pm$4.5 & 13.5$\pm$2.7 & 16.2$\pm$3.2 & 21.8$\pm$4.8 & 15.5$\pm$3.5 & 17.5$\pm$3.9\\
{\bf 2459722.575231} & 11.7$\pm$3.3 & 12.8$\pm$3.5 & 13.6$\pm$3.6 & 12.5$\pm$5.3 & 10.4$\pm$3.9 & 13.0$\pm$2.6 & 16.5$\pm$3.2 & 20.6$\pm$4.5 & 14.5$\pm$3.3 & 13.4$\pm$3.6\\
{\bf 2459723.556030} & 14.0$\pm$3.9 & 12.1$\pm$3.3 & 12.6$\pm$3.3 & 11.5$\pm$4.9 & 10.0$\pm$3.7 & 12.0$\pm$2.4 &  14.2$\pm$2.8 & 19.2$\pm$4.2 & 15.2$\pm$3.4 & 10.9$\pm$2.5\\
{\bf 2459724.573322} & 10.8$\pm$3.0 & 12.8$\pm$3.6 & 12.4$\pm$3.3 & 12.3$\pm$5.2 & 12.0$\pm$4.4 & 11.8$\pm$2.4 & 14.2$\pm$2.8 & 19.7$\pm$4.3 & 14.6$\pm$3.3 & 12.4$\pm$2.8\\
{\bf 2459732.545174} & 3.1$\pm$0.9 & 6.8$\pm$1.9  & 4.1$\pm$1.1   &      \nodata & 5.0$\pm$1.9  & 5.9$\pm$1.2 & 5.0$\pm$1.0 & 8.7$\pm$2.0 & 6.0$\pm$1.4 & 8.6$\pm$2.0\\
{\bf 2459735.531713} & 0.4$\pm$0.1 & 11.5$\pm$3.2 &  8.1$\pm$2.2  & 5.2$\pm$2.3  & 8.2$\pm$3.1 & 8.1$\pm$1.7 & 7.4$\pm$1.5 & 12.3$\pm$2.8 & 5.7$\pm$1.3 & 4.9$\pm$1.1\\
{\bf 2459736.539005} & 3.5$\pm$1.0 & 11.9$\pm$3.3 & 10.8$\pm$2.9  & 9.5$\pm$4.1  & 9.2$\pm$3.4 & 9.3$\pm$1.9 & 10.8$\pm$2.1 & 15.7$\pm$3.5 & 9.6$\pm$2.2 & 6.5$\pm$1.5\\
\enddata
\end{deluxetable*}
\end{longrotatetable}

\begin{deluxetable*}{cccc}
\tablecaption{Mass of the accretion-clumps\label{tab:massAccClumps}}
\tablewidth{0pt}
\tablehead{
\colhead{Epoch} & \colhead{Telescope} & \colhead{Clump Mass} & \colhead{Clump Mass} \\
\colhead{JD} & \colhead{} & \colhead{(10$^{-12}M_{\odot}$)} & \colhead{(10$^{-5}$M$_{MOON}$)} 
}
\decimalcolnumbers
\startdata
2459651 & TMMT & 0.20 & 0.56  \\
2459654 & TMMT & 0.22 & 0.57  \\
2459655 & TMMT & 3.27 & 8.93  \\
2459658 & TMMT & 5.19 & 14.12  \\
2459659 & TMMT & 1.82 & 5.0  \\
2459661 & TMMT & 4.32 & 11.82  \\
2459662 & TMMT & 7.11 & 19.21 \\
2459664 & TMMT & 4.71 & 12.59  \\
2459666 & TMMT & 0.77 & 2.02  \\
2459668 & TMMT & 0.48 & 1.23  \\
2459692 & TMMT & 0.96 & 2.59  \\
2459739 & TMMT & 0.04 & 0.12 \\
\enddata
\tablecomments{The estimated masses of the clumps are tabulated in the units of 10$^{-12}M_{\odot}$ and 10$^{-5}$M$_{MOON}$.}
\end{deluxetable*}

\appendix 

\section{The Period of EX Lupi}\label{App:EXLupiPeriod}
\begin{figure}
    \centering
    \includegraphics[width=18cm,height=10cm]{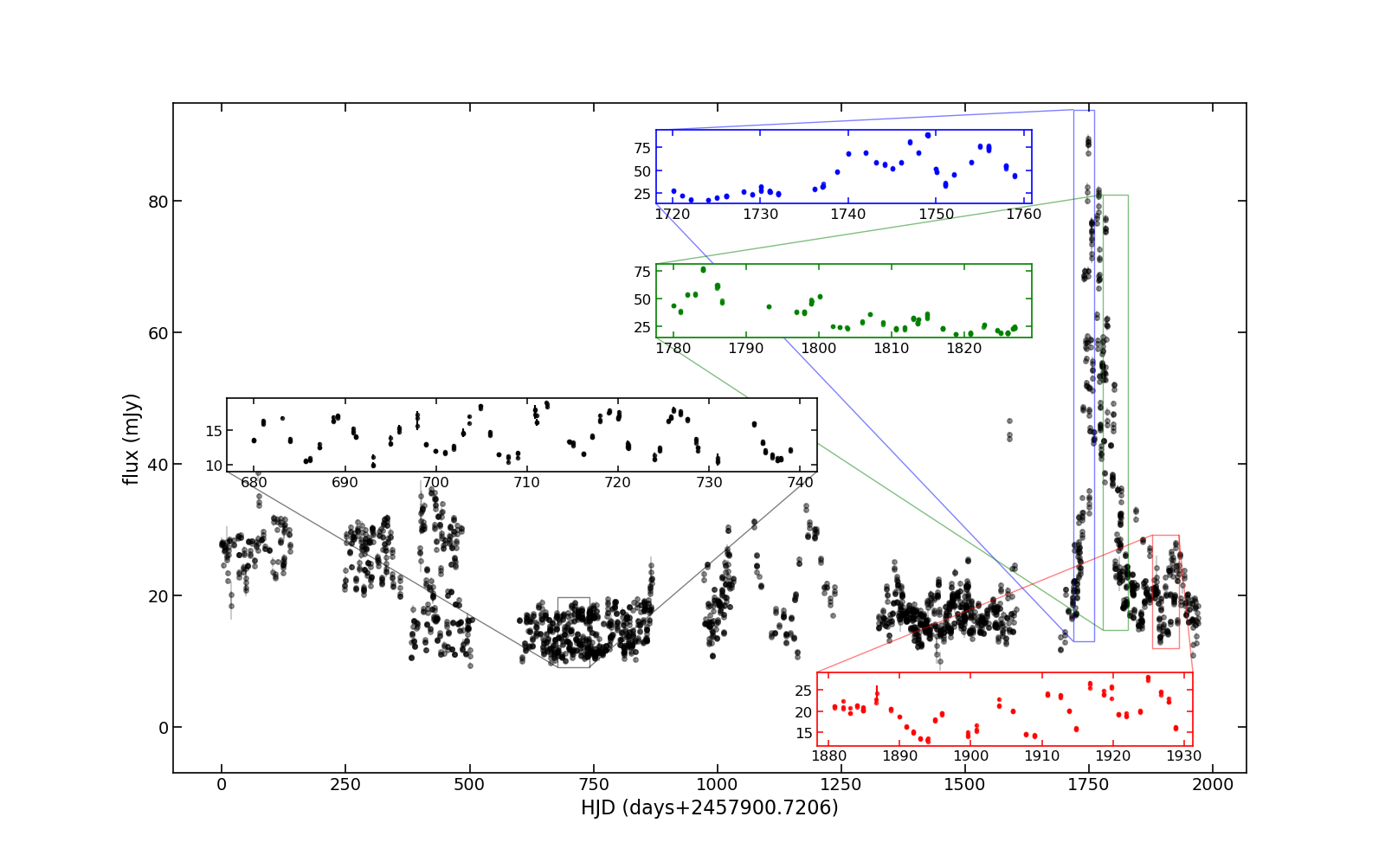}
    \caption{EX Lupi lightcurve over a baseline of 6 years. Insets of periodic oscillations in the photometry are shown. The Y-axis is in mJy while the X-axis is in HJD. The two middle insets (green and blue) during the rise and fall of the outburst show photometric oscillations overlaid on the increase and decrease of the photometric continuum.}
    \label{fig:ASASN_insetLC}
\end{figure}

EX Lupi shows a very strong periodic signal in the lightcurves (for example, see Figure \ref{fig:ASASN_insetLC}). \citet{2014A&A...561A..61K} reported a 7.417 $\pm$ 0.001 day periodicity in the absorption line RV of the EX Lupi. The phase folded RV showed a stable periodicity over a time span of 5 years, starting from July 2007 to July 2012 \citep{2014A&A...561A..61K}. This time span includes $\Delta$$V$ $\sim$ 5 mag outburst in 2008. \citet{2015A&A...580A..82S} calculated the emission line narrow-component RV periodogram by fitting the higher excitation lines. The periodicity varied between a range of values, but most of them are closer to the period of 7.417 days \citep[Table A.2 in][]{2015A&A...580A..82S}. High cadence multiband photometry collected just before the 2022 March outburst showed the periods in a range of 7.0 - 7.2 days \citep{2022RNAAS...6...52K}. If the rotation period of EX Lupi has reduced over time, then the measured phase increase leading up to the outburst would have been a steady increase in time.

\begin{figure} 
    \centering
    \includegraphics[scale=0.43]{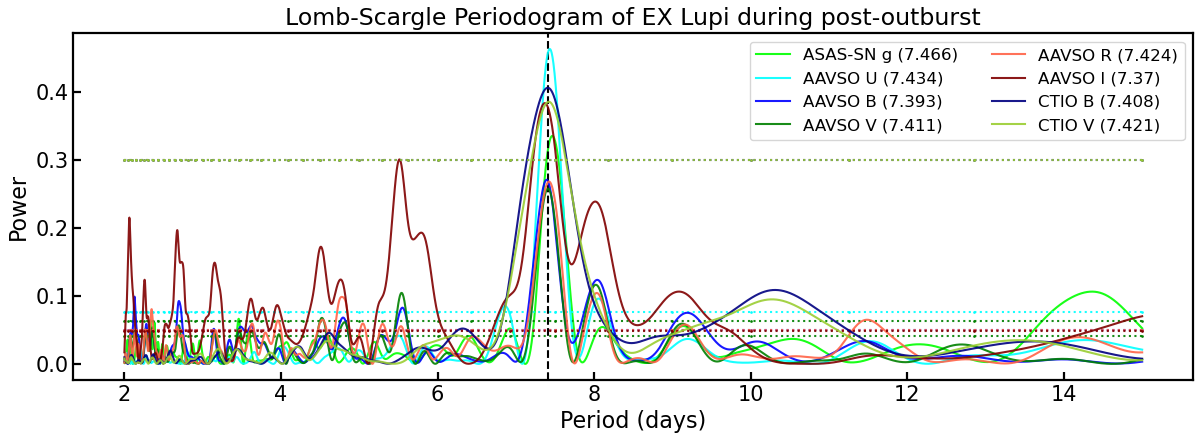}
    \caption{Periodogram for post-outburst multiband photometric data. The dotted horizontal lines correspond to a level of 10 \% False-Alarm-Probability for corresponding data. The legend shows the estimated period from frequency corresponding to maximum power. The black dashed vertical line shows the location of the period = 7.417 days.}
    \label{fig:Periodogram}
\end{figure}

However, there is no systematic phase increase till the onset of the 2022 March outburst (see Figure \ref{fig:phaseVariation_AllLC}). Even if we consider an extreme mass accretion rate of 10$^{-6}M_{\odot}$/year over 14 years, it would change the period only to 7.406 days (under an assumption that material falls from a radius of 5$R_{*}$ onto the star without losing angular momentum). Thus, the spinning-up of the star on this timescale is not explainable by the angular momentum gain. Further, we calculated the Lomb-Scargle periodogram on multi-observatory, multiband post-outburst photometric data that gave period values in the range 7.36 - 7.47 days (see Figure \ref{fig:Periodogram}). The phase is estimated in the lightcurves covering JD=2459702-2459850 (the outburst peaked around JD = 2459655). Any linear trend in the lightcurves is removed by fitting and subtracting a straight line. These period values are spread around the 7.417 days, as shown by a dashed vertical line in Figure \ref{fig:Periodogram}. We did not find any systematic period change in the multiband lightcurves. \citet{2023A&A...678A..88C} also did a period analysis on the EX Lupi's lightcurves during the states of quiescence and outburst. The authors found a period of 7.4 days during the state of dimming of the outburst and the post-outburst. However, the authors found an equally probable period of 7.2 - 8.2 days during the rise of the outburst. Periodogram analysis done by \citet{2023ApJ...957..113W} on AAVSO multiband and ASAS-SN g-band lightcurves showed a rotation period of 7.52-7.62 days during the state of outburst (MJD = 59670 - 59725) which later varied to 7.37-7.40 days in the post-outburst state (MJD = 59725 - 59875). Since they are all consistent with each other, we adopted 7.417 $\pm$ 0.001 days by \citet{2014A&A...561A..61K} as the rotation period of the EX Lupi.

\section{Spectral Line Fittings}\label{App:LineFittings}

We selected some spectral lines and matched them with the lines found by \citet{2012A&A...544A..93S} during the outburst of the year 2008. These spectral lines are then modelled with a sum of multiple Gaussians and a constant of the form:
\begin{equation}
    \sum_{i=1}^{N} A_{i}e^{-\frac{(x-\mu_{i})^{2}}{2\sigma_{i}^{2}}} + C_{i} ~,
\end{equation}
where $N$ is the number of Gaussians used for fitting the spectral lines. $\mu_{i}$ and $\sigma_{i}$ denote line center and line width, respectively. $C_{i}$ is a constant. We fitted with $N$ Gaussians based on the visual inspections of the spectral lines.

\begin{figure*}
\gridline{\fig{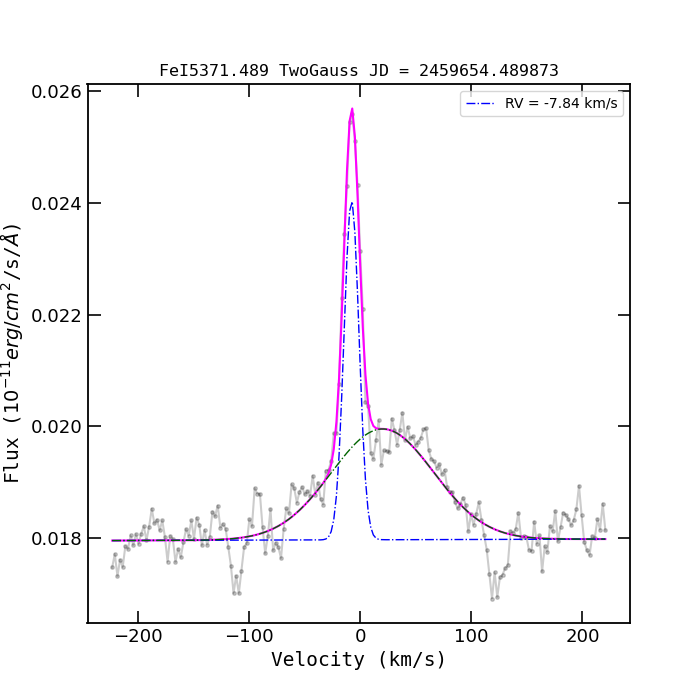}{0.3\textwidth}{(Fe I 5371)}
          \fig{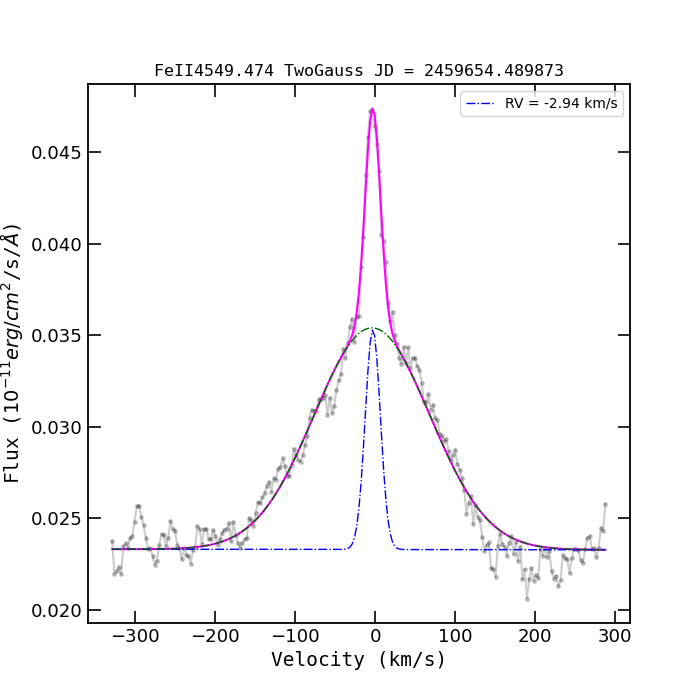}{0.3\textwidth}{(Fe II 4549)}
          \fig{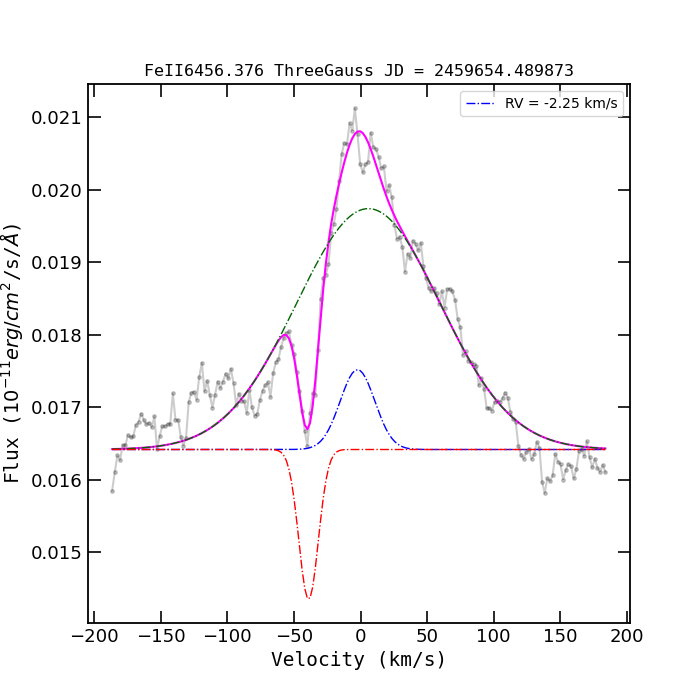}{0.3\textwidth}{(Fe II 6456)}
          }
\gridline{\fig{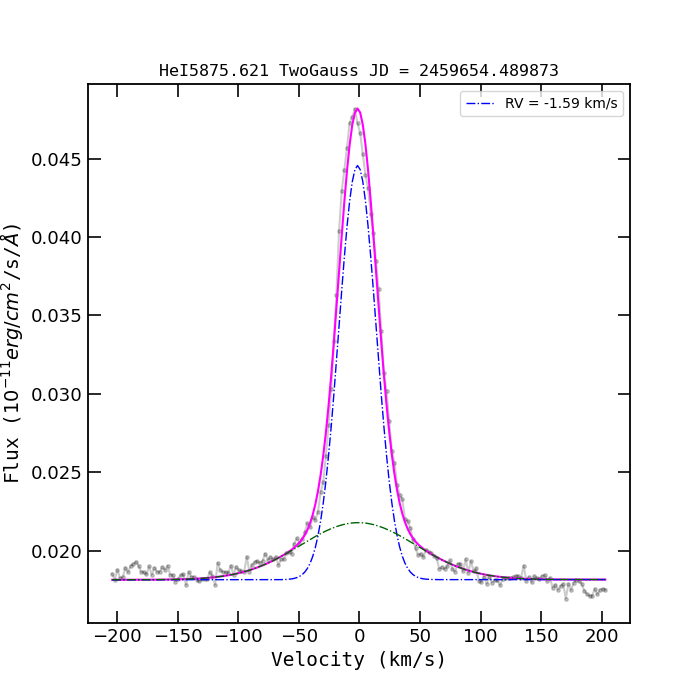}{0.3\textwidth}{(He I 5875)}
          \fig{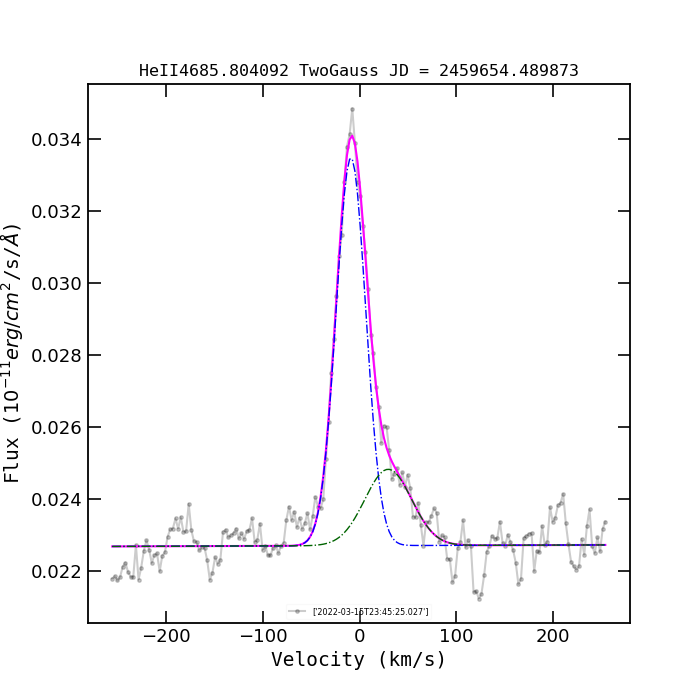}{0.3\textwidth}{(He II 4685)}
          \fig{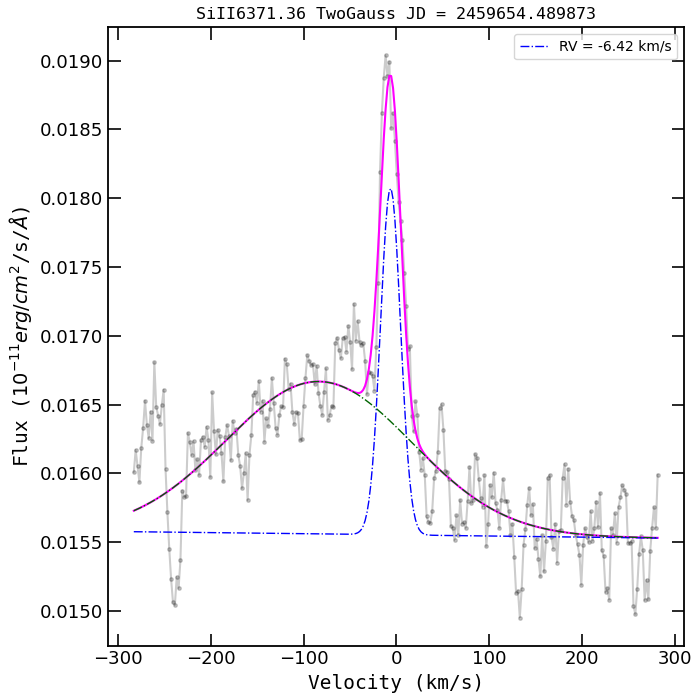}{0.3\textwidth}{(Si II 6371)}
          }
\gridline{\fig{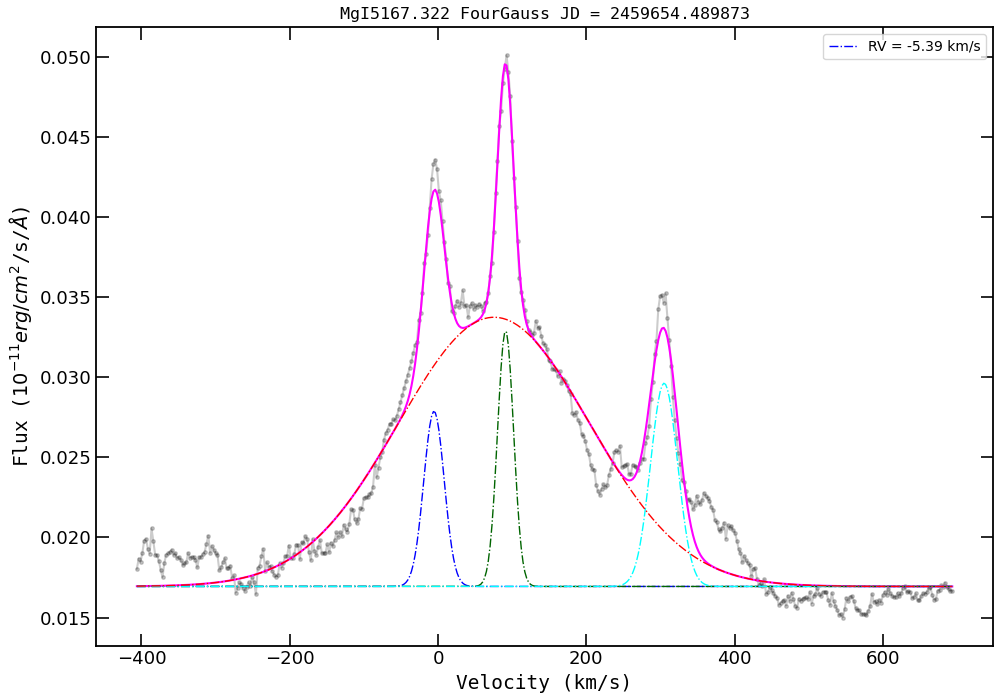}{0.5\textwidth}{(Mg I 5167,  Fe I 5169 and Mg I 5172)}}
\caption{Multi Gaussians fit to the EX Lupi spectral lines are plotted. The pink color curve is the combined fit of the spectral line, while the blue color represents the NC of the line and green represents the BC (or the redshifted, for He II). All the spectral lines correspond to the peak of the outburst, JD = 2459654.4899. The spectra were calibrated before fitting as discussed in Section \ref{subsec:HRS-SALT}. The last profile plot is modelled as a sum of four Gaussians: blue represents Mg I 5167, green represents Fe I 5169 and cyan represents Mg I 5172 and red represents the broad component.
\label{fig:LineFit}}
\end{figure*}

Lines like Fe I, Fe II, Mg I, Si II, and He I are fairly easily recognizable as composed of a symmetric NC, dominating the central part and BC, raising the wings of the spectral lines during the outburst. Hence, these lines are modelled very well with a sum of two Gaussians and generally one Gaussian (representing NC) suffices during the state of quiescence. But, the He II line is asymmetric (see Figure \ref{fig:LineFit}). Generally, the blue part rises steeply while the red part is extended to larger velocity/wavelengths. Two Gaussians are used to model the He II lines. The first Gaussian models the NC, and the second models the red side wing. The Gaussian representing the narrow component is larger in amplitude and narrower, while the second Gaussian is shallower and wider and is red-shifted with respect to the NC. The mean of the NC Gaussian is taken as the RV of the He II line. Mg I 5167 \r{A} and Mg I 5172 \r{A} are close to a Fe I 5169 \r{A} line; the trio of them sit over a BC in the outburst state spectra (see bottom panel in Figure \ref{fig:LineFit}). These lines are modelled with a sum of four Gaussians, three NCs assigned for each spectral line and a BC. Fe II 6456 \r{A} is modelled with a sum of three Gaussians: NC + BC + inverted Gaussian at the bluer wavelength to address an absorption feature (see the top right panel in Figure \ref{fig:LineFit}). All the spectral lines are modelled iteratively for each epoch, using the Lmfit-module \citep{2016ascl.soft06014N}. The phase-folded RV showed a periodic variation. The phase of the periodically varying RV is plotted in Figure \ref{fig:RV_phasechange}. Though the RV inferred from the multi-Gaussian fit revealed a periodic nature, the line width of the NC Gaussian did not show any periodic modulation.

The Ca II IRT lines (8498 \r{A}, 8542 \r{A}, and 8662 \r{A}) are hard to model. They seem to be composed of three Gaussians: blue-shifted broad emission line, central NC, and red-shifted BC. We modelled the Ca II IRT lines by a sum of three Gaussians to estimate the velocity of the blue-shifted wings on the three epochs (see Section \ref{subsec:CaII_blueWings}). The three epochs where a highly blue-shifted feature is seen in the Ca II IRT line are shown in Figure  \ref{fig:CaII_PolarPlot}. Detailed line profile modelling and analysis are beyond the scope of this paper and are planned in our upcoming work.

In Figure \ref{fig:LineFit}, we show a sample of multi-Gaussian fit to the spectral lines. 

\section{Alternate hypotheses for phase change: Activation of a new accretion funnel}\label{app:otherPhasechangeHypothesis}

The phases of the wavebands with shorter effective wavelengths ($U$ and $B$) being almost constant and at larger longitudes could also be a consequence of the permanent reconstruction of some new accretion funnel at the observed longitude. During the rise of the outburst, it is possible that some global changes happened in the star-disk system, which led to the flow of infalling matter streams along new paths of lowest energy. The new funnels that form at new azimuthal positions can stay there for a significant time, depending upon the decay timescale of the changes. 
A completely new funnel can be activated at a different azimuthal location under two possible scenarios: 1) permanent reconstruction of the magnetic field and/or 2) a tilt in the inner disk caused by some cloudlet capture. \\

\textit{Permanent reconstruction of the magnetic field:}

If the magnetic field gets reconstructed during the outburst, then the phase of the lightcurve can change. This phase change would be completely random and will remain permanent owing to the new magnetic field configuration. The observations have previously shown a varying magnetic field configuration \citep{2022AAS...24040613J,2023MNRAS.526.4627F}, with suggestions of varying convection processes inside the star. No such varying magnetic field reconstruction has been inferred in EX Lupi for almost one and a half decades otherwise it would have appeared in terms of lightcurve phase change. It is highly unlikely that the intrinsic parameters governing the magnetic field structure could change during the 2022 March outburst but remain intact during the gigantic outburst of the year 2008. This hypothesis opens too many questions than solving the existing ones. \\

\textit{Tilt in the inner disk:}

Another probable reason for the permanent change in the hotspot's azimuth could be the tilt of the inner disk rotation axis with respect to the pre-outburst inner disk rotation axis. A stream of clouds falling onto the circumstellar disk could cause a tilt in the disk. This tilt could be completely random depending upon the angular momentum, energy, impact parameter, etc. of the infalling stream. Thus, a phase change of the 112$\degree$ could be a random event rather than a systematic forward motion of the hotspot along the rotation as suggested by \citet{10.1093/mnras/stt945}. One of the solutions to the misaligned short-period extrasolar planets is suggested to be a consequence of the tilt in the circumstellar disk caused by the capture of passing by a stream of cloud/gas \citep{2011MNRAS.417.1817T}. \citet{2019A&A...628A..20D} suggested the possibility of gravitational instability in the disk caused by the cloudlet capture, eventually leading to an FU Orionis-like event. The authors further suggested that the timescale of EX Lupi-like event is too small to be explained by a long-time event of cloudlet capture. 
\citet{2021MNRAS.506..372R} did 3D simulations on the accreting stars with a tilted inner disk with respect to the stellar rotation axis. The new tilted inner disk evolves on its viscous timescale after the initial tilt settlements. In this scenario, the phase will not return for a very long time. 
This hypothesis also fails to explain the differential phase evolution in AAVSO multiband data in the post-outburst state.

\section{Accretion rate estimation from photometry}\label{App:PhotoMassAccRate}

\begin{figure} 
    \centering
    \includegraphics[width=18cm,height=10cm]{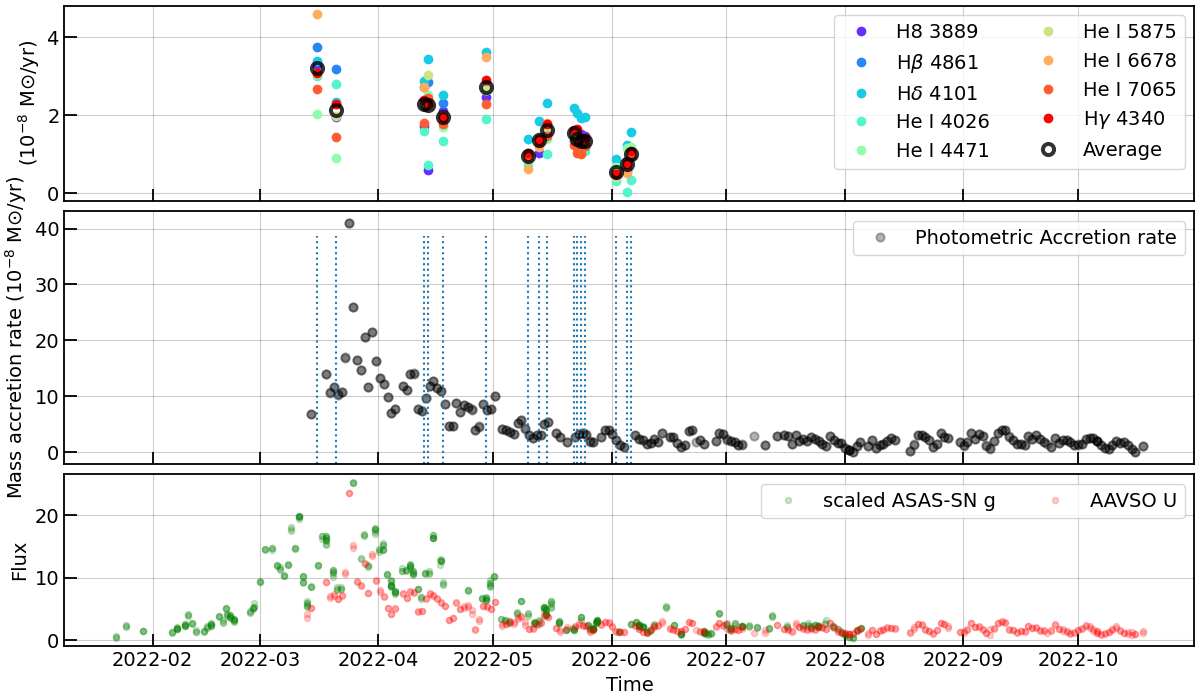}
    \caption{\textit{Top panel}: The accretion rate inferred from the emission lines are shown by colored dots while the average accretion rates are plotted by black circles. \textit{Middle panel}: The mass accretion rate inferred from the photometry is shown by black dots. The blue dotted vertical lines are the epochs of our HRS-SALT spectra observation, shown in the upper panel. \textit{Bottom panel}: The scaled AAVSO $U$-band and the scaled ASAS-SN $g$-band lightcurves are plotted to show the correspondence of the epochs of the outburst.}
    \label{fig:MassAccRate}
\end{figure}

We calculated the mass of the accretion clumps (see Section \ref{subsec:TMMT_clumpy}) by estimating the mass accretion rate from the photometry under simplistic assumptions. The AAVSO $U, B, V, R$ and $I$-band lightcurves are converted into the flux unit using \citet{1998A&A...333..231B}. Then the sets of AAVSO  multiband photometry are chosen such that each band is observed apart not more than 0.05 a day. The fluxes are multiplied by 4$\pi$d$^{2}_{*}$. The extinction correction is done for $A_{V}$=1.1 mag. Then the upper bound on the stellar flux is estimated as the minimum of the fluxes in each photometric band during the quiescence of the post-outburst. This estimated stellar flux is subtracted from the respective band lightcurves to get an estimate of the hotspot flux. Then, a Blackbody is fitted onto the hotspot flux using the Astropy module in Python. The temperature of the Blackbody fit is kept fixed at 10000 K following \citet{2016ARA&A..54..135H}, while the scale is kept variable with an initial value of $R^{2}_{*}$. The integrated luminosity of the best-fit blackbody is taken as the accretion luminosity, which is then converted to the mass accretion rate using Equation \ref{Equ:accretionRate2}. Figure \ref{fig:MassAccRate} shows the accretion rate estimated with the photometry along with the average accretion rate inferred from the spectroscopic lines. The accretion rate estimated from photometry is larger than that from spectroscopic emission lines by a factor of approximately 5 to 6 (see Figure \ref{fig:MassAccRate}). Figure \ref{fig:MassAccRate} also shows the EX Lupi lightcurve during the outburst for a clear picture of the estimated increased accretion with the brightness enhancements during the outburst.

Similarly, we estimated the mass accretion rate during the `clumpy events' observed in the TMMT $B, V$ and $I$ lightcurves. TMMT $B, V$, and $I$ band lightcurves are converted into flux units and corrected for extinctions of $A_{V}$=1.1 mag. The estimates of the stellar fluxes in the $B, V$ and $I$ bands are the same as that in the AAVSO lightcurves. Then, the mass accretion rates are integrated for the timescales of the clump to get an estimate of the clump's mass (see Table \ref{tab:massAccClumps}). The masses of the clumps range from 1 to 192 $\times$ 10$^{-6}$ Lunar mass.

\section*{Acknowledgement}
We thank the anonymous reviewer for the comments and suggestions that enhanced the quality of the paper. We thank Prof. Suvrath Mahadevan for the discussion on the shape of the archival RV. We also thank Prof. Je\^rome Bouvier for the discussion on the phase change of the lightcurves. We thank the organizers of the \textit{Magnetism and Accretion} conference, held from 16 to 19 Jan 2023 in Cape Town, South Africa, which helped establish collaborations on this project. KS and JPN also acknowledge the travel support to attend this conference through the BRICS Astronomy programme. A part of this work is based on observations made with the Southern African Large Telescope (SALT) with the Large Science Programme on transients 2021-2-LSP-001 (PI: DAHB). Polish participation in SALT is funded by grant No. MEiN 2021/WK/01. KS, JPN and DKO acknowledge the support of the Department of Atomic 
Energy, Government of India, under Project Identification No. RTI 4002. MMR acknowledges support from NSF grant AST-2009820. Resources supporting
this work were provided by the National Aeronautics and Space Administration (NASA) High-End Computing (HEC) Program through NASA Center for Computational Sciences
(NCCS) at Goddard Space Flight Center. JB thanks the support from Programa ICM, ANID, ICN12$\_$009 awarded to the Millennium Institute of Astrophysics (MAS). This paper includes data collected by the TESS mission. Funding for the TESS mission is provided by the NASA's Science Mission Directorate. We acknowledge with thanks the variable star observations from the AAVSO International Database contributed by observers worldwide and used in this research. This work made use of Astropy:\footnote{http://www.astropy.org}, a community-developed core Python package and an ecosystem of tools and resources for astronomy. This research was supported in part by a generous donation (from the Murty Trust) aimed at enabling advances in astrophysics through the use of machine learning. Murty Trust, an initiative of the Murty Foundation, is a not-for-profit organization dedicated to the preservation and celebration of culture, science, and knowledge systems born out of India. The Murty Trust is headed by Mrs. Sudha Murty and Mr. Rohan Murty.



\begin{thebibliography}{}
\expandafter\ifx\csname natexlab\endcsname\relax\def\natexlab#1{#1}\fi
\providecommand{\url}[1]{\href{#1}{#1}}
\providecommand{\dodoi}[1]{doi:~\href{http://doi.org/#1}{\nolinkurl{#1}}}
\providecommand{\doeprint}[1]{\href{http://ascl.net/#1}{\nolinkurl{http://ascl.net/#1}}}
\providecommand{\doarXiv}[1]{\href{https://arxiv.org/abs/#1}{\nolinkurl{https://arxiv.org/abs/#1}}}

\bibitem[{{{\'A}brah{\'a}m} {et~al.}(2009){{\'A}brah{\'a}m}, {Juh{\'a}sz},
  {Dullemond}, {K{\'o}sp{\'a}l}, {van Boekel}, {Bouwman}, {Henning},
  {Mo{\'o}r}, {Mosoni}, {Sicilia-Aguilar}, \& {Sipos}}]{2009Natur.459..224A}
{{\'A}brah{\'a}m}, P., {Juh{\'a}sz}, A., {Dullemond}, C.~P., {et~al.} 2009,
  \nat, 459, 224, \dodoi{10.1038/nature08004}

\bibitem[{{Alcal{\'a}} {et~al.}(2017){Alcal{\'a}}, {Manara}, {Natta}, {Frasca},
  {Testi}, {Nisini}, {Stelzer}, {Williams}, {Antoniucci}, {Biazzo}, {Covino},
  {Esposito}, {Getman}, \& {Rigliaco}}]{2017A&A...600A..20A}
{Alcal{\'a}}, J.~M., {Manara}, C.~F., {Natta}, A., {et~al.} 2017, \aap, 600,
  A20, \dodoi{10.1051/0004-6361/201629929}

\bibitem[{{Ambikasaran} {et~al.}(2015){Ambikasaran}, {Foreman-Mackey},
  {Greengard}, {Hogg}, \& {O'Neil}}]{2015ITPAM..38..252A}
{Ambikasaran}, S., {Foreman-Mackey}, D., {Greengard}, L., {Hogg}, D.~W., \&
  {O'Neil}, M. 2015, IEEE Transactions on Pattern Analysis and Machine
  Intelligence, 38, 252, \dodoi{10.1109/TPAMI.2015.2448083}

\bibitem[{Aspin {et~al.}(2010)Aspin, Reipurth, Herczeg, \& Capak}]{Aspin_2010}
Aspin, C., Reipurth, B., Herczeg, G.~J., \& Capak, P. 2010, The Astrophysical
  Journal Letters, 719, L50, \dodoi{10.1088/2041-8205/719/1/L50}

\bibitem[{{Audard} {et~al.}(2014){Audard}, {{\'A}brah{\'a}m}, {Dunham},
  {Green}, {Grosso}, {Hamaguchi}, {Kastner}, {K{\'o}sp{\'a}l}, {Lodato},
  {Romanova}, {Skinner}, {Vorobyov}, \& {Zhu}}]{2014prpl.conf..387A}
{Audard}, M., {{\'A}brah{\'a}m}, P., {Dunham}, M.~M., {et~al.} 2014, in
  Protostars and Planets VI, ed. H.~{Beuther}, R.~S. {Klessen}, C.~P.
  {Dullemond}, \& T.~{Henning}, 387--410,
  \dodoi{10.2458/azu_uapress_9780816531240-ch017}

\bibitem[{{Bachetti} {et~al.}(2010){Bachetti}, {Romanova}, {Kulkarni},
  {Burderi}, \& {di Salvo}}]{2010MNRAS.403.1193B}
{Bachetti}, M., {Romanova}, M.~M., {Kulkarni}, A., {Burderi}, L., \& {di
  Salvo}, T. 2010, \mnras, 403, 1193, \dodoi{10.1111/j.1365-2966.2010.16203.x}

\bibitem[{{Bailer-Jones} {et~al.}(2021){Bailer-Jones}, {Rybizki}, {Fouesneau},
  {Demleitner}, \& {Andrae}}]{2021AJ....161..147B}
{Bailer-Jones}, C.~A.~L., {Rybizki}, J., {Fouesneau}, M., {Demleitner}, M., \&
  {Andrae}, R. 2021, \aj, 161, 147, \dodoi{10.3847/1538-3881/abd806}

\bibitem[{{Batalha} {et~al.}(1996){Batalha}, {Stout-Batalha}, {Basri}, \&
  {Terra}}]{1996ApJS..103..211B}
{Batalha}, C.~C., {Stout-Batalha}, N.~M., {Basri}, G., \& {Terra}, M.~A.~O.
  1996, \apjs, 103, 211, \dodoi{10.1086/192275}

\bibitem[{{Bateson} {et~al.}(1991){Bateson}, {McIntosh}, \&
  {Brunt}}]{1991PVSS...16...49B}
{Bateson}, F.~M., {McIntosh}, R., \& {Brunt}, D. 1991, Royal Astronomical
  Society of New Zealand Publications of Variable Star Section, 16, 49

\bibitem[{{Bessell} {et~al.}(1998){Bessell}, {Castelli}, \&
  {Plez}}]{1998A&A...333..231B}
{Bessell}, M.~S., {Castelli}, F., \& {Plez}, B. 1998, \aap, 333, 231

\bibitem[{{Blinova} {et~al.}(2016){Blinova}, {Romanova}, \&
  {Lovelace}}]{2016MNRAS.459.2354B}
{Blinova}, A.~A., {Romanova}, M.~M., \& {Lovelace}, R.~V.~E. 2016, \mnras, 459,
  2354, \dodoi{10.1093/mnras/stw786}

\bibitem[{{Bohlin} {et~al.}(2014){Bohlin}, {Gordon}, \&
  {Tremblay}}]{2014PASP..126..711B}
{Bohlin}, R.~C., {Gordon}, K.~D., \& {Tremblay}, P.~E. 2014, \pasp, 126, 711,
  \dodoi{10.1086/677655}

\bibitem[{{Bouvier} {et~al.}(2007){Bouvier}, {Alencar}, {Harries},
  {Johns-Krull}, \& {Romanova}}]{2007prpl.conf..479B}
{Bouvier}, J., {Alencar}, S.~H.~P., {Harries}, T.~J., {Johns-Krull}, C.~M., \&
  {Romanova}, M.~M. 2007, in Protostars and Planets V, ed. B.~{Reipurth},
  D.~{Jewitt}, \& K.~{Keil}, 479, \dodoi{10.48550/arXiv.astro-ph/0603498}

\bibitem[{{Bramall} {et~al.}(2010){Bramall}, {Sharples}, {Tyas}, {Schmoll},
  {Clark}, {Luke}, {Looker}, {Dipper}, {Ryan}, {Buckley}, {Brink}, \&
  {Barnes}}]{2010SPIE.7735E..4FB}
{Bramall}, D.~G., {Sharples}, R., {Tyas}, L., {et~al.} 2010, in Society of
  Photo-Optical Instrumentation Engineers (SPIE) Conference Series, Vol. 7735,
  Ground-based and Airborne Instrumentation for Astronomy III, ed. I.~S.
  {McLean}, S.~K. {Ramsay}, \& H.~{Takami}, 77354F, \dodoi{10.1117/12.856382}

\bibitem[{Buckley(2017)}]{buckley_2017}
Buckley, D. A.~H. 2017, Proceedings of the International Astronomical Union,
  14, 176–180, \dodoi{10.1017/S174392131800251X}

\bibitem[{{Buckley} {et~al.}(2006){Buckley}, {Swart}, \&
  {Meiring}}]{2006SPIE.6267E..0ZB}
{Buckley}, D. A.~H., {Swart}, G.~P., \& {Meiring}, J.~G. 2006, in Society of
  Photo-Optical Instrumentation Engineers (SPIE) Conference Series, Vol. 6267,
  Society of Photo-Optical Instrumentation Engineers (SPIE) Conference Series,
  ed. L.~M. {Stepp}, 62670Z, \dodoi{10.1117/12.673750}

\bibitem[{{Burdonov} {et~al.}(2022){Burdonov}, {Yao}, {Sladkov}, {Bonito},
  {Chen}, {Ciardi}, {Korzhimanov}, {Soloviev}, {Starodubtsev}, {Zemskov},
  {Orlando}, {Romanova}, \& {Fuchs}}]{2022A&A...657A.112B}
{Burdonov}, K., {Yao}, W., {Sladkov}, A., {et~al.} 2022, \aap, 657, A112,
  \dodoi{10.1051/0004-6361/202140997}

\bibitem[{{Caldwell} {et~al.}(2020){Caldwell}, {Tenenbaum}, {Twicken},
  {Jenkins}, {Ting}, {Smith}, {Hedges}, {Fausnaugh}, {Rose}, \&
  {Burke}}]{2020RNAAS...4..201C}
{Caldwell}, D.~A., {Tenenbaum}, P., {Twicken}, J.~D., {et~al.} 2020, Research
  Notes of the American Astronomical Society, 4, 201,
  \dodoi{10.3847/2515-5172/abc9b3}

\bibitem[{{Campbell-White} {et~al.}(2021){Campbell-White}, {Sicilia-Aguilar},
  {Manara}, {Matsumura}, {Fang}, {Frasca}, \&
  {Roccatagliata}}]{2021MNRAS.507.3331C}
{Campbell-White}, J., {Sicilia-Aguilar}, A., {Manara}, C.~F., {et~al.} 2021,
  \mnras, 507, 3331, \dodoi{10.1093/mnras/stab2300}

\bibitem[{{Cardelli} {et~al.}(1989){Cardelli}, {Clayton}, \&
  {Mathis}}]{1989ApJ...345..245C}
{Cardelli}, J.~A., {Clayton}, G.~C., \& {Mathis}, J.~S. 1989, \apj, 345, 245,
  \dodoi{10.1086/167900}

\bibitem[{{Cieza} {et~al.}(2016){Cieza}, {Casassus}, {Tobin}, {Bos},
  {Williams}, {Perez}, {Zhu}, {Caceres}, {Canovas}, {Dunham}, {Hales},
  {Prieto}, {Principe}, {Schreiber}, {Ruiz-Rodriguez}, \&
  {Zurlo}}]{2016Natur.535..258C}
{Cieza}, L.~A., {Casassus}, S., {Tobin}, J., {et~al.} 2016, \nat, 535, 258,
  \dodoi{10.1038/nature18612}

\bibitem[{{Comer{\'o}n}(2008)}]{2008hsf2.book..295C}
{Comer{\'o}n}, F. 2008, in Handbook of Star Forming Regions, Volume II, ed.
  B.~{Reipurth}, Vol.~5, 295

\bibitem[{{Crause} {et~al.}(2014){Crause}, {Sharples}, {Bramall}, {Schmoll},
  {Clark}, {Younger}, {Tyas}, {Ryan}, {Brink}, {Strydom}, {Buckley},
  {Wilkinson}, {Crawford}, \& {Depagne}}]{2014SPIE.9147E..6TC}
{Crause}, L.~A., {Sharples}, R.~M., {Bramall}, D.~G., {et~al.} 2014, in Society
  of Photo-Optical Instrumentation Engineers (SPIE) Conference Series, Vol.
  9147, Ground-based and Airborne Instrumentation for Astronomy V, ed. S.~K.
  {Ramsay}, I.~S. {McLean}, \& H.~{Takami}, 91476T, \dodoi{10.1117/12.2055635}

\bibitem[{{Cruz-S{\'a}enz de Miera} {et~al.}(2023){Cruz-S{\'a}enz de Miera},
  {K{\'o}sp{\'a}l}, {Abrah{\'a}m}, {Claes}, {Manara}, {Wendeborn},
  {Fiorellino}, {Giannini}, {Nisini}, {Sicilia-Aguilar}, {Campbell-White},
  {Alcal{\'a}}, {Banzatti}, {Szab{\'o}}, {Lykou}, {Antoniucci}, {Varga},
  {Siwak}, {Park}, {Nagy}, \& {Kun}}]{2023A&A...678A..88C}
{Cruz-S{\'a}enz de Miera}, F., {K{\'o}sp{\'a}l}, {\'A}., {Abrah{\'a}m}, P.,
  {et~al.} 2023, \aap, 678, A88, \dodoi{10.1051/0004-6361/202347063}

\bibitem[{{Donati} {et~al.}(1999){Donati}, {Collier Cameron}, {Hussain}, \&
  {Semel}}]{1999MNRAS.302..437D}
{Donati}, J.~F., {Collier Cameron}, A., {Hussain}, G.~A.~J., \& {Semel}, M.
  1999, \mnras, 302, 437, \dodoi{10.1046/j.1365-8711.1999.02095.x}

\bibitem[{{Dullemond} {et~al.}(2019){Dullemond}, {K{\"u}ffmeier}, {Goicovic},
  {Fukagawa}, {Oehl}, \& {Kramer}}]{2019A&A...628A..20D}
{Dullemond}, C.~P., {K{\"u}ffmeier}, M., {Goicovic}, F., {et~al.} 2019, \aap,
  628, A20, \dodoi{10.1051/0004-6361/201832632}

\bibitem[{{Espaillat} {et~al.}(2021){Espaillat}, {Robinson}, {Romanova},
  {Thanathibodee}, {Wendeborn}, {Calvet}, {Reynolds}, \&
  {Muzerolle}}]{2021Natur.597...41E}
{Espaillat}, C.~C., {Robinson}, C.~E., {Romanova}, M.~M., {et~al.} 2021, \nat,
  597, 41, \dodoi{10.1038/s41586-021-03751-5}

\bibitem[{{Finociety} {et~al.}(2023){Finociety}, {Donati}, {Cristofari},
  {Moutou}, {Cadieux}, {Cook}, {Artigau}, {Baruteau}, {Debras}, {Fouqu{\'e}},
  {Bouvier}, {Alencar}, {Delfosse}, {Grankin}, {Carmona}, {Petit},
  {K{\'o}sp{\'a}l}, \& {The Sls/Spice Consortium}}]{2023MNRAS.526.4627F}
{Finociety}, B., {Donati}, J.~F., {Cristofari}, P.~I., {et~al.} 2023, \mnras,
  526, 4627, \dodoi{10.1093/mnras/stad3012}

\bibitem[{{Ghosh} \& {Lamb}(1978)}]{1978ApJ...223L..83G}
{Ghosh}, P., \& {Lamb}, F.~K. 1978, \apjl, 223, L83, \dodoi{10.1086/182734}

\bibitem[{{Goto} {et~al.}(2011){Goto}, {Reg{\'a}ly}, {Dullemond}, {van den
  Ancker}, {Brown}, {Carmona}, {Pontoppidan}, {{\'A}brah{\'a}m}, {Blake},
  {Fedele}, {Henning}, {Juh{\'a}sz}, {K{\'o}sp{\'a}l}, {Mosoni},
  {Sicilia-Aguilar}, {Terada}, {van Boekel}, {van Dishoeck}, \&
  {Usuda}}]{2011ApJ...728....5G}
{Goto}, M., {Reg{\'a}ly}, Z., {Dullemond}, C.~P., {et~al.} 2011, \apj, 728, 5,
  \dodoi{10.1088/0004-637X/728/1/5}

\bibitem[{{Gras-Vel{\'a}zquez} \& {Ray}(2005)}]{2005A&A...443..541G}
{Gras-Vel{\'a}zquez}, {\`A}., \& {Ray}, T.~P. 2005, \aap, 443, 541,
  \dodoi{10.1051/0004-6361:20042397}

\bibitem[{{Gregory} {et~al.}(2014){Gregory}, {Holzwarth}, {Donati}, {Hussain},
  {Montmerle}, {Alecian}, {Alencar}, {Argiroffi}, {Audard}, {Bouvier},
  {Damiani}, {G{\"u}del}, {Huenemoerder}, {Kastner}, {Maggio}, {Sacco}, \&
  {Wade}}]{2014IAUS..302...44G}
{Gregory}, S.~G., {Holzwarth}, V.~R., {Donati}, J.~F., {et~al.} 2014, in
  Magnetic Fields throughout Stellar Evolution, ed. P.~{Petit}, M.~{Jardine},
  \& H.~C. {Spruit}, Vol. 302, 44--45, \dodoi{10.1017/S1743921314001689}

\bibitem[{{Gullbring} {et~al.}(1996){Gullbring}, {Barwig}, {Chen}, {Gahm}, \&
  {Bao}}]{1996A&A...307..791G}
{Gullbring}, E., {Barwig}, H., {Chen}, P.~S., {Gahm}, G.~F., \& {Bao}, M.~X.
  1996, \aap, 307, 791

\bibitem[{{Hales} {et~al.}(2018){Hales}, {P{\'e}rez}, {Saito}, {Pinte}, {Knee},
  {de Gregorio-Monsalvo}, {Dent}, {L{\'o}pez}, {Plunkett}, {Cort{\'e}s},
  {Corder}, \& {Cieza}}]{2018ApJ...859..111H}
{Hales}, A.~S., {P{\'e}rez}, S., {Saito}, M., {et~al.} 2018, \apj, 859, 111,
  \dodoi{10.3847/1538-4357/aac018}

\bibitem[{{Hamann} \& {Persson}(1992)}]{1992ApJS...82..247H}
{Hamann}, F., \& {Persson}, S.~E. 1992, \apjs, 82, 247, \dodoi{10.1086/191715}

\bibitem[{{Hambsch}(2012)}]{2012JAVSO..40.1003H}
{Hambsch}, F.~J. 2012, \jaavso, 40, 1003

\bibitem[{{Hartmann} {et~al.}(2016){Hartmann}, {Herczeg}, \&
  {Calvet}}]{2016ARA&A..54..135H}
{Hartmann}, L., {Herczeg}, G., \& {Calvet}, N. 2016, \araa, 54, 135,
  \dodoi{10.1146/annurev-astro-081915-023347}

\bibitem[{{Herbig}(1977)}]{1977ApJ...217..693H}
{Herbig}, G.~H. 1977, \apj, 217, 693, \dodoi{10.1086/155615}

\bibitem[{{Herbig}(2007)}]{2007AJ....133.2679H}
---. 2007, \aj, 133, 2679, \dodoi{10.1086/517494}

\bibitem[{{Herbig} {et~al.}(2001){Herbig}, {Aspin}, {Gilmore}, {Imhoff}, \&
  {Jones}}]{2001PASP..113.1547H}
{Herbig}, G.~H., {Aspin}, C., {Gilmore}, A.~C., {Imhoff}, C.~L., \& {Jones},
  A.~F. 2001, \pasp, 113, 1547, \dodoi{10.1086/324420}

\bibitem[{{Herbst} {et~al.}(1994){Herbst}, {Herbst}, {Grossman}, \&
  {Weinstein}}]{1994AJ....108.1906H}
{Herbst}, W., {Herbst}, D.~K., {Grossman}, E.~J., \& {Weinstein}, D. 1994, \aj,
  108, 1906, \dodoi{10.1086/117204}

\bibitem[{{Johns-Krull} {et~al.}(2022){Johns-Krull}, {Prato}, {Stahl}, {Tang},
  {Llama}, {Jaffe}, \& {Mace}}]{2022AAS...24040613J}
{Johns-Krull}, C., {Prato}, L., {Stahl}, A., {et~al.} 2022, in American
  Astronomical Society Meeting Abstracts, Vol.~54, American Astronomical
  Society Meeting Abstracts, 406.13

\bibitem[{{Johns-Krull} {et~al.}(1999){Johns-Krull}, {Valenti}, {Hatzes}, \&
  {Kanaan}}]{1999ApJ...510L..41J}
{Johns-Krull}, C.~M., {Valenti}, J.~A., {Hatzes}, A.~P., \& {Kanaan}, A. 1999,
  \apjl, 510, L41, \dodoi{10.1086/311802}

\bibitem[{{Johns-Krull} {et~al.}(2001){Johns-Krull}, {Valenti}, {Piskunov},
  {Saar}, \& {Hatzes}}]{2001ASPC..248..527J}
{Johns-Krull}, C.~M., {Valenti}, J.~A., {Piskunov}, N.~E., {Saar}, S.~H., \&
  {Hatzes}, A.~P. 2001, in Astronomical Society of the Pacific Conference
  Series, Vol. 248, Magnetic Fields Across the Hertzsprung-Russell Diagram, ed.
  G.~{Mathys}, S.~K. {Solanki}, \& D.~T. {Wickramasinghe}, 527

\bibitem[{{Johnstone} {et~al.}(2014){Johnstone}, {Jardine}, {Gregory},
  {Donati}, \& {Hussain}}]{2014MNRAS.437.3202J}
{Johnstone}, C.~P., {Jardine}, M., {Gregory}, S.~G., {Donati}, J.~F., \&
  {Hussain}, G. 2014, \mnras, 437, 3202, \dodoi{10.1093/mnras/stt2107}

\bibitem[{{Kanodia} \& {Wright}(2018)}]{2018RNAAS...2....4K}
{Kanodia}, S., \& {Wright}, J. 2018, Research Notes of the American
  Astronomical Society, 2, 4, \dodoi{10.3847/2515-5172/aaa4b7}

\bibitem[{{Kniazev} {et~al.}(2016{\natexlab{a}}){Kniazev}, {Gvaramadze}, \&
  {Berdnikov}}]{2016MNRAS.459.3068K}
{Kniazev}, A.~Y., {Gvaramadze}, V.~V., \& {Berdnikov}, L.~N.
  2016{\natexlab{a}}, \mnras, 459, 3068, \dodoi{10.1093/mnras/stw889}

\bibitem[{{Kniazev} {et~al.}(2016{\natexlab{b}}){Kniazev}, {Gvaramadze}, \&
  {Berdnikov}}]{2016arXiv161200292K}
---. 2016{\natexlab{b}}, arXiv e-prints, arXiv:1612.00292,
  \dodoi{10.48550/arXiv.1612.00292}

\bibitem[{{Koenigl}(1991)}]{1991ApJ...370L..39K}
{Koenigl}, A. 1991, \apjl, 370, L39, \dodoi{10.1086/185972}

\bibitem[{{K{\'o}sp{\'a}l} {et~al.}(2020){K{\'o}sp{\'a}l}, {Donati}, {Bouvier},
  \& {{\'A}brah{\'a}m}}]{2020IAUGA..30..125K}
{K{\'o}sp{\'a}l}, {\'A}., {Donati}, J.~F., {Bouvier}, J., \& {{\'A}brah{\'a}m},
  P. 2020, in IAU General Assembly, 125--125, \dodoi{10.1017/S1743921319003788}

\bibitem[{{K{\'o}sp{\'a}l} {et~al.}(2022){K{\'o}sp{\'a}l}, {Fiorellino},
  {{\'A}brah{\'a}m}, {Giannini}, \& {Nisini}}]{2022RNAAS...6...52K}
{K{\'o}sp{\'a}l}, {\'A}., {Fiorellino}, E., {{\'A}brah{\'a}m}, P., {Giannini},
  T., \& {Nisini}, B. 2022, Research Notes of the American Astronomical
  Society, 6, 52, \dodoi{10.3847/2515-5172/ac5e3c}

\bibitem[{{K{\'o}sp{\'a}l} {et~al.}(2014){K{\'o}sp{\'a}l}, {Mohler-Fischer},
  {Sicilia-Aguilar}, {{\'A}brah{\'a}m}, {Cur{\'e}}, {Henning}, {Kiss},
  {Launhardt}, {Mo{\'o}r}, \& {M{\"u}ller}}]{2014A&A...561A..61K}
{K{\'o}sp{\'a}l}, {\'A}., {Mohler-Fischer}, M., {Sicilia-Aguilar}, A., {et~al.}
  2014, \aap, 561, A61, \dodoi{10.1051/0004-6361/201322428}

\bibitem[{Kramida {et~al.}(2022)Kramida, {Yu.~Ralchenko}, Reader, \& {and NIST
  ASD Team}}]{NIST_ASD}
Kramida, A., {Yu.~Ralchenko}, Reader, J., \& {and NIST ASD Team}. 2022, {NIST
  Atomic Spectra Database (ver. 5.10), [Online]. Available:
  {\tt{https://physics.nist.gov/asd}} [2023, August 21]. National Institute of
  Standards and Technology, Gaithersburg, MD.}

\bibitem[{{Kulkarni} \& {Romanova}(2008)}]{2008MNRAS.386..673K}
{Kulkarni}, A.~K., \& {Romanova}, M.~M. 2008, \mnras, 386, 673,
  \dodoi{10.1111/j.1365-2966.2008.13094.x}

\bibitem[{{Kulkarni} \& {Romanova}(2009)}]{2009MNRAS.398..701K}
---. 2009, \mnras, 398, 701, \dodoi{10.1111/j.1365-2966.2009.15186.x}

\bibitem[{Kulkarni \& Romanova(2013)}]{10.1093/mnras/stt945}
Kulkarni, A.~K., \& Romanova, M.~M. 2013, Monthly Notices of the Royal
  Astronomical Society, 433, 3048, \dodoi{10.1093/mnras/stt945}

\bibitem[{{Kurosawa} \& {Romanova}(2013)}]{2013MNRAS.431.2673K}
{Kurosawa}, R., \& {Romanova}, M.~M. 2013, \mnras, 431, 2673,
  \dodoi{10.1093/mnras/stt365}

\bibitem[{{Lamzin}(1998)}]{1998ARep...42..322L}
{Lamzin}, S.~A. 1998, Astronomy Reports, 42, 322,
  \dodoi{10.48550/arXiv.1303.4066}

\bibitem[{{Long} {et~al.}(2011){Long}, {Romanova}, {Kulkarni}, \&
  {Donati}}]{2011MNRAS.413.1061L}
{Long}, M., {Romanova}, M.~M., {Kulkarni}, A.~K., \& {Donati}, J.~F. 2011,
  \mnras, 413, 1061, \dodoi{10.1111/j.1365-2966.2010.18193.x}

\bibitem[{{Long} {et~al.}(2012){Long}, {Romanova}, \&
  {Lamb}}]{2012NewA...17..232L}
{Long}, M., {Romanova}, M.~M., \& {Lamb}, F.~K. 2012, \na, 17, 232,
  \dodoi{10.1016/j.newast.2011.08.001}

\bibitem[{{Long} {et~al.}(2005){Long}, {Romanova}, \&
  {Lovelace}}]{2005ApJ...634.1214L}
{Long}, M., {Romanova}, M.~M., \& {Lovelace}, R.~V.~E. 2005, \apj, 634, 1214,
  \dodoi{10.1086/497000}

\bibitem[{{Lorenzetti} {et~al.}(2009){Lorenzetti}, {Larionov}, {Giannini},
  {Arkharov}, {Antoniucci}, {Nisini}, \& {Di Paola}}]{2009ApJ...693.1056L}
{Lorenzetti}, D., {Larionov}, V.~M., {Giannini}, T., {et~al.} 2009, \apj, 693,
  1056, \dodoi{10.1088/0004-637X/693/2/1056}

\bibitem[{{Martin} {et~al.}(2019){Martin}, {Reichart}, {Dutton}, {Maples},
  {Berger}, {Ghigo}, {Haislip}, {Shaban}, {Trotter}, {Barnes}, {Paggen}, {Gao},
  {Salemi}, {Langston}, {Bussa}, {Duncan}, {White}, {Heatherly}, {Karlik},
  {Johnson}, {Reichart}, {Foster}, {Kouprianov}, {Mazlin}, \&
  {Harvey}}]{2019ApJS..240...12M}
{Martin}, J.~R., {Reichart}, D.~E., {Dutton}, D.~A., {et~al.} 2019, \apjs, 240,
  12, \dodoi{10.3847/1538-4365/aad7c1}

\bibitem[{{McGinnis} {et~al.}(2020){McGinnis}, {Bouvier}, \&
  {Gallet}}]{2020MNRAS.497.2142M}
{McGinnis}, P., {Bouvier}, J., \& {Gallet}, F. 2020, \mnras, 497, 2142,
  \dodoi{10.1093/mnras/staa2041}

\bibitem[{{McLaughlin}(1946)}]{1946AJ.....52..109M}
{McLaughlin}, D.~B. 1946, \aj, 52, 109, \dodoi{10.1086/105935}

\bibitem[{{Monson} {et~al.}(2017){Monson}, {Beaton}, {Scowcroft}, {Freedman},
  {Madore}, {Rich}, {Seibert}, {Kollmeier}, \&
  {Clementini}}]{2017AJ....153...96M}
{Monson}, A.~J., {Beaton}, R.~L., {Scowcroft}, V., {et~al.} 2017, \aj, 153, 96,
  \dodoi{10.3847/1538-3881/153/3/96}

\bibitem[{{Nardiello} {et~al.}(2019){Nardiello}, {Borsato}, {Piotto},
  {Colombo}, {Manthopoulou}, {Bedin}, {Granata}, {Lacedelli}, {Libralato},
  {Malavolta}, {Montalto}, \& {Nascimbeni}}]{2019MNRAS.490.3806N}
{Nardiello}, D., {Borsato}, L., {Piotto}, G., {et~al.} 2019, \mnras, 490, 3806,
  \dodoi{10.1093/mnras/stz2878}

\bibitem[{{Newville} {et~al.}(2016){Newville}, {Stensitzki}, {Allen}, {Rawlik},
  {Ingargiola}, \& {Nelson}}]{2016ascl.soft06014N}
{Newville}, M., {Stensitzki}, T., {Allen}, D.~B., {et~al.} 2016, {Lmfit:
  Non-Linear Least-Square Minimization and Curve-Fitting for Python},
  Astrophysics Source Code Library, record ascl:1606.014.
\newblock \doeprint{1606.014}

\bibitem[{{Romanova} {et~al.}(2021){Romanova}, {Koldoba}, {Ustyugova},
  {Blinova}, {Lai}, \& {Lovelace}}]{2021MNRAS.506..372R}
{Romanova}, M.~M., {Koldoba}, A.~V., {Ustyugova}, G.~V., {et~al.} 2021, \mnras,
  506, 372, \dodoi{10.1093/mnras/stab1724}

\bibitem[{{Romanova} \& {Owocki}(2015)}]{2015SSRv..191..339R}
{Romanova}, M.~M., \& {Owocki}, S.~P. 2015, \ssr, 191, 339,
  \dodoi{10.1007/s11214-015-0200-9}

\bibitem[{{Romanova} {et~al.}(2002){Romanova}, {Ustyugova}, {Koldoba}, \&
  {Lovelace}}]{2002ApJ...578..420R}
{Romanova}, M.~M., {Ustyugova}, G.~V., {Koldoba}, A.~V., \& {Lovelace},
  R.~V.~E. 2002, \apj, 578, 420, \dodoi{10.1086/342464}

\bibitem[{{Romanova} {et~al.}(2004){Romanova}, {Ustyugova}, {Koldoba}, \&
  {Lovelace}}]{2004ApJ...610..920R}
---. 2004, \apj, 610, 920, \dodoi{10.1086/421867}

\bibitem[{{Romanova} {et~al.}(2003){Romanova}, {Ustyugova}, {Koldoba}, {Wick},
  \& {Lovelace}}]{2003ApJ...595.1009R}
{Romanova}, M.~M., {Ustyugova}, G.~V., {Koldoba}, A.~V., {Wick}, J.~V., \&
  {Lovelace}, R.~V.~E. 2003, \apj, 595, 1009, \dodoi{10.1086/377514}

\bibitem[{{Shappee} {et~al.}(2014){Shappee}, {Prieto}, {Grupe}, {Kochanek},
  {Stanek}, {De Rosa}, {Mathur}, {Zu}, {Peterson}, {Pogge}, {Komossa}, {Im},
  {Jencson}, {Holoien}, {Basu}, {Beacom}, {Szczygie{\l}}, {Brimacombe},
  {Adams}, {Campillay}, {Choi}, {Contreras}, {Dietrich}, {Dubberley},
  {Elphick}, {Foale}, {Giustini}, {Gonzalez}, {Hawkins}, {Howell}, {Hsiao},
  {Koss}, {Leighly}, {Morrell}, {Mudd}, {Mullins}, {Nugent}, {Parrent},
  {Phillips}, {Pojmanski}, {Rosing}, {Ross}, {Sand}, {Terndrup}, {Valenti},
  {Walker}, \& {Yoon}}]{2014ApJ...788...48S}
{Shappee}, B.~J., {Prieto}, J.~L., {Grupe}, D., {et~al.} 2014, \apj, 788, 48,
  \dodoi{10.1088/0004-637X/788/1/48}

\bibitem[{{Sicilia-Aguilar} {et~al.}(2015){Sicilia-Aguilar}, {Fang},
  {Roccatagliata}, {Collier Cameron}, {K{\'o}sp{\'a}l}, {Henning},
  {{\'A}brah{\'a}m}, \& {Sipos}}]{2015A&A...580A..82S}
{Sicilia-Aguilar}, A., {Fang}, M., {Roccatagliata}, V., {et~al.} 2015, \aap,
  580, A82, \dodoi{10.1051/0004-6361/201525970}

\bibitem[{{Sicilia-Aguilar} {et~al.}(2012){Sicilia-Aguilar}, {K{\'o}sp{\'a}l},
  {Setiawan}, {{\'A}brah{\'a}m}, {Dullemond}, {Eiroa}, {Goto}, {Henning}, \&
  {Juh{\'a}sz}}]{2012A&A...544A..93S}
{Sicilia-Aguilar}, A., {K{\'o}sp{\'a}l}, {\'A}., {Setiawan}, J., {et~al.} 2012,
  \aap, 544, A93, \dodoi{10.1051/0004-6361/201118555}

\bibitem[{{Sicilia-Aguilar} {et~al.}(2023){Sicilia-Aguilar}, {Campbell-White},
  {Roccatagliata}, {Desira}, {Gregory}, {Scholz}, {Fang}, {Cruz-Saenz de
  Miera}, {K{\'o}sp{\'a}l}, {Matsumura}, \&
  {{\'A}brah{\'a}m}}]{2023MNRAS.526.4885S}
{Sicilia-Aguilar}, A., {Campbell-White}, J., {Roccatagliata}, V., {et~al.}
  2023, \mnras, 526, 4885, \dodoi{10.1093/mnras/stad3029}

\bibitem[{{Sipos} {et~al.}(2009){Sipos}, {{\'A}brah{\'a}m}, {Acosta-Pulido},
  {Juh{\'a}sz}, {K{\'o}sp{\'a}l}, {Kun}, {Mo{\'o}r}, \&
  {Setiawan}}]{2009A&A...507..881S}
{Sipos}, N., {{\'A}brah{\'a}m}, P., {Acosta-Pulido}, J., {et~al.} 2009, \aap,
  507, 881, \dodoi{10.1051/0004-6361/200911641}

\bibitem[{{Spruit} {et~al.}(1995){Spruit}, {Stehle}, \&
  {Papaloizou}}]{1995MNRAS.275.1223S}
{Spruit}, H.~C., {Stehle}, R., \& {Papaloizou}, J.~C.~B. 1995, \mnras, 275,
  1223, \dodoi{10.1093/mnras/275.4.1223}

\bibitem[{{Takasao} {et~al.}(2022){Takasao}, {Tomida}, {Iwasaki}, \&
  {Suzuki}}]{2022ApJ...941...73T}
{Takasao}, S., {Tomida}, K., {Iwasaki}, K., \& {Suzuki}, T.~K. 2022, \apj, 941,
  73, \dodoi{10.3847/1538-4357/ac9eb1}

\bibitem[{{Thies} {et~al.}(2011){Thies}, {Kroupa}, {Goodwin}, {Stamatellos}, \&
  {Whitworth}}]{2011MNRAS.417.1817T}
{Thies}, I., {Kroupa}, P., {Goodwin}, S.~P., {Stamatellos}, D., \& {Whitworth},
  A.~P. 2011, \mnras, 417, 1817, \dodoi{10.1111/j.1365-2966.2011.19390.x}

\bibitem[{{Wang} {et~al.}(2023){Wang}, {Herczeg}, {Liu}, {Fang}, {Johnstone},
  {Lee}, {Walter}, {Hambsch}, {Contreras Pe{\~n}a}, {Lee}, {Millward},
  {Pearce}, {Monard}, \& {Zhou}}]{2023ApJ...957..113W}
{Wang}, M.-T., {Herczeg}, G.~J., {Liu}, H.-G., {et~al.} 2023, \apj, 957, 113,
  \dodoi{10.3847/1538-4357/acf2f4}

\bibitem[{{White} {et~al.}(2020){White}, {K{\'o}sp{\'a}l}, {Hughes},
  {{\'A}brah{\'a}m}, {Akimkin}, {Banzatti}, {Chen}, {Cruz-S{\'a}enz de Miera},
  {Dutrey}, {Flock}, {Guilloteau}, {Hales}, {Henning}, {Kadam}, {Semenov},
  {Sicilia-Aguilar}, {Teague}, \& {Vorobyov}}]{2020ApJ...904...37W}
{White}, J.~A., {K{\'o}sp{\'a}l}, {\'A}., {Hughes}, A.~G., {et~al.} 2020, \apj,
  904, 37, \dodoi{10.3847/1538-4357/abbb94}

\bibitem[{{Zanni} \& {Ferreira}(2009)}]{2009A&A...508.1117Z}
{Zanni}, C., \& {Ferreira}, J. 2009, \aap, 508, 1117,
  \dodoi{10.1051/0004-6361/200912879}

\bibitem[{{Zhou} \& {Herczeg}(2022)}]{2022ATel15271....1Z}
{Zhou}, L., \& {Herczeg}, G.~J. 2022, The Astronomer's Telegram, 15271, 1

\end{thebibliography}



\end{document}